\documentclass[journal]{IEEEtran}

\ifCLASSINFOpdf
\else
\fi

\usepackage{setspace}
\usepackage{bbm}
\usepackage[cmex10]{amsmath}
\usepackage{amssymb}
\usepackage{cite}
\usepackage{graphicx}
\usepackage{array,color}
\usepackage{amsmath}
\allowdisplaybreaks
\usepackage{stfloats}
\usepackage{graphicx}
\usepackage{epstopdf}
\usepackage{subfigure}
\usepackage{tabularx}
\usepackage{epsfig,epsf,color,balance,cite}
\usepackage{algorithmic}
\usepackage{algorithm}
\usepackage{url}
\usepackage{bm}
\usepackage{multirow}

\usepackage{amsthm}

\ifodd 1
\usepackage{soul}
\usepackage{color}
\setstcolor{red}

\newcommand{\del}[1]{\st{#1}} 

\newcommand{\com}[1]{\textbf{ (COMMENT: #1)}} 
\newcommand{\response}[1]{\textbf{\color{green} (RESPONSE: #1)}} 
\else

\newcommand{\del}[1]{}

\newcommand{\com}[1]{}
\newcommand{\comg}[1]{}
\newcommand{\response}[1]{}
\fi

\title{\LARGE {A Survey on Channel Estimation and Practical Passive Beamforming  Design for Intelligent Reflecting Surface Aided Wireless Communications}}

\author{ 
	Beixiong Zheng,~\IEEEmembership{Member,~IEEE}, Changsheng You,~\IEEEmembership{Member,~IEEE}, \\
	Weidong Mei,~\IEEEmembership{Member,~IEEE}, and Rui Zhang,~\IEEEmembership{Fellow,~IEEE} 
	
	\thanks{This work was supported in part by NUS Research Grants XXXX and XXXX. \emph{(Corresponding author: Rui Zhang.)}
		
		B. Zheng, W. Mei, and R. Zhang are with the Department of Electrical and Computer Engineering, National University of Singapore, Singapore 117583 (e-mail: \{elezbe, wmei, elezhang\}@nus.edu.sg).
		
		C. You is with the Department of Electrical and Electronic Engineering, Southern University of Science and Technology (SUSTech), Shenzhen 518055, China (e-mail: youcs@sustech.edu.cn). He was with the Department of Electrical and Computer Engineering, National University of Singapore, Singapore 117583.
	}
}

\begin{document}
\markboth{IEEE Communications Surveys \& Tutorials, vol.XX, No. XX, XXX 2022}{SKM: My IEEE article}
\maketitle
\vspace{-0.3cm}
\begin{abstract}
Intelligent reflecting surface (IRS) has emerged as a key enabling technology to realize smart and reconfigurable radio environment for wireless communications, by digitally controlling the signal reflection via a large number of passive reflecting elements in real time.
Different from conventional wireless communication techniques that only adapt to but have no or limited control over dynamic wireless channels, IRS provides a new and cost-effective means to combat the wireless channel impairments in a proactive manner.
However, despite its great potential, IRS faces new and unique challenges in its efficient integration into wireless communication systems, especially its channel estimation and passive beamforming design under various practical hardware constraints.
In this paper, we provide a comprehensive survey on the up-to-date research in IRS-aided wireless communications,
with an emphasis on the promising solutions to tackle practical design issues.
Furthermore, we discuss new and emerging IRS architectures and applications as well as their practical design problems to motivate future research.
\end{abstract}
\IEEEpeerreviewmaketitle

\section{Introduction}\label{Introduction}
\subsection{Motivation of IRS and Prior Work}\label{Motivation}
\IEEEPARstart{W}{hile} the fifth-generation (5G) wireless communication system is being rapidly rolled out and deployed globally, wireless communication researchers from both industry and academia have started planning ahead for the beyond-5G as well as the next-/sixth-generation (6G) wireless networks.
It is foreseen that with the drastic growth of mobile subscribers and wireless devices as well as the rapid emergence of new wireless applications such as augmented/mixed/virtual reality (AR/MR/VR), industrial automation, tactile Internet, etc., 5G may encounter both capacity and performance limitations in supporting the high demand for massive/ubiquitous connectivity in the forthcoming era of Internet-of-Everything (IoE), thus motivating intensive research on 6G technologies to meet such demands \cite{saad2019vision,letaief2019roadmap,zhang20196g}.
Recently, as an early initiative for 6G research, the International Telecommunication Union (ITU) has launched a focus group called Technologies for Network 2030 (FG NET-2030) to identify and study fundamental challenges, use cases, and key technologies to pave the way for 6G wireless communications \cite{network2030}.
It is generally known that the main 5G services \cite{andrews2014will,boccardi2014five,shafi20175g} include: 1) enhanced mobile broadband (eMBB) to provide high data rates up to 1 Gigabit per second (Gbps) for mobile users; 2) ultra-reliable low-latency communication (URLLC) to achieve the reliability no less than $99.999$\% at the millisecond (ms)-level latency; and 3) massive machine-type communication (mMTC) to simultaneously connect a large
number of devices (in the scale of $10^6$~{\rm devices/km}$^2$) in the Internet-of-Things (IoT) network.
As compared to 5G, 6G is anticipated to not only substantially boost the network performance in all the key aspects (e.g.,  data rate, latency, energy efficiency, reliability), but also undergo a fundamental paradigm shift from supporting connected people/things to enabling connected intelligence with the integrated functions of communication, computing, control, sensing, and learning.
Specifically, the representative key performance indicators (KPIs) advocated for 6G are summarized as follows \cite{saad2019vision,letaief2019roadmap,zhang20196g,tataria20216g,jiang2021road,lee20206g}:
\begin{itemize}
	\item {\bf Peak data rate}: $\ge 1$ Terabit per second (Tbps) for both indoor and outdoor connections (under ideal wireless propagation conditions), which is 100-1000 times that of 5G;
	\item {\bf User-experienced data rate}: $\ge$1 Gbps for downlink and $\ge$500 Mbps for uplink, which is about 10 times that of 5G;
	\item {\bf Bandwidth}: up to 10 GHz can be supported in millimeter-wave (mmWave) frequency bands, while up to 100 GHz can be reached in terahertz (THz) and visible light frequency bands;
	\item {\bf Energy efficiency}: 10-100 times that of 5G to achieve the green communication network and reduce the overall network energy consumption;
	\item {\bf Spectral efficiency}: 5 times that of 5G to utilize the available frequency spectrum more efficiently by adopting advanced multi-antenna and modulation techniques;
	\item {\bf Connection density}: $\approx 10^7$~{\rm devices/km}$^2$ to meet the high demand for massive connectivity in IoE and enhanced mMTC, which is about 10 times that of 5G;
	\item {\bf Reliability}: $\ge 99.999 99$\% to support more enhanced URLLC as compared to 5G;
	\item {\bf Air-interface latency}: $\le 100~{\rm\mu s}$ to support more enhanced URLLC than 5G and emerging IoE applications like AR/MR/VR;
	\item {\bf Positioning accuracy}: at centimeter (cm) level in three-dimensional (3D) space to meet the much stronger demand in various vertical and industrial applications, as compared to 5G with the required positioning accuracy at meter (m) level in two-dimensional (2D) space;
	\item {\bf Mobility management}: support high-mobility communications with the maximal speed of  $1,000$~{\rm km/h} for high-speed trains and even aircraft.
\end{itemize}

The main KPIs targeted for 5G and 6G are compared in Table \ref{KPI}.

\begin{table*}[]
	\centering
	\caption{Comparison of Main KPIs between 5G and 6G \cite{saad2019vision,letaief2019roadmap,zhang20196g,tataria20216g,jiang2021road,lee20206g}.}\label{KPI}
	\vspace{-0.3cm}
	\resizebox{\textwidth}{!}{
		\begin{tabular}{|m{0.7cm}<{\centering}|m{1.8cm}<{\centering}|m{3.7cm}<{\centering}|m{2cm}<{\centering}|m{2cm}<{\centering}|m{2cm}<{\centering}|m{2cm}<{\centering}|m{2cm}<{\centering}|m{2.5cm}<{\centering}|m{2cm}<{\centering}|m{2cm}<{\centering}|}
			\hline
			{\bf KPIs} & {\bf Peak data rate} & {\bf User-experienced data rate}                                                            & {\bf Maximum bandwidth} & {\bf Energy efficiency} & {\bf Spectral efficiency} & {\bf Connection density}                & {\bf Reliability}     & {\bf Air-interface latency} & {\bf Positioning accuracy} & {\bf Mobility}          \\ \hline\hline
			{\bf 5G} & $\ge 1$ Gbps   & \begin{tabular}[c]{@{}l@{}}Downlink: $\ge$0.1 Gbps\\ Uplink: $\ge$50 Mbps\end{tabular}                                                                                     & 1 GHz    &  10-100$\times$ over 4G                 & 3$\times$ over 4G                   & $10^6{\rm/km}^2$ & $99.999$\%   & $1~{\rm ms}$       & m-level (2D)             & $500~{\rm km/h}$ \\ \hline
			{\bf 6G} & $\ge 1$ Tbps   & \begin{tabular}[c]{@{}l@{}}Downlink: $\ge$1 Gbps\\ Uplink: $\ge$500 Mbps\end{tabular} & 100 GHZ         & 10-100$\times$ over 5G   & 5$\times$ over 5G          & $10^7{\rm/km}^2$ & $99.99999$\% & 10-100 ${\rm\mu s}$  & cm-level (3D)            & $1000~{\rm km/h}$ \\ \hline
		\end{tabular}
	}
\end{table*}

\begin{figure*}[!t]
	\centering
	\includegraphics[width=7in]{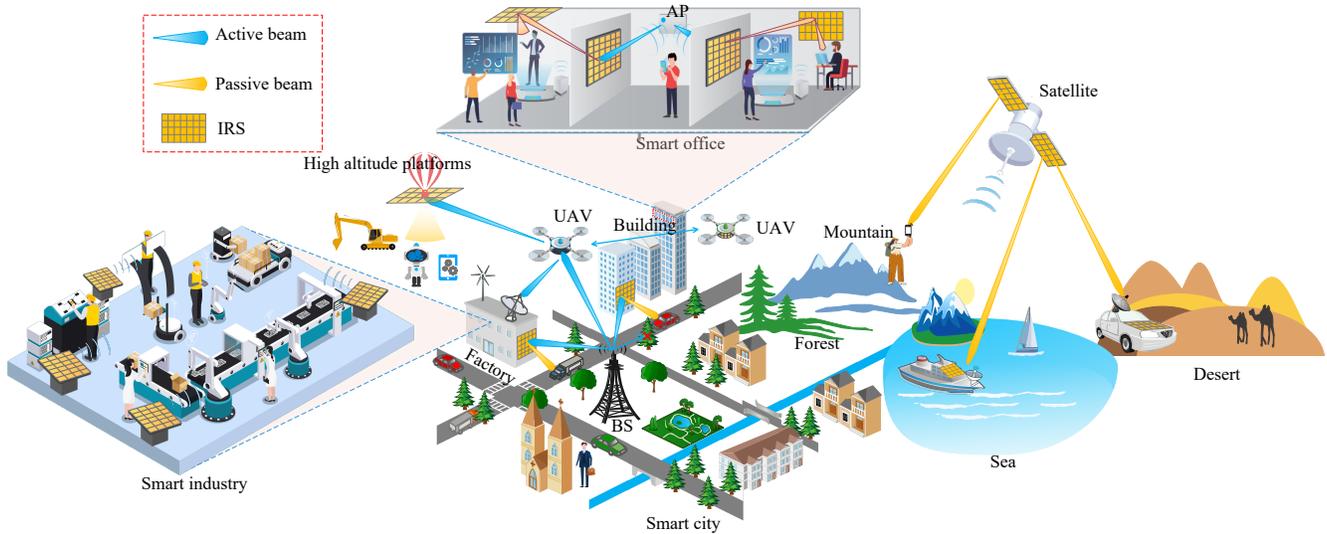}
	\caption{{Illustration of IRS application/deployment scenarios in future wireless networks.}}
	\label{6Gnetwork}
\end{figure*}

Over the last decade, significant efforts have been devoted to developing a variety of enabling
technologies for 5G \cite{andrews2014will,boccardi2014five,shafi20175g}, such as massive multiple-input multiple-output (MIMO), mmWave communication,
and network densification. Although these technologies have led to significant performance improvement to realize the KPIs of 5G, they face more severe difficulties in implementation due to the ever-increasing energy consumption and hardware cost.
For example, massive MIMO entails a large number of active antennas/radio-frequency (RF) chains to achieve high spectral efficiency, which, however, incurs high energy consumption and hardware cost that may hinder its larger-scale deployment in the future. Moreover, although mmWave communication benefits from its large available bandwidth for achieving high capacity, it is more susceptible to blockage and absorption loss in practice. As such, more costly antennas/RF chains and sophisticated array signal processing are needed for mmWave communication systems to compensate for the high propagation loss, which may not be a scalable  solution to 6G.
Last but not the least, as a key enabler of 5G, network densification is an effective means to enhance the network coverage and capacity by adding more cell sites. However, with the continuous addition of active nodes such as small-cell base stations (BSs), access points (APs), relays, and distributed active antennas invoked by network densification, the overall network energy consumption and deployment/maintenance cost will dramatically increase. Moreover, excessive network densification also results in other practical issues such as more severe intra-/inter-system interference, complicated resource management, and rate-demanding backhaul requirement, which are challenging to solve in practice.
Apart from the aforementioned drawbacks of existing 5G technologies, one ultimate bottleneck to achieving extremely high-capacity and ultra-reliable wireless communications lies in the random and largely uncontrollable wireless propagation environment, which causes undesired channel fading and signal attenuation/distortion that are detrimental to wireless system performance.
Substantial research efforts have been dedicated to developing various wireless communication techniques such as
adaptive modulation/coding, diversity and adaptive equalization, power/rate control, and active beamforming to either compensate for the wireless channel fading or adapt to dynamic channel conditions \cite{tse2005fundamentals,goldsmith2005wireless}. Since these techniques are typically applied at wireless transceivers that have no control over the random wireless propagation environment, they cannot always guarantee the stringent quality-of-service (QoS) requirements of 6G for uninterrupted/ubiquitous connectivity.
In view of the above issues and challenges, more research endeavors are pressingly needed to devise new, innovative, and even disruptive wireless technologies that can fulfill the KPIs of 6G in a cost-effective and sustainable manner.
Intelligent reflecting surface (IRS)\footnote{Note that in the current literature, there are other equivalent terms of IRS, such as reconfigurable intelligent surface (RIS) \cite{Renzo2019Smart,Renzo2020Smart}, large intelligent surface/antennas (LISA) \cite{liang2019large}, smart reflect-array \cite{tan2018enabling}, to name a few, all of which share the basic principle of employing passive and tunable reflecting surfaces to achieve smart and reconfigurable wireless environment.} has recently emerged as a promising new approach for enabling smart and reconfigurable wireless environment cost-effectively \cite{wu2019intelligent,wu2021intelligent,wu2019towards}.
Specifically, IRS is a digitally-controlled metasurface composed of a large number of passive reflecting elements that consume ultra-low power in tuning the phase shifts and/or amplitudes of the incident signals to the IRS in a programmable manner.
As such, different from conventional wireless communication techniques employed at transceivers,
IRS is able to directly reshape the wireless propagation channel in favor of signal transmission (e.g., boosting the
received signal power at intended users and/or suppressing the interference at unintended users \cite{wu2019intelligent,wu2021intelligent,wu2019towards}), thus providing an innovative and cost-effective means to realize the 6G KPIs.
In fact, IRS has a great potential in achieving a wide variety of appealing functionalities to provide promising performance gains for 6G, including 
\begin{enumerate}
\item {\bf Extended coverage} to support ubiquitous connectivity and ultra-high data rates by establishing a virtual line-of-sight (LoS) link to bypass signal blockage between transceivers;
\item {\bf Channel power and rank improvement} to increase the spatial multiplexing gain and spectral efficiency by adding  more controllable signal paths between transceivers in multi-antenna/broadband communications;
\item {\bf Channel statistics/distribution refinement} to achieve ultra-reliable communications by e.g., converting the Rayleigh/fast fading channels (in high mobility scenarios) into their Rician/slow fading counterparts; 
\item {\bf Interference mitigation} to enhance the user-experienced QoS by effectively nulling/canceling the co-channel/inter-cell interference while enhancing the desired signal reception quality;
\item {\bf Connection density enhancement} to support massive connectivity in a cost-effective manner without the need of densely deploying power-hungry active BSs/APs;
\item {\bf Positioning accuracy improvement} for the vertical and industrial applications
by providing controlled signal reflection and serving as the reference node for local sensing. 
\end{enumerate}
Besides the above appealing functionalities, IRS is able to achieve passive beamforming in full-duplex mode without incurring processing delay, self-interference, and noise amplification, thus offering competitive advantages over conventional active relays, e.g., half-duplex relay with processing delay as well as full-duplex relay that requires sophisticated self-interference cancellation.
Furthermore, IRS possesses other practical advantages such as low profile, lightweight, and conformal geometry, which facilitate its flexible, transparent, and large-scale deployment in future wireless networks, as shown in Fig.~\ref{6Gnetwork}.

{In Fig.~\ref{6Gnetwork}, we envision various IRS application/deployment scenarios in future wireless networks such as smart cities, smart offices, smart industries, and remote areas. For example, in smart cities, IRS can be coated on the building facade, lamppost, advertising board, and even inside 
high-speed moving vehicles, to support ubiquitous connectivity and ultra-high data rates for devices, people, and sensors. 
In the indoor environment such as smart offices and smart factories, IRS can be mounted on the ceilings, walls,
furniture/platform, and even behind the paintings/decorations, to achieve the enhanced URLLC and coverage as well as
support various vertical and industrial applications. On the other hand, for remote areas such as deserts,
mountains, rural areas, and oceans, the communication coverage by today's terrestrial networks is
still largely unavailable due to the practical difficulty and/or high cost of deploying terrestrial
BSs and backhauls. In view of this limitation, non-terrestrial communications, such as satellite communication and unmanned aerial vehicle (UAV) communication, have become
a promising solution to assist/complement terrestrial communications, aiming to provide global
coverage to support ubiquitous and seamless communications, as illustrated in Fig.~\ref{6Gnetwork}. In this case, IRS can be flexibly mounted on cars, camping tents, house roofs, ships, UAVs, and even on the reverse side of the satellite's solar panels  \cite{zheng2022Satellite},
to support the high-mobility communication, long-distance transmission, and massive connectivity in a cost-effective manner.
In a nutshell, IRS is a disruptive technology that can be applied to a wide range of scenarios for making our current ``dumb"
wireless environment become intelligent to better support massive IoE devices.}

\begin{table*}[!t]
	\centering
	\caption{List of Representative Overview/Survey/Tutorial Papers on IRS/RIS}\label{IRS_papers}
	\vspace{-0.3cm}
	\resizebox{\textwidth}{!}{
		\begin{tabular}{|m{0.7cm}<{\centering}|m{2.7cm}<{\centering}|m{5cm}<{\centering}|m{18cm}|}
			\hline
			{\bf Ref. }                                 & {\bf Publication Year/Type} & {\bf Topics/Theme}                                                                    & \qquad\qquad\qquad\qquad\qquad\qquad\qquad\qquad{\bf Major Contributions}                                                                                                                                                                                                               \\ \hline\hline
			\cite{Renzo2019Smart}                 & 2019/Survey           & Communication-theoretical models \& functionalities                              & Introduce reconfigurable metasurfaces to empower smart radio environments with the corresponding communication-theoretical model, and discuss a variety of functionalities for improving communications, sensing, and computing. \\ \hline
			\cite{Renzo2020Smart}                 & 2020/Survey           & Enabling technologies \& application/usage cases                                & Present a comprehensive overview of the enabling RIS technologies from a communication-theoretic perspective and discuss state-of-the-art application/usage cases.                                                                \\ \hline
			\cite{liang2019large}                 & 2020/Tutorial         & Hardware implementations \& applications                                        & Discuss the hardware implementations, performance gains, and applications of LISA with an emphasis on its differences and relationship with the backscatter communication and reflective relay.                                        \\ \hline
			\cite{wu2021intelligent}              & 2021/Tutorial         & Technical challenges \& system designs                                    & Provide a tutorial overview of IRS-aided wireless communications to address three main technical challenges from a communication theory standpoint.                                                                               \\ \hline
			\cite{wu2019towards}                  & 2020/Overview         & Main applications \& technical challenges                                       & Present an early overview of the potential of IRS for wireless communications, and discuss its basic concept, main applications, and key technical challenges.                                                                       \\ \hline
			\cite{yuan2021reconfigurable}         & 2021/Overview         & Technical challenges                                                               & Briefly summarize three design challenges on RIS, including CSI acquisition, passive information transfer, and reflection optimization.                                                                                               \\ \hline
			\cite{alghamdi2020intelligent}        & 2020/Survey           & System designs \& analytical frameworks                                   & Provide a survey on the optimization methods and analytical frameworks for LISA, and briefly discuss future research directions.                                                                                                   \\ \hline
			\cite{bjornson2021reconfigurable}     & 2021/Tutorial         & System/channel modeling \& signal processing                                   & Introduce the system/channel modeling and provide a tutorial overview of the RIS fundamentals from a signal processing standpoint, including communication, localization, and sensing.                                        \\ \hline
			\cite{gong2020toward}                 & 2020/Survey           & Performance analysis \& applications                                            & Give a literature survey on the performance analysis/metrics of IRS-aided networks and classify state-of-the-art results for IRS applications according to the design objectives and control variables.                       \\ \hline
			\cite{de2020role}                     & 2020/Overview         &  NOMA with IRS                                                                       & Discuss the potential improvements by IRS in NOMA in terms of channel gain, power allocation fairness, covergae range, and energy efficiency.                                                                          \\ \hline 
			\cite{liaskos2018new}                 & 2018/Overview         & Basic concept \& physical architectures                                                & Envision the concept and physical architecture of software-controlled/defined metasurface and discuss the research challenges.                                                                                                    \\ \hline
			\cite{elmossallamy2020reconfigurable} & 2020/Tutorial         & Hardware implementations \& channel modeling                                   & Introduce different hardware implementations of RIS using metasurface and reflectarray, and discuss channel modeling as well as challenges and opportunities in RIS-aided wireless networks.                                     \\ \hline
			\cite{huang2020holographic}           & 2020/Overview         & Hardware architectures \& functionalies/features                                & Introduce the holographic MIMO surface and summarize its hardware architectures, functionalities/features, and communication applications.                                                                               \\ \hline
			\cite{wymeersch2020radio}             & 2020/Overview         & Localization and mapping                                                        & Present the relevant channel models and discuss the benefits of RIS for localization and mapping in terms of improved accuracy and extended coverage.                                                                             \\ \hline
			\cite{liu2021reconfigurable}          & 2021/Survey           & Performance analysis \& system designs & Provide a survey on the performance analysis, beamforming and resource allocation, as well as machine learning for RIS-aided wireless communications and discuss relevant applications.                                                               \\ \hline
			\cite{bjornson2020reconfigurable}     & 2020/Overview         & Key features \& myths                                                           & Review the key features of RIS and raise three myths about the functionality, performance gain, and path loss.                                                                                                                   \\ \hline
			\cite{you2020deploy}                  & {2021/Overview}         & {Deployment stategies}                                                            & {Overview typical deployment strategies for IRS-aided communications and compare their performance.}                                                                                    \\ \hline    
			\cite{mei2021intelligent}                  & {2022/Tutorial}          & {Multi-IRS design and optimization}                                                         & {Provide a tutorial overview of multi-IRS aided wireless networks, with an emphasis on addressing the new challenges due to multi-IRS signal reflection and routing strategy. }                                                                                   \\ \hline    
		\end{tabular}
	}
\end{table*}

The above benefits of IRS have spurred extensive studies on investigating its design and performance in a variety of wireless systems, e.g., orthogonal frequency division multiplexing (OFDM) \cite{yang2020intelligent,zheng2019intelligent,zheng2020fast,zheng2020intelligent,jiang2021joint}, multi-antenna communication \cite{zhang2019capacity,Pan2020Multicell,ozdogan2020Intelligent}, non-orthogonal multiple access (NOMA) \cite{Zheng2020IRSNOMA,yanggang2019intelligent,ding2020simple,mu2019exploiting}.
Moreover, there are more than a dozen of overview/survey/tutorial papers that have disseminated the state-of-the-art results on IRS and its various equivalents for wireless communications with different focuses, including technical challenges \cite{wu2021intelligent,wu2019towards}, system models/designs \cite{wu2021intelligent,wu2019towards,yuan2021reconfigurable,Renzo2019Smart,alghamdi2020intelligent,bjornson2021reconfigurable}, applications \cite{wu2019towards,Renzo2020Smart,gong2020toward,de2020role}, hardware implementations \cite{liang2019large,liaskos2018new,elmossallamy2020reconfigurable,huang2020holographic}, functionalities \cite{Renzo2019Smart,huang2020holographic,wymeersch2020radio}, and performance analysis \cite{gong2020toward,liu2021reconfigurable}, which are summarized in Table~\ref{IRS_papers} for the ease of reference. 
It is noted that the existing works have mainly considered the ideal assumption of perfect channel state information (CSI), which, however, is difficult to achieve in practice, especially for IRSs without signal processing/transmission capabilities. While for some other works (e.g., \cite{liang2019large,liaskos2018new,elmossallamy2020reconfigurable,huang2020holographic}) that address IRS hardware implementations, they mainly focus on the physical modeling, hardware architecture, and characteristics of the metasurface, but without an in-depth discussion of the effects of practical IRS hardware on the system design and communication performance.
In view of the above, a dedicated work that comprehensively surveys the practical design and implementation issues of IRS from a communication viewpoint is still missing, which thus motivates this paper.

\begin{figure*}[!t]
	\centering
	\includegraphics[width=5in]{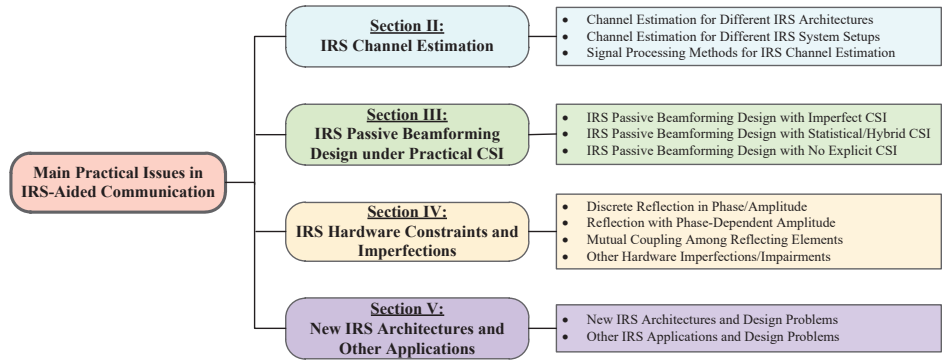}
	\caption{Organization of the paper.}
	\label{Organization}
\end{figure*}
\subsection{Main Practical Issues in IRS-aided Wireless Communications}\label{Issues}
Despite the appealing advantages and great potential of IRS, there are new practical issues and challenges that need to be addressed in IRS-aided wireless communications.
Particularly, for transforming the IRS-aided wireless communication from theory to practice, the main challenges include IRS channel estimation/acquisition, practical IRS passive beamforming/reflection design under imperfect CSI, as well as various hardware constraints and imperfections. In the following, we elaborate these issues in the practical design and implementation of IRS-aided wireless communication systems. 
\begin{itemize}
	\item {\bf IRS channel estimation/acquisition:} To achieve a high degree of control over the wireless propagation environment by IRSs, the acquisition of accurate CSI in IRS-aided wireless communication systems is indispensable, which, however, is practically challenging
	due to the following two main reasons. First, due to the lack of active components for baseband signal processing, the low-cost passive IRS elements can only reflect the incident signals,
	but without the capability to transmit/receive pilot signals for channel estimation as active transceivers in conventional wireless communication systems.
	Second, IRS is typically composed of a vast number of passive elements that in general have different channel coefficients to be estimated in the IRS-associated channels, thus resulting in a substantial increase in the system overhead for IRS channel estimation.
	Besides, different system setups and IRS deployments (e.g., single-/multi-user, single-/multi-IRS, single-/multi-antenna BS/user, low-/high-mobility user, and narrowband/broadband communication) generally impose different requirements on IRS channel estimation/acquisition, as will be discussed in detail in Section~\ref{sec_CE} of this paper.
	\item {\bf Practical IRS passive beamforming/reflection design:} In practice, IRS passive beamforming/reflection optimization highly depends on the available CSI, which, however, is difficult to obtain perfectly. 
	Under different scenarios of CSI availability (e.g., imperfect instantaneous CSI and/or statistical CSI), IRS passive reflection needs to be properly designed jointly with the active transceivers to achieve robust communication performance subjected to CSI errors. Besides the CSI-based IRS reflection design, another practical approach is to design the IRS reflection via passive  beam training with the beams selected from a designed codebook, thus alleviating the need for explicit CSI. However, this approach may require long training time if there is no prior information given on the CSI, as the IRS needs to exhaustively search over a large number of passive beams due to its vast number of reflecting elements. Moreover, different system setups (e.g., single-/multi-user, single-/multi-IRS, single-/multi-antenna BS/user, and narrowband/broadband communication) and design objectives (throughput/rate maximization, outage/delay minimization, etc.) generally lead to different IRS reflection designs, which need to be addressed in a systematic way, as will be detailed in Section~\ref{sec_RD} of this paper.
	\item {\bf Hardware constraints and imperfections:} In the initial investigation of IRS-aided wireless communications, prior works have mainly considered the ideal IRS reflection model for the ease of communication system design and performance optimization. However, IRS faces a variety of hardware constraints and imperfections/impairments in practice that may limit its capability in signal reflection, such as discrete phase-shift/amplitude and phase-dependent amplitude of its reflecting elements, as well as the mutual coupling among them. As such, both IRS channel estimation and passive beamforming/reflection optimization need to cater for the practical IRS reflection models that can accurately capture the IRS hardware constraints and imperfections/impairments so as to maximize its performance gains in practice. This, however, further complicates the design problems, as will be discussed in Section~\ref{Hardware} of this paper in more detail.  
\end{itemize}
\subsection{Objective and Organization}\label{Organization1}

In contrast to the existing overview/survey/tutorial papers on IRS listed in Table~\ref{IRS_papers}, this paper aims to provide a comprehensive and up-to-date survey on the research in IRS-aided wireless communications, with an emphasis to address the practical challenges in IRS channel estimation and passive beamforming/reflection optimization for different channel models and system setups. 
In particular, this paper overviews a wide class of signal processing and communication techniques for resolving the practical issues in IRS channel estimation and passive beamforming/reflection optimization, and studies the effects of various IRS hardware constraints/imperfections on the system design and achievable performance. Moreover, practical design challenges for emerging IRS architectures and other IRS applications in wireless networks are discussed as well to inspire future research.


\begin{table*}[!t]
	\centering
	\caption{List of Main Acronyms.}\label{acronyms}
	\vspace{-0.3cm}
	\resizebox{0.9\textwidth}{!}{
	\begin{tabular}{l|l||l|l}
		\hline
		\textbf{Acronyms} & \qquad\qquad\qquad\textbf{Definition}                         & \textbf{Acronyms} & \qquad\qquad\qquad\textbf{Definition}                        \\ \hline\hline
5G                & Fifth-generation communication system     & LoS               & Line-of-sight                              \\ \hline
6G                & Sixth-generation communication system     & LS                & Least square                               \\ \hline
ADC               & Analog-to-digital converter               & MIMO              & Multiple-input multiple-output             \\ \hline
AO                & Alternating optimization                  & MISO              & Multiple-input single-output               \\ \hline
AoA               & Angle-of-arrival                          & (L)MMSE              & (Linear) Minimum mean-squared-error                  \\ \hline
AoD               & Angle-of-departure                        & mMTC              & Massive machine-type communication         \\ \hline
AP                & Access point                              & mmWave            & Millimeter-wave                            \\ \hline
BCD               & Block coordinate descent                  & MRT               & Maximum ratio transmission                 \\ \hline
BS                & Base station                              & NOMA              & Non-orthogonal multiple access             \\ \hline
CFR               & Channel frequency response                & OFDM              & Orthogonal frequency division multiplexing \\ \hline
CIR               & Channel impulse responses                 & OMP               & Orthogonal matching pursuit                \\ \hline
CNN               & Convolutional neural network              & PHY               & Physical layer                             \\ \hline
CRLB              & Cram\'{e}r-Rao lower bound                & PU                & Primary user                               \\ \hline
CSI               & Channel state information                 & QoS               & Quality-of-service                         \\ \hline
DFT               & Discrete Fourier transform                & RF                & Radio-frequency                            \\ \hline
DNN               & Deep neural network                       & RIS               & Reconfigurable intelligent surface         \\ \hline
eMBB              & Enhanced mobile broadband                 & SCA               & Successive convex approximation            \\ \hline
FDD               & Frequency-division duplexing              & SDR               & Semi-definite relaxation                    \\ \hline
GNN               & Graph neural network                      & SINR              & Signal-to-interference-plus-noise ratio    \\ \hline
IoE               & Internet-of-Everything                    & SISO              & Single-input single-output                 \\ \hline
IoT               & Internet-of-Things                        & SNR               & Signal-to-noise ratio                      \\ \hline
IRS               & Intelligent reflecting surface & SU                & Secondary user                             \\ \hline
ISAC              & Integrated sensing and communication      & SVD               & Singular value decomposition               \\ \hline
ITS               & Intelligent transmitting surface          & TDD               & Time-division duplexing                    \\ \hline
ITU               & International Telecommunication Union     & THz               & Terahertz                                  \\ \hline
KPI               & Key performance indicator                 & URLLC             & Ultra-reliable low-latency communication   \\ \hline
LISA              & Large intelligent surface/antennas        & WPT               & Wireless power transfer                    \\ \hline
	\end{tabular}
}
\end{table*}
As shown in Fig.~\ref{Organization}, the rest of this paper is organized as follows. Section~\ref{sec_CE} presents a comprehensive overview of the state-of-the-art results on channel estimation for IRS-aided wireless communication systems. In Section \ref{sec_RD}, we overview the up-to-date results on IRS passive beamforming/reflection design under different scenarios of CSI availability in practice.  In Section~\ref{Hardware}, we further discuss different IRS hardware constraints and imperfections/impairments as well as the existing results on their modeling and effects on the IRS channel estimation and passive beamforming/reflection design.
Finally, practical design challenges for new IRS architectures and other IRS applications are outlined and discussed in Section~\ref{app}, followed by  
the conclusions drawn in Section~\ref{con}.
For ease of reference, we summarize the definitions of the main acronyms used in this paper in TABLE~\ref{acronyms}.



\section{IRS Channel Estimation}\label{sec_CE}
\begin{figure*}[!t]
	\centering
	\includegraphics[width=5in]{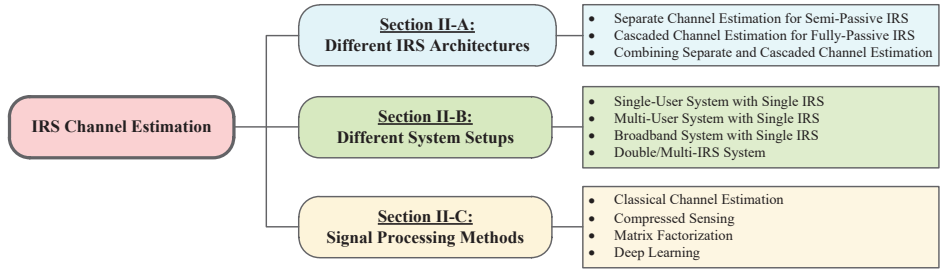}
	\caption{Classification of IRS channel estimation methods.}
	\label{Organization_CE}
\end{figure*}

As shown in Fig.~\ref{6Gnetwork}, IRS can be widely deployed in various system setups to improve the communication performance by properly designing the IRS passive beamforming/reflection.
In particular, to enable the effective IRS passive beamforming/reflection optimization for achieving high-rate and ultra-reliable communications, accurate CSI of the communication environment to be reconfigured by IRSs is essential.
However, as discussed in Section~\ref{Issues}, IRS channel estimation is a practically challenging task, due to the massive number of passive IRS elements without transmitting/receiving capabilities. As such, researchers in wireless communications have devoted  great efforts to devising efficient channel estimation methods for different IRS architectures and system setups, aiming to achieve high channel estimation accuracy with low training overhead.
In this section, we discuss and classify the existing works on IRS channel estimation from three different perspectives, namely, IRS architectures, IRS system setups, and signal processing methods, as outlined in Fig.~\ref{Organization_CE}.
Specifically, our discussion starts from a brief comparison between the channel estimation for two practical IRS architectures by unveiling their basic design principles.
Next, we study the channel estimation for different IRS system setups, with an emphasis on their main differences that need to be accounted for to achieve high channel estimation efficiency.
Last, we discuss the main signal processing methods for IRS channel estimation and highlight their applicable channel models and communication scenarios.
\subsection{Channel Estimation for Different IRS Architectures}
\begin{figure*}
	\centering
	\subfigure[{Separate channel estimation with semi-passive IRS.}]{
		\begin{minipage}[b]{0.4\textwidth}
			\includegraphics[width=0.8\textwidth]{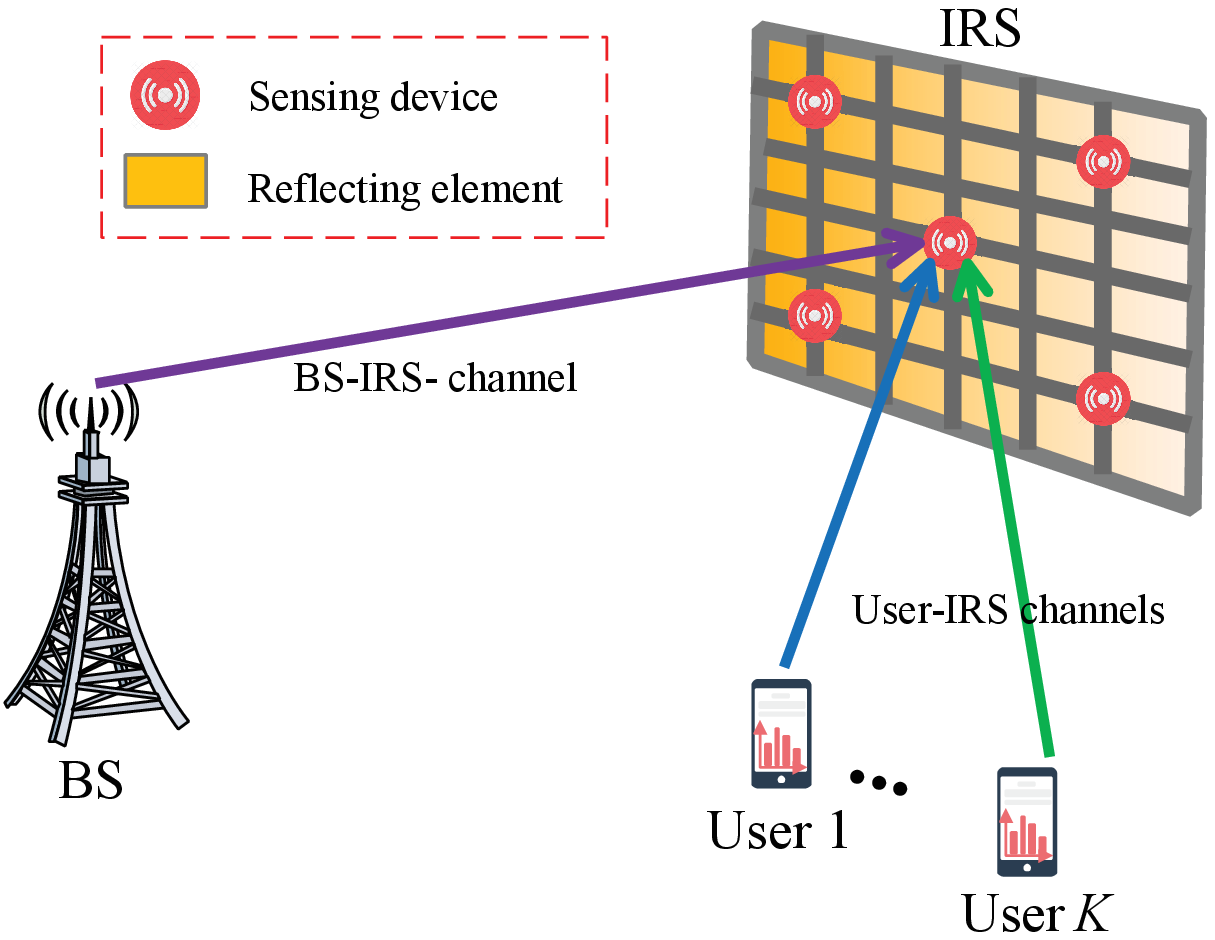}
		\end{minipage}\label{IRS_separateCSI}
	}
	~~
	\subfigure[{Cascaded channel estimation (uplink) with fully-passive IRS.}]{
		\begin{minipage}[b]{0.4\textwidth}
			\includegraphics[width=0.8\textwidth]{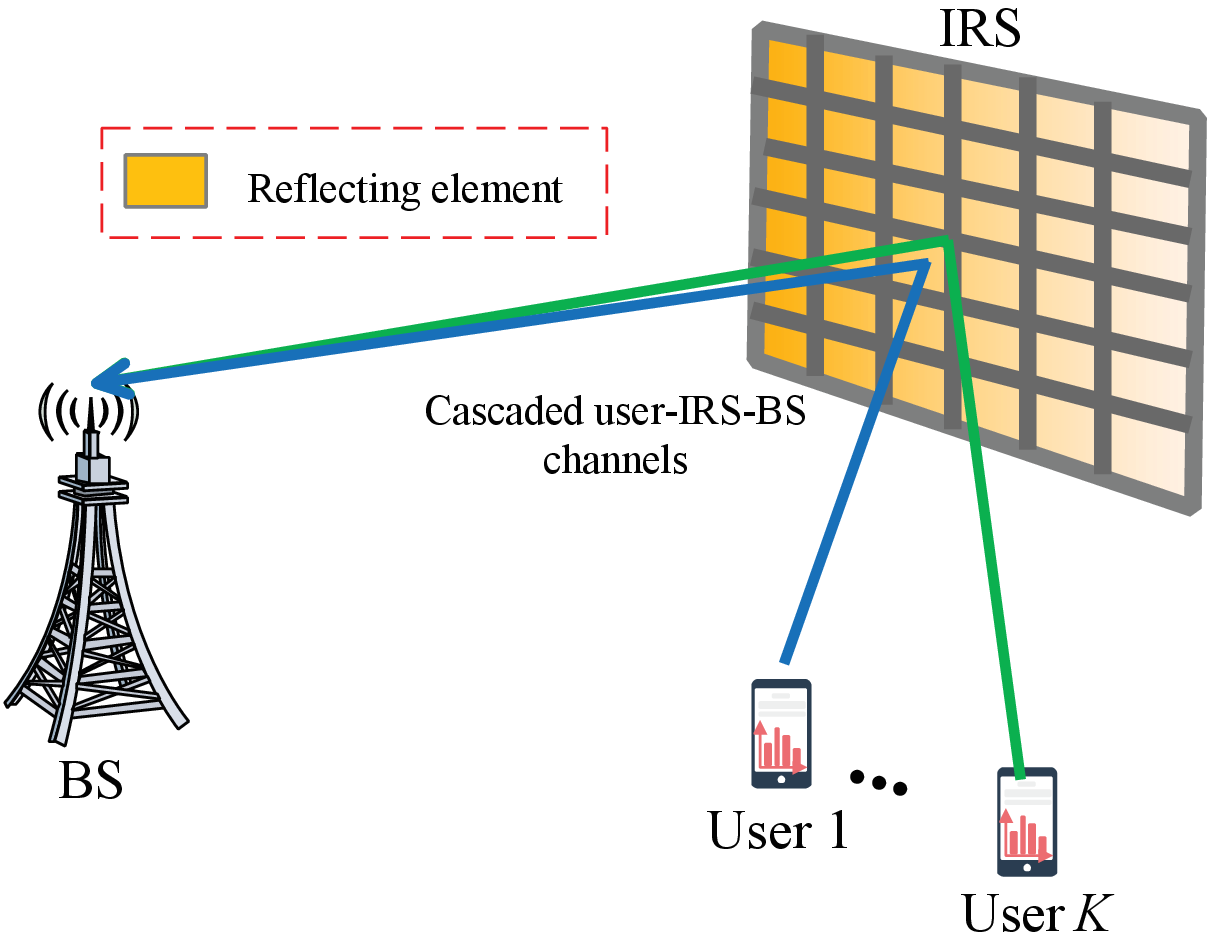}
		\end{minipage}\label{IRS_cascadedCSI}
	}
	\caption{{Comparison of separate versus cascaded channel estimation for IRS.}} \label{IRS_CE}
\end{figure*}

Depending on whether the IRS is mounted with sensing devices or not, there are two basic IRS architectures in practice, termed semi-passive IRS and fully-passive IRS, respectively \cite{wu2021intelligent}.
Accordingly, existing approaches for estimating the IRS-associated channels can be classified into two categories for each of them, namely, ``separate channel estimation" and ``cascaded channel estimation", as illustrated in Fig.~\ref{IRS_CE}. 
{For the purpose of exposition, we primarily consider an IRS-aided narrowband communication system that includes an $M_B$-antenna BS, an $N$-element IRS, and potentially $K$ co-channel users (each of which is equipped with $M_u$ antennas).
Moreover, we let ${\bm G}\in {\mathbb{C}^{ M_B\times N}} $, ${\bm H}_{k}\in {\mathbb{C}^{N\times M_u }}$ and ${\bm D}_{k}\in {\mathbb{C}^{M_B\times M_u }}$ denote the IRS-BS, user $k$-IRS, and user $k$-BS direct
channels, respectively.
The equivalent reflection vector of IRS is denoted by ${\bm \theta}=\left[\theta_1,\theta_2,\ldots,\theta_N\right]^T\in {\mathbb{C}^{N\times 1}}$.
Note that by turning OFF the IRS in the absorbing mode, i.e., ${\bm \theta}={\bm 0}$,
the BS can estimate the direct channels, i.e., $\left\{{\bm D}_{k}\right\}_{k=1}^K$, via
conventional channel estimation methods with orthogonal/sequential pilots sent from different users. As such, we mainly focus on the channel estimation of the IRS-associated channels, i.e., ${\bm G}$ and $\left\{{\bm H}_{k}\right\}_{k=1}^K$.}
In the following, we overview the main pros and cons of these two channel estimation frameworks for IRS-aided wireless communications and discuss the representative works in the literature to unveil their basic design principles.
\subsubsection{Separate Channel Estimation with Semi-Passive IRS} 
For semi-passive IRS mounted with $N_s$ dedicated sensing devices {(with low-power receive RF chains and $N_s\ll N$ in practice)}, the channels from the BS/users to the IRS can be separately estimated at these sensing devices based on the pilots sent by the BS/users, as illustrated in Fig.~\ref{IRS_separateCSI}
This approach works well for the time-division duplexing (TDD) system by exploiting channel reciprocity to acquire the reverse CSI from the IRS to the BS/users, but may be inapplicable to the frequency-division duplexing (FDD) system  (unless active sensors that can both transmit and receive pilot signals are mounted on the IRS, which inevitably incurs higher hardware cost and power consumption). Moreover, to reduce the hardware cost of IRS, only a small number of low-cost sensing devices may be installed on the semi-passive IRS, e.g., with low-resolution analog-to-digital converters (ADCs).
{Specifically, as shown in Fig.~\ref{IRS_separateCSI}, 
we let ${\ddot{\bm G}}\in {\mathbb{C}^{ N_s \times M_B}} $ and ${\ddot{\bm H}}_{k}\in {\mathbb{C}^{ N_s\times M_u}}$ denote the channels from the BS and user $k$ to the $N_s$ sensing devices, respectively. 
Moreover, let ${\bm X}_B\in {\mathbb{C}^{ M_B \times T}}$ and ${\bm X}_k\in {\mathbb{C}^{ M_u \times T}}$ denote the pilot sequences transmitted by the BS and user~$k$, with $T$ being the number of pilot symbols during the channel sensing period.
Accordingly,
the signal received at the $N_s$ sensing devices is given by
\begin{align}\label{Semi}
{\bm Y}_s={\mathbb Q}\left(\sqrt{P_B}{\ddot{\bm G}} {\bm X}_B+\sum_{k=1}^{K}\sqrt{P_u}{\ddot{\bm H}}_{k} {\bm X}_k+ {\bm V}_s\right),
\end{align}
where ${\mathbb Q}\left(\cdot \right)$ denotes the quantization function that depends on the resolution of ADCs,
$P_B$ and $P_u$ represent the transmit power at the BS and users, respectively,
and ${\bm V}_s\in {\mathbb{C}^{ N_s\times T}}$ is the additive white Gaussian noise (AWGN) at the sensing devices.
Thus, the key challenge of the above separate channel estimation at IRS lies in how to construct the accurate CSI from the BS/users to IRS reflecting elements based on the signal measurement given in \eqref{Semi} at a limited number of low-cost sensing devices.}
As such, efficient signal processing methods are needed to interpolate/extrapolate the {limited amount of measurement data in ${\bm Y}_s$} for constructing the full CSI from the BS/users to IRS reflecting elements, {i.e., ${\bm G}$ and $\left\{{\bm H}_{k}\right\}_{k=1}^K$,} by exploiting some common channel properties such as low-rank, sparsity, and spatial correlation, e.g., in mmWave \cite{heath2016overview} or THz \cite{guan2019channel} frequency bands. Nonetheless, further investigation on the channel modeling and estimation is needed to enable more accurate CSI construction by accounting for the distortion in the measurement data due to quantization error, ambient noise, circuit non-linearity, etc.%

For semi-passive IRS, a handful of preliminary separate channel estimation schemes have been proposed.
Specifically, with an L-shaped channel sensing array mounted on IRS,
a low-complexity separate channel estimation scheme was proposed in \cite{chen2021low} to estimate the BS-IRS and user-IRS channels separately in terms of the angle-of-arrival (AoA) and path gain.
{Moreover, with randomly distributed sensing devices on IRS,
separate channel estimation schemes based on different signal processing tools such as compressed sensing \cite{jian2020modified,liu2020deep,lin2021tensor,taha2021enabling} and deep learning \cite{taha2021enabling,taha2019deep,taha2020deep} were proposed to estimate the separate CSI from the BS/users to IRS for the narrowband communication.}
Later, by exploiting the angular-domain sparsity of the mmWave MIMO channel under the broadband setup, efficient channel estimation schemes were proposed to improve training efficiency \cite{liu2020deep,lin2021tensor}.
Despite the above studies,
a systematic study on the pilot sequence design at the BS/users, optimal deployment of sensing devices on IRS, and efficient channel sensing/construction algorithm is still needed to achieve high IRS channel estimation accuracy at low hardware cost and with short channel sensing time.
\subsubsection{Cascaded Channel Estimation with Fully-Passive IRS} \label{Rep_cascaded}
On the other hand, for fully-passive IRS without sensing devices, {the BS-IRS and user-IRS channels, i.e., ${\bm G}$ and $\left\{{\bm H}_{k}\right\}_{k=1}^K$,} cannot be estimated separately in general. Instead, only the cascaded user-IRS-BS (BS-IRS-user) channel can be estimated at one endpoint of the communication system, e.g., the BS with higher signal processing capability, as illustrated in Fig.~\ref{IRS_cascadedCSI}. 
 Unlike the separate channel estimation, the cascaded channel estimation applies to both TDD and FDD systems. Specifically, the cascaded CSI in the TDD system can be estimated in one direction and used in both directions by leveraging the uplink-downlink channel reciprocity; while additional feedback of the cascaded CSI is needed in the FDD system for both uplink and downlink communications. Moreover, as compared to the separate channel estimation, the cascaded channel estimation is more practically appealing due to the much lower hardware cost
 and energy consumption at the IRS as it does not need active sensing devices. 

{As shown in Fig.~\ref{IRS_cascadedCSI},
we consider the uplink channel training with the users sending pilot signals, and thus the signal received at the BS during time slot $t$ is given by
\begin{align}\label{Fully}
{\bm y}_B^{(t)}=\sum_{k=1}^{K}\sqrt{P_u}{\bm G} {\bm \Theta}^{(t)} {\bm H}_{k} {\bm x}_k^{(t)} + {\bm v}_B^{(t)},
\end{align}
where ${\bm x}_k^{(t)}\in {\mathbb{C}^{ M_u \times 1}}$ denotes the pilot signal transmitted by user $k$,
${\bm \Theta}^{(t)}=\text{diag} \left({\bm \theta}^{(t)}\right)$ denotes the diagonal reflection matrix of the IRS during time slot $t$, and ${\bm v}_B^{(t)}\in {\mathbb{C}^{ M_B \times 1}}$ is the AWGN vector at the BS. 
According to the property of the Khatri-Rao product, we have 
\begin{align}\label{cascaded}
\text{vec} \left( {\bm G} {\bm \Theta}^{(t)} {\bm H}_{k}\right)= \underbrace{{\bm H}_{k}^T \odot {\bm G}}_{{\vec{\bm H}}_{k}}
 {\bm \theta}^{(t)}, \qquad k=1,\ldots K,
\end{align}
where ${\vec{\bm H}}_{k}\in {\mathbb{C}^{ M_B M_u \times N}}$ denotes the cascaded channel of user~$k$,
$\text{vec} \left( \cdot \right)$ denotes the vectorization operation,
 and $\odot$ stands for the Khatri-Rao product. Furthermore, by exploiting the Kronecker product, we can rewrite \eqref{Fully} in a compact form as
\begin{align}\label{Fully2}
{\bm y}_B^{(t)}=\sum_{k=1}^{K}\sqrt{P_u} \left( \left({\bm x}_k^{(t)}\right)^T \otimes {\bm I}_{M_B} \right){\vec{\bm H}}_{k} {\bm \theta}^{(t)} + {\bm v}_B^{(t)},
\end{align}
where $\otimes$ stands for the Kronecker product.
From \eqref{cascaded} and \eqref{Fully2}, we can observe that due to the cascade of the user-IRS and IRS-BS channels, the number of channel coefficients to be estimated in the cascaded channels $\left\{{\vec{\bm H}}_{k}\right\}_{k=1}^K$ is generally larger than that in the separate channels ${\bm G}$ and $\left\{{\bm H}_{k}\right\}_{k=1}^K$, thus incurring 
higher training overhead and imposing new challenges in practice.}

There have been substantial works on cascaded channel estimation in recent years. Due to space limits, we only discuss some representative works to unveil its basic design principles in the following, while leaving the more comprehensive overview of them in the following subsections.
First, a simple and straightforward scheme is to estimate the cascaded channel associated with each IRS element successively based on the channel measurement at the receiver. This can be achieved by adopting the ON/OFF training reflection pattern at the IRS with pilot symbols sent from the transmitter, as studied in \cite{yang2020intelligent,mishra2019channel}. 
Later, the full-ON IRS training reflection pattern based on some special matrices (e.g., the discrete Fourier transform (DFT) matrix, Hadamard matrix, and circulant matrix generated by Zadoff-Chu sequence) was developed for the cascaded channel estimation in \cite{zheng2019intelligent,jensen2020optimal,zheng2020fast}. It was shown that the channel estimation accuracy can be significantly improved by exploiting the full IRS aperture gain with the full-ON IRS training reflection pattern. 
Furthermore, the training reflection pattern at the IRS was jointly designed with the pilot sequence at the transmitter in \cite{zheng2020fast,kang2020intelligent,zheng2020intelligent}, aiming to achieve perfect orthogonality over the IRS-reflected signals for minimizing the channel estimation error.
In addition, different algorithms based on some well-known signal processing methods such as least square (LS)/linear minimum mean-squared-error (LMMSE) \cite{zheng2019intelligent,jensen2020optimal,zheng2020fast,kang2020intelligent,zheng2020intelligent}, compressed sensing \cite{ardah2021trice,jia2020high,he2021channel,he2020channel,ma2020joint,lin2020Channel,mirza2021channel,he2019cascaded,deepak2021channel,liu2020asymptotic}, and deep learning \cite{gao2021deep,xu2021ordinary,kundu2021channel,kundu2020deep,li2020channel} were proposed to resolve both the direct and/or cascaded channels at the receiver.
To sum up, the cascaded channel estimation for IRS hinges on how to jointly design {the pilot sequence $\left\{{\bm x}_k^{(t)}\right\}_{k=1}^K$ at the transmitter (users), training reflection pattern ${\bm \theta}^{(t)}$ at the IRS, and signal processing algorithms at the receiver (BS)} to accurately estimate both the direct and cascaded channels with the minimum training overhead.

\subsubsection{Comparison/Combination of Separate and Cascaded Channel Estimation}

As summarized in Table~\ref{Architecture}, the separate and cascaded channel estimation approaches have their respective pros and cons.
In \cite{schroeder2020passive}, the authors made an initial comparison between the separate and cascaded channel estimation approaches based on the atomic norm minimization, which revealed the higher estimation accuracy of the cascaded channel estimation with less hardware cost and energy consumption. Nevertheless, it is noted that the CSI errors and cost-performance trade-offs of
the separate and cascaded channel estimation approaches highly depend on the adopted signal processing techniques, channel models, training designs, as well as hardware constraints, which deserve a more thorough comparison in the future.
\begin{table*}[!t]
	\centering
	\caption{{Comparison of IRS Channel Estimation with Different IRS Architectures}}\label{Architecture}
	\vspace{-0.3cm}
	\resizebox{\textwidth}{!}{
	\begin{tabular}{|m{2cm}<{\centering}|m{2cm}<{\centering}|m{2.8cm}<{\centering}|m{1.8cm}<{\centering}|m{1.8cm}<{\centering}|m{1.8cm}<{\centering}|m{4.5cm}<{\centering}|m{1.8cm}<{\centering}|m{2.5cm}<{\centering}|m{2.5cm}<{\centering}|}
		\hline
		{\bf IRS   Architecture} &
		{\bf Channel Estimation Framework} &
		{\bf Hardware Cost and Power   Consumption} &
		{\bf Processing Node} &
		{\bf Duplexing Mode} &
		{\bf CSI Type} &
		{\bf Technical Challenges} &
		{\bf Training Overhead} &
		{\bf Channel Model} &
		{\bf Representative Work} \\ \hline\hline
		{\bf Semi-passive   IRS} &
		Separate channel estimation &
		Moderate cost and power   consumption with a few sensing devices &
		Sensing devices at IRS &
		TDD &
		Separate CSI from BS/users to IRS &
		Constructing the full IRS-associated CSI based on the limited measurement data at sensing devices &
		Moderate &
		Sparse/Low-rank/Highly correlated &
		\cite{taha2019deep,taha2020deep,taha2021enabling,jian2020modified,alexandropoulos2020hardware,liu2020deep,lin2021tensor} \\ \hline
		{\bf Fully-passive IRS} &
		Cascaded channel estimation &
		Very low cost and power   consumption with passive reflecting elements only &
		BS/users &
		TDD \& FDD &
		Cascaded CSI of BS-IRS-users &
		A large number of channel coefficients to be estimated &
		High &
		A wide
		variety of
		channel
		models &
		\cite{yang2020intelligent,mishra2019channel,zheng2019intelligent,jensen2020optimal,zheng2020fast,kang2020intelligent,zheng2020intelligent,ardah2021trice,jia2020high,he2021channel,he2020channel,ma2020joint,lin2020Channel,mirza2021channel,he2019cascaded,deepak2021channel,liu2020asymptotic,gao2021deep,xu2021ordinary,kundu2021channel,kundu2020deep,li2020channel} \\ \hline
	\end{tabular}
}
\end{table*}

\begin{figure}[!t]
	\centering
	\includegraphics[width=2.3in]{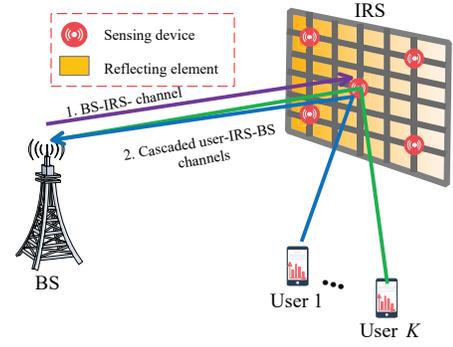}
	\caption{{Hybrid/combined separate and cascaded channel estimation.}}
	\label{IRS_jointCSI}
\end{figure}
In the current literature, the separate and cascaded channel estimation approaches are mainly investigated in a separate manner.
Thus, it is also an interesting direction to combine them to reap their joint benefits and thereby
achieve higher channel estimation efficiency in future work. For instance, it is noted that
{the BS-IRS channel ${\bm G}$} is generally high-dimensional (due to the multi-antenna BS) but quasi-static (due to the fixed locations of the BS and the IRS), {while the IRS-user channels $\left\{{\bm H}_{k}\right\}_{k=1}^K$} are more dynamic (due to the user mobility) but low-dimensional (due to the much fewer antennas employed at each user than the BS in practice) \cite{hu2021two,guan2021anchor,zheng2020intelligent}. 
As such, {the quasi-static BS-IRS channel ${\bm G}$} can be efficiently estimated at the sensing devices on the IRS by first adopting the separate channel estimation approach; whereas {the dynamic IRS-user channels $\left\{{\bm H}_{k}\right\}_{k=1}^K$} can be then estimated/tracked in real time at the BS via the cascaded channel estimation approach by leveraging prior knowledge of the quasi-static BS-IRS channel. 
Thus, this hybrid/combined channel estimation method, as illustrated in Fig.~\ref{IRS_jointCSI}, has the potential of further reducing the real-time training overhead as compared to their separate uses and improving their applicability for both TDD and FDD systems. 
However, how to effectively design and materialize this approach is still an open problem, which deserves further investigation in the future.

\subsection{Channel Estimation for Different IRS System Setups}
Since different system setups and IRS deployments (e.g., single/multi-user, single/multi-IRS, single-/multi-antenna BS/user, low-/high-mobility user, and narrowband/broadband communication) generally impose different requirements on the CSI, customized channel estimation schemes are thus needed for achieving high training efficiency and estimation accuracy.
In this subsection, we discuss different IRS channel estimation schemes under different IRS system setups, as outlined in Fig.~\ref{Organization_CE}.
Specifically, we first consider the single-IRS case and discuss its channel estimation schemes for the single-user and multi-user setups, respectively, in the narrowband system. Then, we extend the discussion to the broadband system over frequency-selective fading channels, with more channel coefficients to be estimated.
Last, we consider the emerging double-/multi-IRS systems in the presence of inter-IRS channels that make the channel estimation problem even more challenging to solve.

\subsubsection{Single-User System with Single IRS}\label{Sparsity}
For the single-user system with one single IRS, the effective channel is the superposition of the direct channel and reflected channels associated with a large number of IRS elements. 
{Specifically, for the single-user case with $K=1$ (where the user index $k$ is dropped without causing any confusion), the cascaded channel can be expanded as ${\vec{\bm H}}\triangleq \left[{\vec{\bm h}}_{1},{\vec{\bm h}}_{2},\ldots ,{\vec{\bm h}}_{N}\right]$, and thus the received signal at the BS in \eqref{Fully2} with $K=1$ can be expressed as
\begin{align}
{\bm y}_B^{(t)}&= \sqrt{P_u} \left( \left({\bm x}^{(t)}\right)^T \otimes {\bm I}_{M_B} \right) {\vec{\bm H}} {\bm \theta}^{(t)} + {\bm v}_B^{(t)}\label{Fully2.5}\\
&= \sqrt{P_u}\left( \left({\bm x}^{(t)}\right)^T \otimes {\bm I}_{M_B} \right) \left(\sum_{n=1}^{N} {\vec{\bm h}}_{n} {\theta}_n^{(t)} 
\right) + {\bm v}_B^{(t)}.\label{Fully3}
\end{align}
Based on \eqref{Fully3}, the training overhead for resolving the full CSI of $\left\{{\vec{\bm h}}_{n}\right\}_{n=1}^N$ is generally proportional to the number of reflecting elements $N$ and thus may incur a long estimation delay.}
As such, how to effectively reduce the training overhead for the IRS-aided single-user system becomes a critical problem, which has been extensively studied in the literature, as discussed below.

Based on the antenna configuration in the downlink, existing works on IRS channel estimation for
the single-user (point-to-point) system can be classified into single-input single-output (SISO), multiple-input single-output (MISO), and MIMO setups.
Some early works have tackled the IRS channel estimation problem under the SISO and MISO setups, where the direct/cascaded CSI was estimated at one or multiple antennas of the BS independently based on the
IRS training reflection pattern and the pilot signals sent by the single-antenna user \cite{mishra2019channel,yang2020intelligent,zheng2019intelligent,zheng2020fast,jensen2020optimal}.
However, these IRS channel estimation schemes cannot be effectively extended to the more general MIMO setup, since the pilot signals from multiple transmit antennas are intricately coupled at multiple receive antennas, thus calling for more advanced signal processing techniques to perform joint channel estimation.
By exploiting the channel sparsity and low-rank properties (e.g., in the mmWave or THz frequency bands), 
various cascaded channel estimation schemes based on compressed sensing were proposed for the IRS-aided single-user MIMO system \cite{ardah2021trice,jia2020high,he2021channel,he2020channel,ma2020joint,lin2020Channel,mirza2021channel,he2019cascaded,deepak2021channel,liu2020asymptotic}.
In addition, deep learning has emerged as a promising method to learn the direct/cascaded CSI efficiently from training pilots/data in the single-user MISO \cite{gao2021deep,xu2021ordinary,kundu2021channel,kundu2020deep} and MIMO \cite{li2020channel} systems.
In \cite{de2021channel,zegrar2020reconfigurable,he2019cascaded}, matrix factorization/decomposition was exploited to 
reduce the high dimensionality of the cascaded MIMO channel, so as to ease the channel estimation in the IRS-aided single-user MIMO system.

Besides, the channel estimation problem has been studied for other IRS-aided single-user communication systems.
Specifically, the authors of \cite{abeywickrama2021channel} considered an IRS-aided backscatter communication system and proposed an efficient LS-based channel estimation scheme under the SISO setup.
To support the high-mobility communication aided by IRS with massive passive elements, various channel estimation protocols and schemes were developed for the SISO \cite{huang2020transforming,huang2021transforming}, MISO \cite{mao2021channel,cai2021downlink}, and MIMO \cite{sun2020channel,zegrar2020general} setups to track the direct/cascaded CSI efficiently. 
{However, the IRS channel estimation design for the doubly-selective (or time-varying multipath) channel remains a very challenging problem, which deserves further investigation.}
\subsubsection{Multi-User System with Single IRS}\label{Fac_SU}
\begin{figure}[!t]
	\centering
	\includegraphics[width=3.5in]{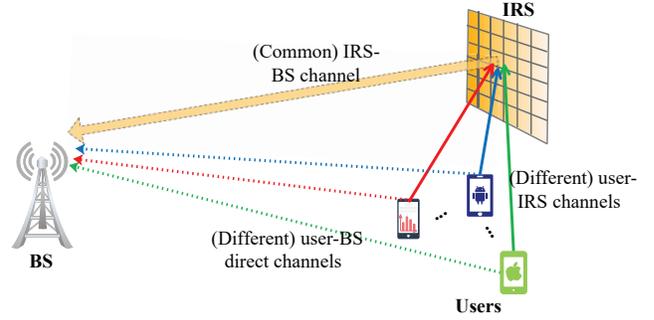}
	\caption{Uplink channel estimation for IRS-aided multi-user communications.}
	\label{common_CH}
\end{figure}
{Next, we consider the channel estimation for the IRS-aided multi-user system with the received signal model given in \eqref{Fully2},} where multiple users are served by a common IRS (or equivalently, multiple distributed IRSs at a given site). {Note that it is practically inefficient to directly apply the channel estimation schemes developed for the single-user case to the multi-user case (e.g., apply the user-by-user successive channel estimation \cite{nadeem2020intelligent,mishra2020passive,buzzi2021ris}), for which the received signal model in \eqref{Fully2} reduces to
\begin{align}\label{Fully_MU}
{\bm y}_{B,k}^{(t)}=\sqrt{P_u} \left( \left({\bm x}^{(t)}\right)^T \otimes {\bm I}_{M_B} \right){\vec{\bm H}}_{k} {\bm \theta}^{(t)} + {\bm v}_B^{(t)}, 
\end{align}
with $k=1,\ldots K$.
Similar to the single-user case given in \eqref{Fully2.5}, the cascaded CSI ${\vec{\bm H}}_{k}$ for each user is estimated separately based on each ${\bm y}_{B,k}^{(t)}$ in \eqref{Fully_MU} without the co-channel interference and thus
 the resultant training time will scale with the number of users and may become practically unaffordable if the number of users $K$ is large.} As such, to achieve high training efficiency, it is crucial to develop more efficient designs for the multi-user pilot sequence, IRS training reflection pattern, as well as channel estimation algorithm tailored for the IRS-aided multi-user system.
In the following, we provide an overview of the recent advances in this line of work.

{For the IRS-aided multi-user system illustrated in Fig.~\ref{common_CH}, all the users share the same (common) IRS-BS channel ${\bm G}$ in their respective cascaded user-IRS-BS channels $\left\{{\vec{\bm H}}_{k}\right\}_{k=1}^K$.
By leveraging this fact, the authors in \cite{wang2020channel} proposed to first estimate the cascaded CSI of one typical user (say, ${\vec{\bm H}}_{1}$ for user~1) at the BS, based on which the cascaded CSI of the remaining users (i.e., $\left\{{\vec{\bm H}}_{k}\right\}_{k=2}^K$) was then estimated with significantly reduced training overhead in the IRS-aided multi-user MISO system.} Similarly, the common IRS-BS channel property was also exploited in \cite{zheng2020intelligent}  to enhance the training efficiency as well as accommodate more users concurrently for channel estimation in the IRS-aided multi-user OFDM system. The channel estimation performance of \cite{wang2020channel} was later improved in \cite{weiyi2021channel} by jointly estimating the direct and cascaded channels in the IRS-aided multi-user MISO system. Moreover, as previously discussed, {the common IRS-BS channel ${\bm G}$} is typically quasi-static in practice. By exploiting this useful property, the authors in \cite{hu2021two} and \cite{guan2021anchor} proposed to resolve {the common (quasi-static) IRS-BS channel ${\bm G}$ first}, which was then leveraged as prior knowledge for {estimating different IRS-user channels $\left\{{\bm H}_{k}\right\}_{k=1}^K$ in real time} with reduced training overhead in the IRS-aided multi-user MISO system.

As for the cascaded channel estimation with fully-passive IRS,
various schemes have been proposed for the IRS-aided multi-user MISO system, by leveraging different signal processing techniques such as matrix factorization/decomposition \cite{de2021channel,weili2021channel}, compressed sensing \cite{liu2020matrix,he2021semi,shtaiwi2021channel}, and deep learning \cite{elbir2020deep,liu2020deepresidual}.
Note that those channel estimation schemes were typically applied to the uplink channel training, for which the BS needs to jointly estimate the direct/cascaded CSI from multiple users.
To simplify the joint channel estimation in the uplink, the authors in \cite{kang2020intelligent,weili2021channel} considered the downlink channel training where 
each user estimates its {individual direct/cascaded CSI (i.e., ${\bm D}_{k}$ and ${\vec{\bm H}}_{k}$) in parallel} based on the broadcast pilot signals from the BS that are reflected by the IRS with different training patterns. However, it is worth pointing out that for the multi-user downlink channel estimation, each user still needs to feed back its direct/cascaded CSI to a central processing unit (e.g., the BS), thus inevitably incurring high CSI feedback overhead.
On the other hand, with semi-passive IRS, separate channel estimation schemes based on sparse Bayesian learning \cite{jian2020modified} and canonical polyadic decomposition tensors \cite{lin2021tensor} were proposed for the multi-user MISO system over the flat-fading and frequency-selective fading channels, respectively. In particular, the (common) IRS-BS channel and (different) user-IRS channels are estimated in parallel at the dedicated sensing devices on the IRS.

\subsubsection{Double/Multi-IRS System}
\begin{figure*}[!t]
	\centering
	\includegraphics[width=6.0in]{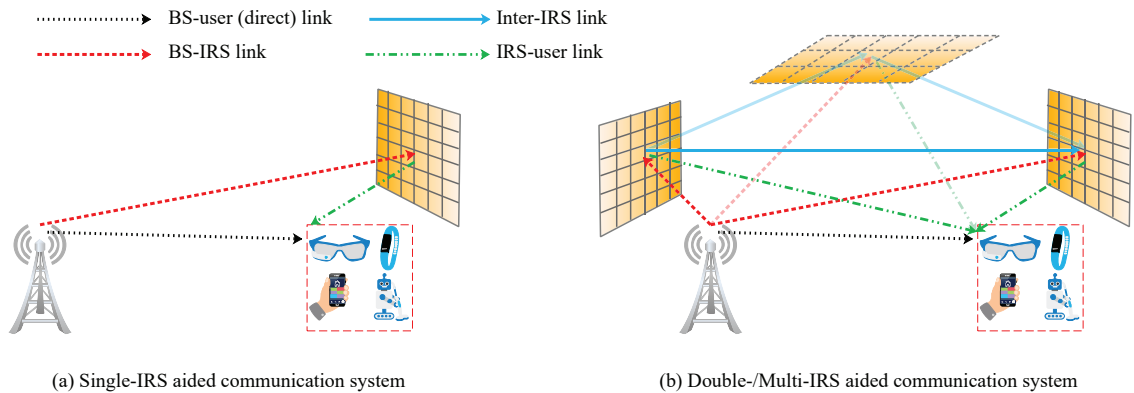}
	\caption{IRS-aided communication system with different IRS deployments.}
	\label{multiIRS}
\end{figure*}
Prior works on IRS channel estimation mainly considered the wireless system aided by one or more distributed IRSs with single signal reflection only (see Fig.~\ref{multiIRS}(a)), while ignoring the inter-IRS signal reflection for design simplicity.
Recently, the great potential of the cooperative passive beamforming over the inter-IRS channel has been uncovered in the double-/multi-IRS aided communication system \cite{zheng2021double,you2020wireless,han2020cooperative,mei2021multi}, which provides a higher-order passive beamforming gain 
than its single-IRS counterpart. Nevertheless, despite the more pronounced
passive beamforming gain, it gives rise to new and unique challenges in channel estimation for the double-/multi-IRS aided system. 
{For example, let us consider the double-IRS aided communication system where a new IRS (labeled as IRS~$2$) is deployed near the BS in addition to the conventional single IRS (labeled as IRS~$1$) deployed near the users \cite{zheng2021double}. As such, besides ${\bm G}$ and ${\bm H}_{k}$ for the pre-existing IRS~$1$-BS and user $k$-IRS~$1$ channels, we let ${\tilde{\bm G}}$ and ${\tilde{\bm H}}_{k}$
denote the IRS~$2$-BS and user $k$-IRS~$2$ channels due to the newly added IRS. Furthermore, there exists the inter-IRS channel between IRS~1 and IRS~2, which is denoted by ${\bm S}$.
Thus, the effective channel between user~$k$ and the BS is the superimposition of the double-reflection link, the two single-reflection links, and the direct link, which is given by
\begin{align}\label{Double}
{\bm E}_k=\underbrace{{\tilde{\bm G}} {\tilde{\bm \Theta}} {\bm S}{\bm \Theta} {\bm H}_{k}}_{\rm Double-reflection}+ \underbrace{{\tilde{\bm G}} {\tilde{\bm \Theta}} {\tilde{\bm H}}_{k} +{\bm G} {\bm \Theta} {\bm H}_{k}}_{\rm Single-reflection}+\underbrace{{\bm D}_{k}}_{\rm Direct},
\end{align}
where ${\bm \Theta}=\text{diag} \left({\bm \theta}\right)$ and ${\tilde{\bm \Theta}}=\text{diag} \left({\tilde{\bm \theta}}\right)$ denote the diagonal reflection matrices of IRS~1 and IRS~2, respectively.
Hence, as shown in \eqref{Double} and Fig.~\ref{multiIRS}(b), the IRS-aided links with different numbers of signal reflections are intricately coupled and also entail more channel coefficients for estimation, which renders existing techniques for single-IRS channel estimation inapplicable.}
In the following, we review the up-to-date research results for the channel estimation in the double-/multi-IRS aided system.

In \cite{han2020cooperative}, the authors considered two semi-passive IRSs for estimating their separate channels with the BS/user in the double-IRS aided single-user SISO system, where the inter-IRS channel is assumed to be LoS and simply determined by the geometry relationship between the two IRSs.
 Later, the double-IRS channel estimation with two fully-passive IRSs was investigated in \cite{you2020wireless} for the single-user SISO case. In particular, by assuming the blocked direct and single-reflection links, the authors in \cite{you2020wireless} proposed two effective channel estimation schemes for estimating the double-reflection channels over the two IRSs in the cases of general and LoS-dominant inter-IRS channels, respectively.
To achieve practically low training overhead, the authors in \cite{zheng2020uplink} proposed an efficient channel
estimation scheme based on an ON/OFF training reflection pattern to acquire the cascaded CSI of both the single- and double-reflection links in the double-IRS aided multi-user MISO system. In addition, to overcome the error propagation issue and
reflection power loss due to the ON/OFF reflection control in \cite{zheng2020uplink}, the authors in \cite{zheng2021efficient} proposed another efficient channel estimation scheme with the always-ON training reflection to jointly estimate the cascaded CSI of the single- and double-reflection links. This thus significantly improves the channel estimation accuracy by exploiting the full-reflection power gain in the double-IRS aided multi-user MISO system. In
particular, both double-IRS channel estimation schemes proposed in \cite{zheng2020uplink} and \cite{zheng2021efficient} are able to achieve low training overhead (which is comparable to that of the single-IRS counterpart), by exploiting the intrinsic relationship between the
single- and double-reflection channels as well as that among multiple users.
However, for the multi-IRS aided system with more than two cooperative IRSs, it remains unknown how to accurately and efficiently estimate the full CSI on all the involved single-/double-/multi-reflection links due to the drastically increasing number of coupled channel coefficients, which is worthy of further investigation in the future.

\subsubsection{Broadband System with Single IRS}
Moreover, it is important to consider channel estimation for the IRS-aided broadband system over frequency-selective fading channels, as today's wireless systems are typically broadband and OFDM-based.
{Different from the narrowband system where the cascaded channel is the ``product" of the user-IRS and IRS-BS (single-tap) channels as shown in \eqref{cascaded}, the cascaded channel in the broadband system becomes  the ``convolution" of the user-IRS and IRS-BS (multi-tap) channels. 
Specifically, let $L_h$ and $L_g$ denote the numbers of delayed taps in the time-domain channel impulse responses (CIRs) for the user-IRS and IRS-BS channels, respectively. For the purpose of exposition, we consider the single-user SISO setup, i.e., $K=M_B=M_u=1$.
Accordingly, we let ${\bm h}_{n}\in {\mathbb{C}^{L_h \times 1}}$ and ${\bm g}_n \in {\mathbb{C}^{L_g \times 1}}$ denote the time-domain CIRs from IRS element $n$ to the user and BS, respectively.
Thus, the effective cascaded channel from the user to the BS via each IRS element $n$ can be
expressed as the convolution of the user-IRS channel, the IRS reflection coefficient, and the IRS-BS channel, which is given by
\begin{align}\label{cascaded_OFDM}
{\bm g}_n  \ast \theta_n \ast {\bm h}_{n}=\theta_n {\bm g}_n  \ast {\bm h}_{n}= \theta_n {\bm q}_n, \qquad n=1,\ldots N,
\end{align}
where ${\bm q}_n\triangleq  {\bm g}_n  \ast {\bm h}_{n} \in {\mathbb{C}^{(L_h+L_g-1) \times 1}}$ denotes the cascaded user-IRS-BS channel (without taking the effect of IRS phase shift yet) associated with IRS element $n$ and $\ast$ denotes the convolution operation.
Due to the multi-path effect in the broadband system, it is expected that significantly more channel coefficients need to be estimated for the cascaded user-IRS-BS channels $\left\{{\bm q}_n\right\}_{n=1}^{N}$, as compared to the narrowband system.} Furthermore, for the OFDM-based broadband system, the passive IRS reflection is frequency-flat and hence affects the channel frequency response (CFR) at each OFDM subcarrier identically, thus lacking the design flexibility over different subcarriers.
Consequently, the channel estimation schemes designed for IRS-aided narrowband communication become inapplicable to their broadband counterparts in practice, which thus calls for more effective solutions. 
In the following, we review the existing results on the broadband IRS channel estimation.

For the first time, the authors in \cite{yang2020intelligent} and \cite{zheng2019intelligent} considered the IRS-aided single-user OFDM system and proposed the comb-type pilot schemes with the ON/OFF and full-ON training reflection designs for the IRS, respectively. Later, to reduce the large training delay arising from the long OFDM symbol duration, the authors in \cite{zheng2020fast} proposed two efficient schemes to accelerate the broadband channel estimation by redesigning the OFDM pilot symbol structures and IRS training reflection patterns.
Furthermore, the authors in \cite{zheng2020intelligent} proposed an efficient multi-user channel estimation scheme for the IRS-aided OFDM system, by multiplexing the pilot symbols of multiple users in the frequency domain over a large number of OFDM subcarriers.
By exploiting the common channel sparsity over different subcarriers in the mmWave frequency band, a distributed orthogonal matching pursuit (OMP) algorithm was proposed in \cite{wan2020Broadband} for the IRS-aided MISO-OFDM system to jointly estimate the direct and cascaded broadband channels of multiple users, where the BS-IRS channel is assumed to be LoS dominant and exploited as prior knowledge.
In \cite{wan2021terahertz}, the authors proposed a two-stage broadband channel estimation scheme for the IRS-aided THz massive MIMO system, consisting of a downlink coarse channel estimation stage and an uplink finer-grained channel estimation stage for the multi-user case.
In the IRS-aided OFDM system, deep learning-based schemes adopting the convolutional neural network (CNN) \cite{liu2020deep2} and federated learning \cite{elbir2020federated} were proposed to learn the direct/cascaded CSI under the MISO and MIMO setups, respectively.
On the other hand, for semi-passive IRS,
the authors in \cite{taha2021enabling} and \cite{taha2019deep} proposed the separate channel estimation schemes based on deep learning and compressed sensing for the single-user SISO-OFDM system over mmWave channels, respectively. 
Moreover, by leveraging the deep denoising neural network and exploiting the angular-domain sparsity of the mmWave channel, the authors in \cite{liu2020deep} developed a separate channel estimation scheme for the IRS-aided MIMO-OFDM system.

 \begin{table*}[!t]
 	\centering
 	\caption{{IRS Channel Estimation for Different System Setups}}\label{Setups}
 	\vspace{-0.3cm}
 	\resizebox{\textwidth}{!}{
 		\begin{tabular}{|m{1.8cm}<{\centering}|m{2cm}<{\centering}|m{1.8cm}<{\centering}|m{22.5cm}|}
 			\hline
 			{\bf IRS   Deployment} &
 			{\bf   System Setup} &
 			{\bf Antenna Setup (Downlink)} &
 			\qquad\qquad\qquad\qquad\qquad\qquad\qquad\qquad\qquad\qquad\qquad\qquad{\bf Main Results} \\ \hline\hline
 			\multirow{34}{*}{\bf Single-IRS} &
 			\multirow{10}{*}{\begin{tabular}[c]{@{}l@{}}Single-user,\\      narrowband\end{tabular}} &
 			SISO &
 			\begin{tabular}[c]{@{}l@{}}$\bullet$   (Progressive) cascaded channel estimation under discrete phase shift model   \cite{you2020channel,you2020intelligent}\\      $\bullet$ Cascaded channel estimation for IRS-aided backscatter communication   \cite{abeywickrama2021channel}\\      $\bullet$ Cascaded channel estimation based on compressed sensing by exploiting the channel sparsity with the derived CRLB \cite{noh2020training}\\      $\bullet$ Cascaded channel estimation for the high-mobility case with vehicle-side IRS   \cite{huang2020transforming,huang2021transforming}\\      $\bullet$ Separate channel estimation via a single RF chain   at IRS \cite{alexandropoulos2020hardware}\end{tabular} \\ \cline{3-4} 
 			&
 			&
 			MISO &
 			\begin{tabular}[c]{@{}l@{}}$\bullet$   Cascaded channel estimation using ON/OFF training reflection pattern  at IRS \cite{mishra2019channel}\\      $\bullet$ Cascaded channel estimation using full-ON training reflection pattern at IRS   \cite{jensen2020optimal}\\      $\bullet$ Cascaded channel estimation based on compressed sensing by exploiting sparsity of mmWave channels   \cite{wang2020compressed}\\      $\bullet$ Cascaded channel estimation based on deep learning techniques \cite{gao2021deep,xu2021ordinary,kundu2021channel,kundu2020deep}\\      $\bullet$ Cascaded channel estimation based on Kalman filter for the high-mobility case   \cite{mao2021channel,cai2021downlink}\end{tabular} \\ \cline{3-4} 
 			&
 			&
 			MIMO &
 			\begin{tabular}[c]{@{}l@{}}$\bullet$   Cascaded channel estimation based on compressed sensing by exploiting low-rank/sparse channels  \cite{ardah2021trice,jia2020high,he2021channel,he2020channel,ma2020joint,lin2020Channel,mirza2021channel,he2019cascaded,deepak2021channel,liu2020asymptotic}\\      $\bullet$ Cascaded channel estimation based on matrix factorization/decomposition   techniques \cite{de2021channel,zegrar2020reconfigurable,he2019cascaded}\\      $\bullet$ Cascaded channel estimation based on deep learning for THz channels   \cite{li2020channel}\\      $\bullet$ Cascaded channel estimation for the high-mobility case with fixed-position IRS   \cite{sun2020channel,zegrar2020general}\end{tabular} \\ \cline{2-4} 
 			&
 			\multirow{2}{*}{\begin{tabular}[c]{@{}l@{}}Multi-user,\\      narrowband\end{tabular}} &
 			MISO &
 			\begin{tabular}[c]{@{}l@{}}$\bullet$   Straightforward user-by-user (successive) cascaded channel estimation   \cite{nadeem2020intelligent,mishra2020passive,buzzi2021ris}\\      $\bullet$ Cascaded channel estimation by exploiting common IRS-BS channel   \cite{wang2020channel} and additional channel sparsity \cite{chen2019channel,weixiuhong2021channel2}\\      $\bullet$ Cascaded channel estimation based on dual-link (BS-IRS-BS) reflection   \cite{hu2021two} and anchor nodes \cite{guan2021anchor} to resolve common IRS-BS \\channel (offline) and IRS-user channel (online) sequentially\\      $\bullet$ Cascaded channel estimation based on LMMSE criterion in the downlink   \cite{kang2020intelligent}\\      $\bullet$ Cascaded channel estimation based on matrix factorization/decomposition   \cite{de2021channel,weili2021channel} and additional channel sparsity    \cite{liu2020matrix,he2021semi,shtaiwi2021channel}\\      $\bullet$ Cascaded channel estimation based on deep learning techniques using convolutional neural network  \cite{elbir2020deep,liu2020deepresidual}\\ 	$\bullet$   Separate channel estimation based on sparse Bayesian learning \cite{jian2020modified}\end{tabular} \\ \cline{2-4} 
 			&
 			\multirow{2}{*}{\begin{tabular}[c]{@{}l@{}}Single-user,\\      broadband\end{tabular}} &
 			SISO &
 			\begin{tabular}[c]{@{}l@{}}$\bullet$   Cascaded channel estimation using ON/OFF training reflection pattern and element-grouping   strategy at IRS \cite{yang2020intelligent}\\      $\bullet$ Cascaded channel estimation using DFT-based training reflection pattern and   element-grouping strategy at IRS \cite{zheng2019intelligent}\\      $\bullet$ Fast cascaded channel estimation using (sampling-wise) full-ON training   reflection pattern (using circulant matrix) at IRS   \cite{zheng2019intelligent}\\      $\bullet$ Separate channel estimation based on deep learning/compressed sensing for mmWave channels   \cite{taha2019deep,taha2021enabling}\end{tabular} \\ \cline{3-4} 
 			&
 			&
 			MISO &
 			$\bullet$   Cascaded channel estimation based on single convolutional neural network to reduce the training complexity \cite{liu2020deep2} \\ \cline{3-4} 
 			&
 			&
 			MIMO &
 			$\bullet$   Separate channel estimation based on deep denoising neural network  for mmWave channels   \cite{liu2020deep} \\ \cline{2-4} 
 			&
 			\multirow{6}{*}{\begin{tabular}[c]{@{}l@{}}Multi-user,\\      broadband\end{tabular}} &
 			SISO &
 			\begin{tabular}[c]{@{}l@{}}$\bullet$   Cascaded channel estimation by exploiting common IRS-BS channel and optimal training design   \cite{zheng2020intelligent}\\      $\bullet$ Cascaded channel estimation accounting for phase-dependent amplitude  variation \cite{yang2021channel} and   carrier frequency offset \cite{jeong2021low}\end{tabular} \\ \cline{3-4} 
 			&
 			&
 			MISO &
 			\begin{tabular}[c]{@{}l@{}}$\bullet$   Cascaded channel estimation based on OMP algorithm by exploiting common channel sparsity\\
 				 over different   subcarriers \cite{wan2020Broadband}\\      $\bullet$ Separate channel estimation based on canonical polyadic decomposition tensors   \cite{lin2021tensor}\end{tabular} \\ \cline{3-4} 
 			&
 			&
 			MIMO &
 			\begin{tabular}[c]{@{}l@{}}$\bullet$   Cascaded channel estimation based on deep learning techniques using federated learning   \cite{elbir2020federated}\\      $\bullet$ Cascaded channel estimation based on compressed sensing by exploiting dual sparsity of THz MIMO   channels \cite{wan2021terahertz}\end{tabular} \\ \hline
 			\multirow{5}{*}{\bf Multi-IRS} &
 			\multirow{3}{*}{\begin{tabular}[c]{@{}l@{}}Single-user,\\      narrowband\end{tabular}} &
 			SISO &
 			$\bullet$   Double-IRS channel estimation under Rician fading inter-IRS channel \cite{you2020wireless} \\ \cline{3-4} 
 			&
 			&
 			MISO &
 			$\bullet$   Double-IRS channel estimation by exploiting common BS-IRS channel as well as channel   relationship between single and double-reflection links   \cite{zheng2021efficient,zheng2020uplink} \\ \cline{2-4} 
 			&
 			\begin{tabular}[c]{@{}l@{}}Multi-user,\\      narrowband\end{tabular} &
 			MISO &
 			$\bullet$   Double-IRS channel estimation by exploiting common BS-IRS and inter-IRS channels as well as   channel relationship among different users   \cite{zheng2021efficient,zheng2020uplink} \\ \hline
 		\end{tabular}
 	}
 \end{table*}

In Table \ref{Setups}, we summarize the up-to-date research works on IRS channel estimation according to different system setups.
It is noted that since the double-/multi-IRS channel estimation is still in its infancy, more research endeavor along this direction is desired for devising more efficient channel estimation schemes.
Furthermore, the existing works on the double-/multi-IRS channel estimation only considered the narrowband setup over flat-fading channels, while their extensions to the broadband systems over frequency-selective fading channels remain open and deserve further studies. 


\subsection{Signal Processing Methods for IRS Channel Estimation}
Under different IRS channel models,
substantial research efforts have been devoted to designing efficient IRS channel estimation schemes based on various signal processing methods, such as LS/LMMSE, compressed sensing, matrix factorization, and deep learning in the current literature.
In this subsection, we classify different IRS channel estimation schemes according to their adopted signal processing methods as shown in Fig.~\ref{Organization_CE}, and provide in-depth discussions on their working principles as well as applicable models/scenarios. 
\subsubsection{Classical Channel Estimation}\label{LS/LMMSE}
As one classical approach to approximate the solution of over-determined linear models, the LS/LMMSE method has been widely applied to the pilot-based channel estimation in the IRS-aided system owing to its low complexity in practical implementation. 
{For the LS/LMMSE-based IRS channel estimation, the number of observations/measures generally needs to be no less than that of unknown channel parameters to avoid ambiguity.
For example, let us consider the received signal model in \eqref{Fully2.5} for the single-user case, which can be further expressed as (by exploiting the property of the Kronecker product)
\begin{align}\label{Fully_LS}
{\bm y}_B^{(t)}&= \sqrt{P_u} \Big( \underbrace{\left( {\bm \theta}^{(t)} \otimes {\bm x}^{(t)}\right)^T \otimes {\bm I}_{M_B}}_{{\bm F}^{(t)} } \Big){\vec{\bm h}} + {\bm v}_B^{(t)},
\end{align}
where ${\vec{\bm h}}\triangleq\text{vec}\left( {\vec{\bm H}} \right)$ is the cascaded channel vector and ${\bm F}^{(t)} \in {\mathbb{C}^{ M_B \times M_B M_u N}}$ can be regarded as the observation matrix for ${\vec{\bm h}}$. Apparently, the cascaded channel vector ${\vec{\bm h}}$ is underdetermined in \eqref{Fully_LS} and thus more observations over multiple pilot symbols are needed.
Let $T$ be the number of pilot symbols during the channel training period, and by stacking the received signal vectors $\left\{{\bm y}_B^{(t)}\right\}_{t=1}^T$ into ${\bm y}_B$, we have
\begin{align}\label{rec_pilot2}
\underbrace{\begin{bmatrix}
	{\bm y}_B^{(1)}\\
	\vdots\\
	{\bm y}_B^{(T)}\end{bmatrix}}_{{\bm y}_B}
=\sqrt{P_u}
\underbrace{\begin{bmatrix}
	{\bm F}^{(1)}\\
	\vdots\\
	{\bm F}^{(T)}\end{bmatrix}}_{{\bm F}}
{\vec{\bm h}}
+\underbrace{\begin{bmatrix}
	{\bm v}_B^{(1)}\\
	\vdots\\
	{\bm v}_B^{(T)}\end{bmatrix}}_{{\bm v}_B},
\end{align}
where ${\bm F} \in {\mathbb{C}^{ M_B T \times M_B M_u N}}$ is the overall observation matrix that depends on the IRS training reflection pattern $\left\{{\bm \theta}^{(t)}\right\}_{t=1}^T$ and the transmit pilot sequence $\left\{{\bm x}^{(t)}\right\}_{t=1}^T$. Note that to uniquely estimate ${\vec{\bm h}}$, ${\bm F}$ needs to be of full column rank, which requires $T\ge M_u N$. As such, given that ${\bm F}$ is of full column rank, there are two classic approaches to estimate ${\vec{\bm h}}$,
elaborated as follows.

{\bf LS Estimation:} The LS-based channel estimation is formulated as
\begin{align}
{\hat{\bm h}}_{{\rm LS}}=\arg \underset{ {\vec{\bm h}} }{\text{min}}~\left\| {\bm y}_B - \sqrt{P_u} {\bm F}{\vec{\bm h}} \right\|^2,
\end{align}
for which the closed-form LS solution is given by 
\begin{align}
{\hat{\bm h}}_{{\rm LS}}=\frac{1}{\sqrt{P_u}} {\bm F}^{\dagger}{\bm y}_B ={\vec{\bm h}}+\frac{1}{\sqrt{P_u}} {\bm F}^{\dagger} {\bm v}_B,
\end{align}
where ${\bm F}^{\dagger}=\left({\bm F}^H {\bm F}\right)^{-1}{\bm F}^H $.

{\bf LMMSE Estimation:} Different from the LS estimation without assuming any prior knowledge, the LMMSE estimation aims to minimize the overall MSE by exploiting the second order statistics of both the channel and the noise. Specifically, the LMMSE-based channel estimation is formulated as
\begin{align}
{\bm W}_{\rm LM}=\arg \underset{ {\bm W} }{\text{min}}~{\mathbb E} \left\{\left\| {\bm W}{\bm y}_B -{\vec{\bm h}} \right\|^2\right\},
\end{align}
with ${\hat{\bm h}}_{{\rm LM}}={\bm W}_{\rm LM}{\bm y}_B$, for which the closed-form LMMSE solution is given by
\begin{align}
{\hat{\bm h}}_{{\rm LM}}=\sqrt{P_u} {\bm R}_{{\vec{\bm h}}} {\bm F}^H
\left(P_u {\bm F} {\bm R}_{{\vec{\bm h}}} {\bm F}^H + \sigma_B^2 {\bm I}_{M_B T} \right)^{-1} {\bm y}_B ,
\end{align}
where we assume ${\mathbb E} \left\{ {\vec{\bm h}} \right\}={\bm 0}$, ${\bm R}_{{\vec{\bm h}}}\triangleq {\mathbb E} \left\{ {\vec{\bm h}}{\vec{\bm h}}^H \right\}$ denotes the spatial correlation matrix of ${\vec{\bm h}}$, and $\sigma_B^2$ is the noise variance at the BS.

It is noted that to ensure the feasibility of the above LS/LMMSE estimation as well as minimize the channel estimation error, the IRS training reflection pattern $\left\{{\bm \theta}^{(t)}\right\}_{t=1}^T$ and transmit pilot sequence $\left\{{\bm x}^{(t)}\right\}_{t=1}^T$ need to be carefully designed in the observation matrix ${\bm F}$. In the following, we provide an overview of the existing works on the LS/LMMSE-based IRS channel estimation for a wide variety of channel models.}

\begin{figure}[!t]
	\centering
	\includegraphics[width=3.3in]{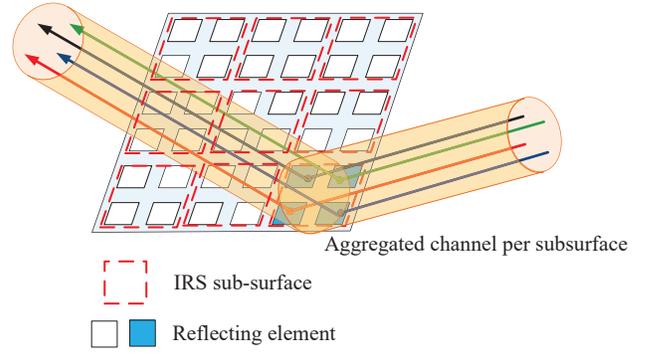}
	\caption{IRS element-grouping.}
	\label{IRSgrouping}
\end{figure}
In \cite{mishra2019channel,yang2020intelligent,mishra2020passive}, the ON/OFF training reflection pattern was adopted at the IRS to facilitate the LS estimation in a decoupled manner. Specifically, the direct user-BS channel is first estimated (e.g., using the conventional LS-based channel estimation) with all the IRS elements turned OFF and the cascaded user-IRS-BS channel is then estimated with each one of the IRS elements turned ON (i.e., with the others set OFF) sequentially over time. 
Although it is simple for implementation, the ON/OFF training reflection pattern suffers from substantial reflection power loss and direct-channel interference, both of which degrade its channel estimation accuracy.
To overcome these issues, the full-ON IRS training reflection pattern (i.e., all the reflecting elements are switched ON throughout the entire channel training time) was later developed in \cite{zheng2019intelligent,jensen2020optimal}, which substantially improved the LS estimation accuracy by leveraging the full IRS aperture gain.
Furthermore, in order to achieve optimal performance for LS/LMMSE estimation subject to the limited training time and transmit power, the IRS training reflection pattern and the transmit pilot sequence were jointly designed in \cite{zheng2020fast,kang2020intelligent,zheng2020intelligent}.
Besides, a novel element-grouping strategy was proposed in \cite{yang2020intelligent,zheng2019intelligent} by properly grouping adjacent IRS elements (typically with sub-wavelength inter-element distance and thus with high channel correlation in practice) into a sub-surface.
{Suppose that the IRS composed of $N$ reflecting elements is divided into ${\bar N}$ subsurfaces, each of which consists of $B=N/{\bar N}$ (assumed to be an integer for convenience) adjacent elements that share a common reflection coefficient as illustrated in Fig.~\ref{IRSgrouping}. Accordingly, the IRS reflection vector can be re-expressed as ${\bm \theta}={\bar{\bm \theta}}\otimes {\bm 1}_{B \times 1}$ and thus the cascaded channel in \eqref{Fully2.5} becomes
\begin{align}
{\vec{\bm H}} {\bm \theta}={\vec{\bm H}}\left({\bar{\bm \theta}}\otimes {\bm 1}_{B \times 1}\right)=\left[{\bar{\bm h}}_{1},{\bar{\bm h}}_{2},\ldots, {\bar{\bm h}}_{{\bar N}} \right] {\bar{\bm \theta}}={\bar {\bm H}} {\bar{\bm \theta}},
\end{align}
where ${\bar{\bm \theta}}\in {\mathbb{C}^{{\bar N} \times 1}}$ represents the IRS grouping reflection vector,
${\bar {\bm H}} \in {\mathbb{C}^{M_B M_u \times {\bar N}}} $ is the effective cascaded channel after the element-grouping strategy, and ${\bar{\bm h}}_{{\bar n}}=\sum_{b=1}^{B} {\vec{\bm h}}_{b+({\bar n}-1)B}$ denotes the 
equivalent aggregated channel of subsurface ${\bar n}$ with ${\bar n}=1,2,\ldots, {\bar N}$.
As a result, it suffices to estimate the equivalent aggregated channel of each subsurface (as illustrated in Fig.~\ref{IRSgrouping}) only, thereby reducing the training overhead by a multiplication factor of $B$.
In addition, by adjusting the size of each subsurface, i.e., $B$, 
the element-grouping strategy provides a flexible trade-off between training overhead/design complexity and passive
beamforming performance, without the need of assuming any specified channel model.}

Besides the designs of the IRS training reflection pattern and transmit pilot sequence, various training protocols have also been proposed to improve the training efficiency of the LS/LMMSE estimation, especially for the multi-user/IRS case that requires efficient coordination.
For example, the reference-user based IRS channel estimation schemes were proposed for the multi-user narrowband and broadband systems in \cite{wang2020channel} and \cite{zheng2020intelligent}, respectively. The key idea is that given an arbitrary user's cascaded channel as the reference CSI, the other users' cascaded channels can be expressed as its lower-dimensional scaled versions and thus efficiently estimated using the LS method at the BS with substantially reduced training overhead.
Moreover, the authors in \cite{hu2021two} proposed a dual-link (BS-IRS-BS) cascaded channel estimation scheme to resolve the common IRS-BS channel via the LS estimation based on the pilot signals sent from the BS and then reflected back by the IRS.
Furthermore, the authors in \cite{guan2021anchor} proposed an anchor-aided cascaded channel estimation scheme to resolve the common IRS-BS channel by leveraging the training and feedback from two dedicated anchor nodes deployed near the IRS.
With the resolved CSI of the common IRS-BS channel, the dynamic IRS-user channels were then efficiently estimated using the LS method with reduced real-time training overhead \cite{hu2021two,guan2021anchor}.
Besides single-IRS channel estimation, the double-IRS channel estimation schemes based on the LS method were also proposed in \cite{you2020wireless,zheng2020uplink,zheng2021efficient} with different training protocols to achieve practically low training overhead.

Due to its low complexity for implementation, the LS/LMMSE-based IRS channel estimation can also be efficiently applied to the highly dynamic wireless environment aided by IRS.
For example, to overcome the long training delay before data transmission, a novel hierarchical training
reflection pattern was proposed in \cite{you2020channel} to progressively resolve the
cascaded CSI via the LS estimation, based on which the passive beamforming design was successively refined.
In \cite{huang2020transforming} and \cite{huang2021transforming}, the authors considered the high-mobility communication aided by the IRS deployed at the vehicle side and proposed a low-complexity LS-based channel estimation scheme to track the channel variation efficiently.
In \cite{mao2021channel}, by exploiting the quasi-static BS-IRS channel and 
modeling the time-variant user-IRS channel as a first-order auto-regressive process, the Kalman filter was employed to track the time variation of cascaded BS-IRS-user channel efficiently in the high-mobility scenario. Furthermore, \cite{cai2021downlink} employed two Kalman filters to track the time-varying direct and cascaded channels in parallel.
 In \cite{zegrar2020general}, with the quasi-static BS-IRS channel estimated using the hierarchical beam searching algorithm,
the extended Kalman filter algorithm was then applied to estimate and track the dynamic user-IRS channel efficiently.

\subsubsection{Compressed Sensing}
For the IRS-aided communication system operating at high frequencies such as mmWave and THz frequency bands, there are only a limited number of scattering paths between the IRS and the BS/user due to the severe path-loss over distance and occasional blockage. {Thus, the IRS-associated channels, i.e., ${\bm G}$ and $\left\{{\bm H}_{k}\right\}_{k=1}^K$, in mmWave/THz frequency bands usually exhibit strong sparsity and low rank in the spatial/angular domain, which can be exploited to reduce the training overhead.
Specifically,
let ${\bm A}_B \in {\mathbb{C}^{M_B \times L_B }}$, ${\bm A}_R \in {\mathbb{C}^{N \times L_R }}$, and ${\bm A}_u \in {\mathbb{C}^{M_u \times L_u }}$ denote the over-complete dictionary matrices whose columns are the array response vectors sampled on the $L_B$, $L_R$, and $L_u$ possible AoA grids at the BS, IRS, and users, respectively.
Based on the geometric channel model, the IRS-BS and user-IRS channels can be expressed as
\begin{align}
\text{vec} \left( {\bm G} \right)&=\left({\bm A}_R^* \otimes {\bm A}_B \right) {\bm \rho}_G,\label{channel_G}\\
\text{vec} \left( {\bm H}_{k} \right)&=\left({\bm A}_u^* \otimes {\bm A}_R \right) {\bm \rho}_k, \label{channel_H}
\end{align}
where ${\bm \rho}_G \in {\mathbb{C}^{L_B L_R \times 1}}$ and ${\bm \rho}_k \in {\mathbb{C}^{L_R L_u \times 1}}$ are the $d_G$-sparse and $d_k$-sparse vectors, respectively, and $d_G$ and $d_k$ are the numbers of spatial paths in ${\bm G}$ and ${\bm H}_{k}$ with $d_G\ll L_B L_R$ and $d_k\ll L_R L_u$, respectively. By substituting \eqref{channel_G} and \eqref{channel_H} into \eqref{cascaded}, the cascaded channel will also exhibit strong sparsity but requires proper sparse representations.}
Accordingly, compressed sensing becomes a promising tool for IRS channel estimation, by exploiting its effectiveness in sensing the sparsity of channels.
In the following, we provide an overview of the recent advances in the compressed-sensing based channel estimation.

For IRS-aided communication systems with sparse channels,
the cascaded channel estimation can be converted into a sparse signal recovery problem \cite{wang2020compressed,ardah2021trice,jia2020high} and then solved via existing compressed sensing approaches efficiently \cite{he2021channel,he2020channel,ma2020joint,lin2020Channel,mirza2021channel}. 
Specifically, it was shown in \cite{chen2019channel} and \cite{weixiuhong2021channel2} that due to the sparsity of the common IRS-BS channel, the cascaded channel matrices of all users share a common row-column-block sparsity structure.  Accordingly, the compressed-sensing based channel estimation schemes were proposed to jointly recover the cascaded CSI for all users with low training overhead \cite{chen2019channel,weixiuhong2021channel2}. The dual sparsity of THz MIMO channels in both the angular and delay domains was later exploited in \cite{wan2021terahertz} to facilitate the compressed-sensing based broadband channel estimation with reduced training overhead. Moreover, the corresponding Cram\'{e}r-Rao lower bound (CRLB) on the estimation error was theoretically characterized in \cite{liu2020asymptotic} and \cite{noh2020training}. 

On the other hand, by exploiting the sparsity in IRS-associated channels, a wide variety of compressed sensing algorithms have been developed for IRS channel estimation.
For example, OMP is a low-complexity algorithm that finds the ``best matching" projections of the received channel measurement,
which has been applied in \cite{wang2020compressed,ardah2021trice,wan2020Broadband} to solve the cascaded channel estimation problem in the beamspace/angular domain.
Moreover, in \cite{he2019cascaded,he2021semi,mirza2021channel,liu2020matrix}, the (approximate) message passing algorithms were applied to solve the cascaded channel estimation problem by performing inference on graphical models in an iterative manner. 
Other compressed-sensing based algorithms such as adaptive grid matching pursuit \cite{jia2020high}, atomic norm minimization \cite{he2021channel}, iterative reweighted method \cite{he2020channel}, iterative atom pruning based subspace pursuit \cite{ma2020joint}, manifold optimization\cite{lin2020Channel}, and sparse Bayesian learning \cite{jian2020modified} have been applied to solve the IRS channel estimation problem.
For the compressed-sensing based algorithms,
better channel estimation performance generally comes at the expense of higher computational complexity and their cost-performance trade-offs deserve further investigation.

\subsubsection{Matrix Factorization/Decomposition}
For fully-passive IRS, the cascaded user-IRS-BS channel is estimated at the BS/users as the product of the user-IRS and IRS-BS channel matrices corrupted by noises (cf. \eqref{Fully} in Section~\ref{Rep_cascaded}), which can be treated as a bilinear channel estimation problem. As compared to the linear channel estimation problem in conventional systems without IRS, the bilinear channel estimation problem is generally more difficult to solve, due to the high-dimensional cascaded channel.
One strategy to overcome this difficulty is to decompose the high-dimensional cascaded channel into a series of lower-dimensional sub-channels that are easier to estimate with lower training overhead.
{However, one key issue in the matrix factorization/decomposition is that there is a scaling
ambiguity for resolving the IRS-BS channel ${\bm G}$ and the user-IRS channel ${\bm H}_{k}$. Specifically, for any invertible $N \times N$ diagonal matrix ${\bm \Lambda}$, we have 
\begin{align}
{\bm G} {\bm \Theta} {\bm H}_{k}={\bm G} {\bm \Lambda} {\bm \Theta} {\bm \Lambda}^{-1} {\bm H}_{k}={\bm G}'{\bm \Theta} {\bm H}'_{k},
\end{align}
where ${\bm G}'\triangleq {\bm G} {\bm \Lambda}$ and ${\bm H}'_{k}\triangleq {\bm \Lambda}^{-1} {\bm H}_{k}$. This implies that based on the received signal model in \eqref{Fully}, it is unable to uniquely resolve the IRS-BS channel ${\bm G}$ and the user-IRS channel ${\bm H}_{k}$ in a separate manner. Fortunately, there is generally no need to address this ambiguity issue when designing the IRS passive beamforming without loss of optimality. In the following, we review the existing works on IRS channel estimation based on the design philosophy of matrix factorization/decomposition.}

In \cite{de2021channel}, the authors proposed two cascaded channel estimation schemes based on a parallel factor tensor modeling of the received signals, which effectively unfolds/decomposes the 3D cascaded MIMO channel into the 2D user-IRS and IRS-BS MIMO channels for achieving efficient channel estimation. 
Moreover, by swapping the roles between the multi-antenna BS and multiple users, the key methods and results based on the parallel factor tensor modeling can be similarly applied to the downlink, where all users estimate their respective cascaded channels with the BS in parallel \cite{weili2021channel}.
By modeling the IRS-aided MIMO channels as the keyhole MIMO channels, a cascaded channel estimation scheme based on singular value decomposition (SVD) was proposed in \cite{zegrar2020reconfigurable}, where the cascaded channel matrix was decomposed into a series of rank-one matrices, each corresponding to one IRS element.
Moreover, under the sparse cascaded MIMO channel model, the sparse matrix factorization and reconstruction problems were investigated in \cite{he2019cascaded,he2021semi,liu2020matrix,shtaiwi2021channel}.

\subsubsection{Deep Learning}
As a powerful tool for tackling nonlinear mapping problems, the deep learning technique can also be applied for IRS channel estimation by learning an approximate mapping function from the (input) training data to the (output) separate/cascaded CSI. {Specifically, one can take the pilot symbols $\left\{{\bm x}_k^{(t)}\right\}_{t=1}^T$, the IRS training reflections $\left\{{\bm \theta}^{(t)}\right\}_{t=1}^T$, and the expected CIRs $\left\{{\vec{\bm H}}_{k}\right\}_{k=1}^K$ as the labeled data for the deep learning process, thereby establishing a fingerprinting database that stores the CSI estimates.}
In \cite{gao2021deep} and \cite{xu2021ordinary}, the synthetic deep neural network (DNN) and CNN were respectively utilized to estimate the cascaded channel with reduced training overhead in real time.
In \cite{kundu2021channel} and \cite{kundu2020deep}, two CNN-based cascaded channel estimation methods were proposed to perform the denoising process and approximate the optimal channel estimation solution based on the minimum mean-squared-error (MMSE) criteria, which outperform their linear channel estimation counterparts.
Moreover, in \cite{li2020channel}, the cascaded channel estimation problem was first formulated as a sparse signal recovery problem for the IRS-aided THz MIMO system and then effectively solved via the deep-learning technique by learning the mapping from the received signals to the cascaded channel path gains.
Based on simulated input signals and the expected output channel vectors, the authors in \cite{elbir2020deep} employed the twin CNN to estimate both the direct and cascaded channels for multiple users.
In addition, the single CNN was employed in \cite{liu2020deep2} to estimate both the direct and cascaded CFRs for the IRS-aided MISO-OFDM system, by designing a proper database to train the CNN in an offline manner.
In \cite{liu2020deepresidual}, the multi-user cascaded channel estimation was first formulated as a denoising problem and then solved under a CNN-based deep residual learning framework for refining the channel coefficients estimated from the noisy pilot-based observations. 
By exploiting the angular-domain sparsity of the mmWave channel, the authors in \cite{liu2020deep} developed a separate channel estimation scheme based on the deep denoising neural network.
Besides CNN and DNN,
other deep learning techniques, e.g., federated/supervised/reinforcement learning    \cite{elbir2020federated,taha2019deep,taha2021enabling,taha2020deep}, have also been applied to acquire the separate/cascaded CSI required by different IRS systems.

\begin{table*}[!t]
	\centering
	\caption{{Different Signal Processing Methods for IRS Channel Estimation}}\label{SP_method}
	\resizebox{\textwidth}{!}{
		\begin{tabular}{|m{3cm}<{\centering}|l|m{4.5cm}<{\centering}|m{5cm}<{\centering}|}
			\hline
			{\bf   Signal Processing Method} &
			\qquad\qquad\qquad\qquad\qquad\qquad\quad {\bf Main Contributions } &
			{~\bf Applicable Channel Model} &
			{\bf Typical Scenario} \\ \hline\hline
			\multirow{10}{*} { \begin{tabular}[c]{@{}l@{}} {\bf ~ Least square/}\\ {\bf minimum mean}\\ ~~{\bf square error} \end{tabular}} &
			\begin{tabular}[c]{@{}l@{}}{\bf   Training design}\\      $\bullet$ ON/OFF training reflection pattern   \cite{mishra2019channel,yang2020intelligent,mishra2020passive}\\      $\bullet$ Full-ON training reflection pattern   \cite{zheng2019intelligent,jensen2020optimal}\\      $\bullet$ Joint pilot sequence and IRS training reflection design   \cite{kang2020intelligent,zheng2020intelligent,zheng2020fast}\\      $\bullet$ IRS element-grouping strategy \cite{yang2020intelligent,zheng2019intelligent}\end{tabular} &
			\multirow{8}{*}{\begin{tabular}[c]{@{}l@{}}$\bullet$ General channel models \\      including flat fading and\\       frequency-selective\\      fading channels;\\$\bullet$ Over-determined linear \\signal model\end{tabular}} &
			\begin{tabular}[c]{@{}l@{}}Various   cases including \\      narrowband/broadband, \\      SISO/SIMO/MISO/MIMO, \\      and single-/multi-user cases\end{tabular} \\ \cline{2-2} \cline{4-4} 
			&
			\begin{tabular}[c]{@{}l@{}}{\bf Training protocol}\\      $\bullet$ Reference-user based cascaded channel estimation   \cite{wang2020channel,zheng2020intelligent,weiyi2021channel}\\      $\bullet$ Dual-link (BS-IRS-BS) cascaded channel estimation \cite{hu2021two}\\      $\bullet$ Anchor-aided channel estimation \cite{guan2021anchor}\\      $\bullet$ Double-IRS channel estimation   \cite{you2020wireless,zheng2020uplink,zheng2021efficient}\end{tabular} &
			&
			Multi-user/IRS case \\ \cline{2-2} \cline{4-4} 
			&
			\begin{tabular}[c]{@{}l@{}}{\bf Adaptive processing}\\      $\bullet$ Progressive channel estimation \cite{you2020channel,you2020intelligent}\\      $\bullet$ Vehicle-side IRS aided channel estimation    \cite{huang2020transforming,huang2021transforming}\\      $\bullet$ Kalman filter   \cite{mao2021channel,cai2021downlink,zegrar2020general}\end{tabular} &
			&
			\begin{tabular}[c]{@{}l@{}}Highly dynamic wireless\\ environment \end{tabular} \\ \hline
			\multirow{8}{*}{\begin{tabular}[c]{@{}l@{}} {\bf  Compressed} \\~~~~{\bf  sensing}   \end{tabular}} &
			\begin{tabular}[c]{@{}l@{}}{\bf   Sparse representation}\\      $\bullet$ Sparsity structure of cascade BS-IRS-user channel   \cite{wang2020compressed,ardah2021trice,jia2020high}\\      $\bullet$ Sparsity structure of the common IRS-BS channel   \cite{chen2019channel,weixiuhong2021channel2}\\      $\bullet$ Dual sparsity in angular and delay domains   \cite{wan2021terahertz}\end{tabular} &
			\multirow{6}{*}{\begin{tabular}[c]{@{}l@{}}$\bullet$ Low-rank/sparse   channel \\      model;\\$\bullet$ Under-determined linear\\ signal model \end{tabular}} &
			\multirow{6}{*}{mmWave   and THz} \\ \cline{2-2}
			&
			\begin{tabular}[c]{@{}l@{}}{\bf Signal processing algorithm}\\      $\bullet$ (Approximate) message passing   \cite{he2019cascaded,he2021semi,mirza2021channel,liu2020matrix}\\      $\bullet$ Orthogonal matching pursuit (OMP)  \cite{wang2020compressed,ardah2021trice,wan2020Broadband}\\      $\bullet$ Other algorithms: atomic norm minimization \cite{he2021channel}, adaptive grid matching pursuit   \cite{jia2020high}, \\       iterative reweighted method   \cite{he2020channel}, iterative atom pruning based subspace \\      pursuit \cite{ma2020joint}, manifold optimization\cite{lin2020Channel},   sparse Bayesian learning \cite{jian2020modified}\end{tabular} &
			&
			\\ \hline
			\multirow{5}{*}{\begin{tabular}[c]{@{}l@{}}~~~~{\bf  Matrix} \\      {\bf  factorization}\end{tabular}} &
			$\bullet$   Tensor modeling \cite{de2021channel,weili2021channel,lin2021tensor} &
			$\bullet$ General   channel model &
			\multirow{4}{*}{MIMO   and multi-user cases} \\ \cline{2-3}
			&
			$\bullet$ Singular value decomposition \cite{zegrar2020reconfigurable} &
			$\bullet$ Keyhole MIMO channel   model &
			\\ \cline{2-3}
			&
			$\bullet$ Bilinear/trilinear matrix factorization \cite{he2019cascaded,he2021semi,liu2020matrix,shtaiwi2021channel} &
			$\bullet$ Sparse channel model &
			\\ \hline
			\multirow{3}{*}{\begin{tabular}[c]{@{}l@{}}~~{\bf  Deep} \\      {\bf  learning}\end{tabular}} &
			$\bullet$   Deep neural network-based training \cite{gao2021deep,liu2020deep,li2020channel} &
			\multirow{3}{*}{$\bullet$ Nonlinear signal model} &
			\multirow{3}{*}{\begin{tabular}[c]{@{}l@{}}Various   cases including\\       narrowband/broadband \\      and single-/multi-user cases\end{tabular}} \\ \cline{2-2}
			&
			$\bullet$ Convolutional neural network-based training \cite{xu2021ordinary,kundu2021channel,kundu2020deep,elbir2020deep,liu2020deepresidual,liu2020deep2} &
			&
			\\ \cline{2-2}
			&
			$\bullet$ Federated/supervised/reinforcement   learning    \cite{elbir2020federated,taha2019deep,taha2021enabling,taha2020deep} &
			&
			\\ \hline
		\end{tabular}
	}
\end{table*}

In Table \ref{SP_method}, we classify the representative works on IRS channel estimation according to their applied signal processing methods.
It is observed that the low-complexity LS/LMMSE method has the broadest range of applicable channel models and scenarios, which thus has been widely used in the literature. 
On the other hand, the compressed sensing, matrix factorization, and deep learning methods have their specific application scenarios, while requiring higher computational complexity as compared to the LS/LMMSE method.
In a nutshell, the performance and applicability of signal processing methods for IRS channel estimation highly depend on the underlying IRS channel models as well as the signal models (e.g., linear vs. nonlinear and over-determined vs. under-determined), which deserve further investigation in the future.


\section{IRS Passive Beamforming Design under Practical CSI}\label{sec_RD}
\begin{figure*}[!t]
	\centering
	\includegraphics[width=5in]{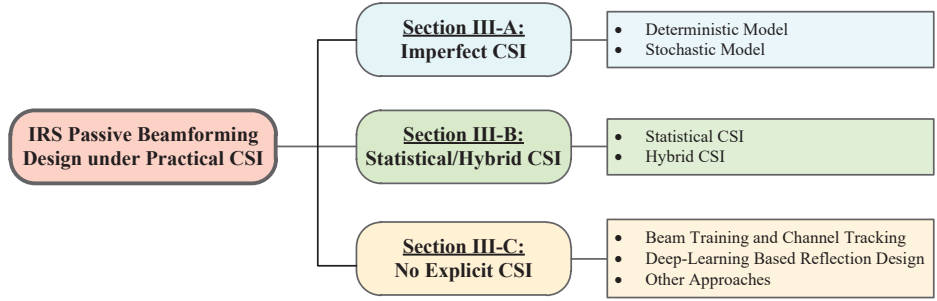}
	\caption{Organization of Section~\ref{sec_RD}.}
	\label{Organization_RD}
\end{figure*}

By utilizing the channel estimation techniques presented in Section~\ref{sec_CE}, the IRS passive beamforming/reflection can be optimized jointly with the BS active beamforming based on the estimated CSI. However, due to various practical factors such as channel aging, limited training/feedback overhead, and noise/interference effect, it is difficult to acquire perfect CSI in practice. In particular, this issue is aggravated in IRS-aided systems due to the extra IRS-associated channels to be estimated. As such, substantial works have looked into practical IRS passive beamforming/reflection design that accounts for different scenarios of CSI availability in various IRS-aided systems, namely, imperfect CSI and statistical/hybrid CSI, which will be discussed in this section (i.e., Sections \ref{imperfect} and \ref{statistical}, respectively). On the other hand, another line of research proposed to circumvent the channel estimation problem in the passive beamforming design, by exploiting (passive) beam training, deep learning, and other promising techniques which require no explicit CSI, as will be presented in Section~\ref{sec_RD}-C. The organization of this section is shown in Fig.~\ref{Organization_RD}. 

\subsection{IRS Passive Beamforming Design with Imperfect CSI}\label{imperfect}
\begin{figure*}[!t]
	\centering
	\includegraphics[width=5.5in]{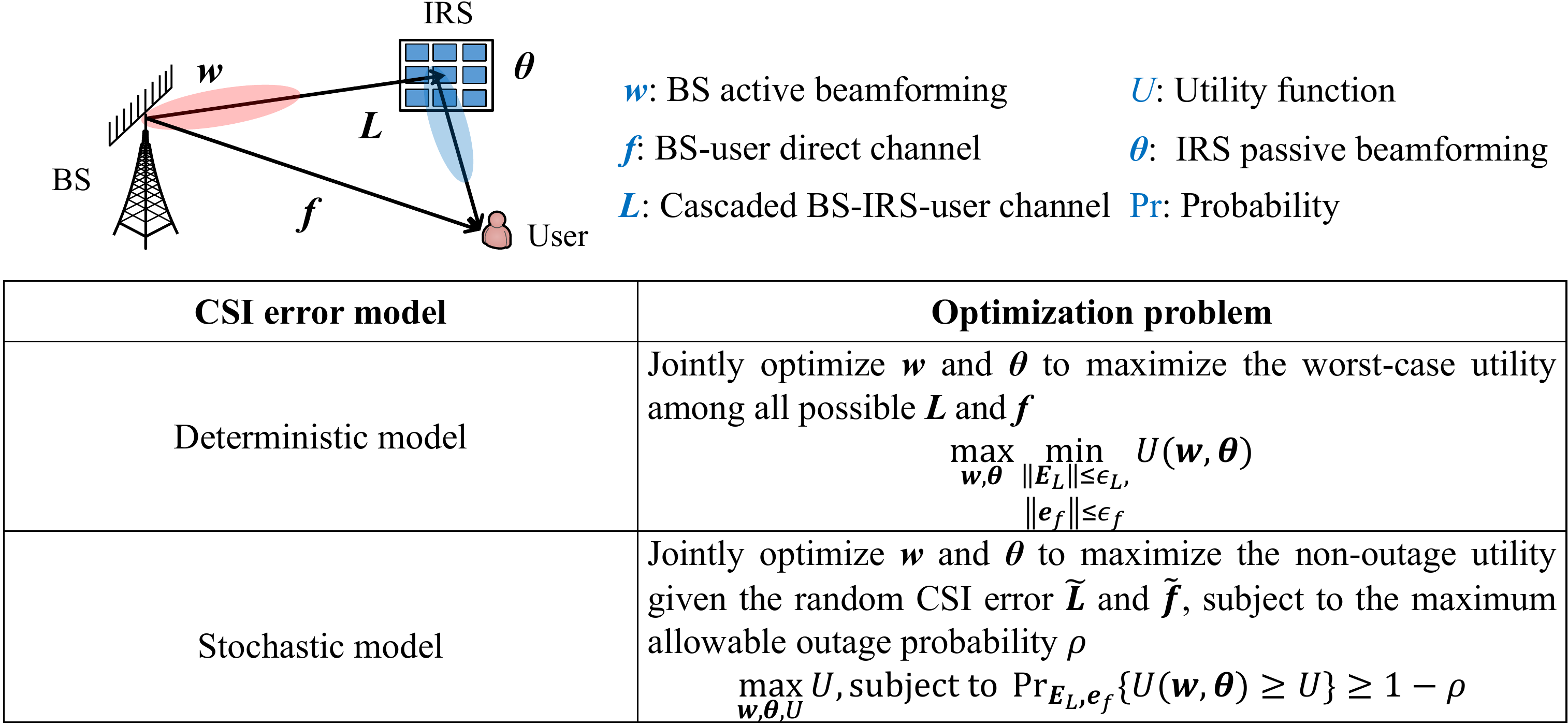}
	\DeclareGraphicsExtensions.\vspace{-6pt}
	\caption{{IRS passive beamforming design under the deterministic and stochastic CSI error models.}}\label{ipCSI}
	\vspace{-9pt}
\end{figure*}

Basically, there are two widely used models to characterize the CSI error due to imperfect channel estimation, namely, the deterministic model and stochastic model. {{As an illustrative example, Fig.\,\ref{ipCSI} depicts an IRS-aided MISO system, where the BS-IRS, IRS-user, and BS-user channels are denoted as ${\bf H}$, ${\bf g}$ and ${\bf f}$, respectively. If the semi-passive IRS is employed, such that the above channels can be individually estimated (see Fig.\,\ref{IRS_CE}), then they are expressed as ${\bf H}=\tilde{\bf H} + {\bf E}_H$, ${\bf g}=\tilde{\bf g} + {\bf e}_g$, and ${\bf f}=\tilde{\bf f} + {\bf e}_f$, where $\tilde{\bf H}$ (${\bf E}_H$), $\tilde{\bf g}$ (${\bf e}_g$) and $\tilde{\bf f}$ (${\bf e}_f$) denote their respective estimate (with CSI error due to the imperfect channel estimation). Alternatively, if the fully passive IRS is employed to estimate the cascaded BS-IRS-user channel ${\bf L} \triangleq {\rm diag}({\bf g}^H){\bf H}$, then it is expressed as ${\bf L}=\tilde{\bf L}+{\bf E}_L$, where $\tilde{\bf L}$ and ${\bf E}_L$ denote its estimate and CSI error, respectively.}}

Then, in the first model, the norm of the CSI error {(e.g., $\lVert {\bf E}_L \rVert$ and $\lVert {\bf e}_f \rVert$ for fully passive IRS) is assumed to be upper-bounded by a set of deterministic values (e.g., $\epsilon_L$ and $\epsilon_f$).} Accordingly, the BS/IRS active/passive beamforming is optimized to ensure the {\it worst-case} performance of a given utility function among all possible CSI, subject to a given maximum norm of CSI errors. While in the second model, in contrast, the CSI error is modeled as a random variable (usually following the complex Gaussian distribution), {e.g., ${\rm vec}({\bf E}_L) \sim {\cal {CN}}(0,\sigma_L^2{\bf I})$ and ${\rm vec}({\bf e}_f) \sim {\cal {CN}}(0,\sigma_f^2{\bf I})$ for fully passive IRS, where ${\rm vec}(\cdot)$ denotes the vectorization, $\sigma_L^2$ and $\sigma_f^2$ denote the variance of CSI error.} Due to the randomly distributed CSI error, the BS/IRS active/passive beamforming is generally optimized to ensure the non-outage performance of a given utility target. {The associated robust beamforming design problems assuming fully passive IRS for both models are shown in Fig.\,\ref{ipCSI}.} In both models, the key design challenge lies in how to achieve a robust beamforming design that caters to an infinite number of possible CSI, either in the worst-case or probabilistic sense. It is noted that the upper bound {(i.e., $\epsilon_L$ and $\epsilon_f$)} and variance {(i.e., $\sigma^2_L$ and $\sigma^2_f$)} of the CSI error capture the uncertainty of CSI in the first and second models, respectively. Generally, the larger the former is, the lower the worst-case/non-outage performance in the first/second model. It is also worth noting that these two CSI error models can be applied to any source of CSI errors, as long as their upper bound or variance can be properly determined. Based on the above modeling, there have been a large number of existing works devoted to the {\it robust} active/passive beamforming design under these two models, as presented below.

\subsubsection{Deterministic Model}
First, by using the advanced deep reinforcement learning technique, the authors in \cite{lin2020deep} proposed a new deep deterministic policy gradient approach for the robust beamforming design in the single-user MISO system aided by an energy-harvesting IRS. In particular, they aimed to minimize the transmit power of a BS, while ensuring both the worst-case reflected and harvested signal power for the IRS, by jointly optimizing its phase shifts and reflection amplitudes. Different from the generic MISO channel model in \cite{lin2020deep}, the authors in \cite{lu2020robust} studied a mmWave secure MISO system under the geometry channel model, considering the uncertainty in the angle and amplitude information on the cascaded BS-IRS-eavesdropper link. To deal with the associated robust beamforming problem, they proposed to approximate the cascaded BS-IRS-eavesdropper channel as a weighted combination of discrete samples and solved the approximate problem via the alternating optimization (AO) and SDR techniques.  

Nonetheless, for the robust beamforming design in the deterministic model, one of the most popular techniques is $\cal S$-procedure, which is able to transform the worst-case objective function or constraints into a tractable form with linear matrix inequalities. In addition, it is usually used jointly with the AO and SDR techniques to tackle the unit-modulus constraints on IRS passive beamforming, as applied in the following works. {In \cite{zhou2020robust,zhou2020framework,yu2021irs}, the authors considered an IRS-assisted MISO broadcast system and the associated robust beamforming design for the BS and IRS. In particular, the authors in \cite{yu2021irs} proposed an improved penalty-based AO algorithm to overcome the non-convergence issue of conventional AO algorithm. In addition to the MISO broadcast systems considered above, the authors in \cite{hao2021robust} considered a more general MIMO-OFDMA THz communication system and jointly optimized the hybrid analog/digital and passive beamforming at the BS and IRS, respectively.}  
{In \cite{yuan2020intelligent}, the authors studied a MISO cognitive radio network and the robust active and passive beamforming at a secondary transmitter and multiple IRSs, respectively, under the imperfect CSI on the links associated with primary users (PUs).} The authors in \cite{xu2020resource} studied further a more complex MISO cognitive radio network, where a full-duplex secondary BS serves multiple secondary users (SUs) in the uplink and downlink at the same time. They aimed to jointly optimize the BS's transmit/receive beamforming in the downlink/uplink, the transmit power of uplink users, as well as the IRS passive beamforming to maximize the weighted sum-rate of all SUs. Several techniques were combined to resolve this problem, including the SDR, successive convex approximation (SCA), and penalty-based approach. 

The robust secure beamforming design has also received considerable attention in the literature by exploiting the similar optimization approaches as above mentioned. Specifically, the authors in \cite{yu2020robust} studied the robust active and passive beamforming designs for a multi-IRS-aided secure MISO broadcast system with multi-antenna eavesdroppers, under the imperfect CSI on the IRS-eavesdropper links. Compared to \cite{yu2020robust}, the authors in \cite{zhang2021robust} further assumed that the BS serves its users via NOMA and the BS-eavesdropper channel is imperfectly known as well. In \cite{hu2020robust}, self-sustainable IRSs were leveraged to improve the physical-layer security of a MISO broadcast system with multiple eavesdroppers, where the IRS can control the ON/OFF status of its reflecting elements for either signal reflection or energy harvesting. An interesting observation made in \cite{hu2020robust} is that with self-sustainable IRSs, the sum rate of all users may be saturated with the increasing number of IRS elements due to hardware constraints. In \cite{li2021robust}, the authors considered an IRS-aided secure UAV-ground communication system, where the robust UAV trajectory and IRS passive beamforming were jointly optimized in the presence of imperfect CSI on the IRS-eavesdropper and UAV-eavesdropper links. 

While the above works have focused on the spectral efficiency, the work \cite{wang2020energy} formulated a secrecy energy efficiency maximization problem for a secure MISO system with multiple eavesdroppers and a friendly jammer. This problem was solved by utilizing the AO algorithm and SDR technique, jointly with the Dinkelbach's method to deal with the fractional objective function of secrecy energy efficiency. It was shown that leveraging IRS helps improve the trade-off between secrecy rate and energy efficiency even with CSI uncertainty.

\begin{table*}[!t]
	\begin{center}
		\renewcommand{\arraystretch}{1.0}
		\caption{IRS Reflection Design with Imperfect CSI}\label{ImpCSI}
		\resizebox{\textwidth}{!}{%
			\large\begin{tabular}{|c|l|l|l|}
				\hline
				\bf{CSI error model} &
				\quad\qquad\quad\;\;\bf{System setup} &
				\quad\qquad\quad\quad\bf{Optimization problem} &
				\quad\qquad\bf{Proposed reflection design algorithm}  \\ \hline\hline
				\multirow{16}{*}{\begin{tabular}[c]{@{}c@{}}{\bf Deterministic error}\end{tabular}}
				&
				\begin{tabular}[c]{@{}l@{}}Single-user MISO with energy-\\harvesting IRS\end{tabular} &
				\begin{tabular}{m{9cm}}{BS transmit power minimization constrained by the worst-case reflected and harvested signal power by the IRS}\end{tabular}&
				Deep deterministic policy gradient algorithm \cite{lin2020deep}\\
				\cline{2-4}
				&
				Single-user mmWave MISO &
				\begin{tabular}{m{9cm}}{Worst-case secrecy rate maximization}\end{tabular} &
				Discrete approximation and AO with SDR \cite{lu2020robust}\\
				\cline{2-4}
				&
				MISO broadcast&
				\begin{tabular}{m{9cm}}{BS transmit power minimization constrained by users' worst-case achievable rates}\end{tabular}&
				Penalty-based AO \cite{zhou2020robust,zhou2020framework,yu2021irs}\\
				\cline{2-4}
				&
				MIMO-OFDMA downlink&
				\begin{tabular}{m{9cm}}{Worst-case user weighted sum-rate maximization}\end{tabular} &
				AO with SDR and SCA\cite{hao2021robust}  \\
				\cline{2-4}
				&
				\begin{tabular}[c]{@{}l@{}}MISO cognitive radio network\\ (half-duplex)\end{tabular}&
				\begin{tabular}{m{9cm}}{Worst-case SU rate maximization constrained by worst-case interference power with PUs}\end{tabular} &
				AO with SDR\cite{yuan2020intelligent}  \\
				\cline{2-4}
				&
				\begin{tabular}[c]{@{}l@{}}MISO cognitive radio network\\ (full-duplex)\end{tabular}&
				\begin{tabular}{m{9cm}}{Worst-case SU rate maximization constrained by worst-case interference power with PUs}\end{tabular} &
				AO, SCA, SDR, and penalty method\cite{xu2020resource} \\
				\cline{2-4}
				&
				Secure MISO broadcast&
				\begin{tabular}{m{9cm}}{User sum rate maximization constrained by the worst-case information leakage to eavesdroppers}\end{tabular} &
				AO with SCA, SDR, and penalty method\cite{yu2020robust}\\
				\cline{2-4}
				&
				\begin{tabular}[c]{@{}l@{}}Secure MISO broadcast with\\ self-sustainable IRS\end{tabular}&
				\begin{tabular}{m{9cm}}{User worst-case sum rate maximization constrained by the worst-case information leakage to eavesdroppers and harvested energy by IRS}\end{tabular} &
				SCA and SDR\cite{hu2020robust}\\
				\cline{2-4}
				&
				Secure MISO-NOMA&
				\begin{tabular}{m{9cm}}{BS transmit power minimization constrained by the worst-case information leakage to eavesdroppers}\end{tabular} &
				AO\cite{zhang2021robust}\\
				\cline{2-4}
				&
				Secure UAV-ground communication&
				\begin{tabular}{m{9cm}}{UAV worst-case average secrecy rate maximization}\end{tabular} &
				AO with SCA\cite{li2021robust}\\
				\cline{2-4}
				&
				Secure jammer-assisted MISO&
				\begin{tabular}{m{9cm}}{Worst-case secrecy energy efficiency maximization constrained by the worst-case secrecy rate}\end{tabular} &
				AO with SDR and Dinkelbach's method\cite{wang2020energy}\\
				\cline{1-4}
				\multirow{14}{*}{\bf{Stochastic error}}
				&
				\multirow{2}{*}{Single-user MISO}
				&
				\begin{tabular}{m{9cm}}{User MSE minimization}\end{tabular}&
				AO and majorization-minimization\cite{zhangjz2020robust}
				\\ \cline{3-4}
				&&
				\begin{tabular}{m{9cm}}{Average user SINR maximization}\end{tabular}&
				Autoencoder-based deep neural network\cite{chu2020label} 
				\\ \cline{2-4}				
				&
				\multirow{2}{*}{MISO broadcast}
				&
				{\begin{tabular}{m{9cm}}{BS transmit power minimization constrained by the maximum outage probability for each user's SINR}\end{tabular}}
				&
				\begin{tabular}[c]{@{}l@{}}
					Constrained stochastic SCA \cite{zhao2021outage}\\ 
					Bernstein inequality and penalty-based AO\cite{zhou2020framework}\\
					Bernstein inequality and AO with SDR\cite{wang2020robust,le2020robust}
				\end{tabular}
				\\ \cline{3-4}
				&&
				{\begin{tabular}{m{9cm}}{Approximate user sum rate maximization}\end{tabular}}&
				Central limit theorem and penalty dual decomposition\cite{omid2021low}
				\\ \cline{2-4}
				&
				MISO mmWave broadcast&
				{\begin{tabular}{m{9cm}}{User sum outage probability minimization}\end{tabular}} &
				Stochastic block gradient descent method\cite{zhou2021stochastic}
				\\ \cline{2-4}
				&
				MISO cognitive radio network&
				{\begin{tabular}{m{9cm}}{Transmit power minimization constrained by the maximum outage probability of interference power with the PU}\end{tabular}} &
				Sphere bounding and AO\cite{zhang2020robust}
				\\ \cline{2-4}
				&
				Secure single-user MISO&
				{\begin{tabular}{m{9cm}}{BS transmit power minimization constrained by the minimum outage probability of each eavesdropper's achievable rate}\end{tabular}} &
				Bernstein inequality and penalty-based AO\cite{hong2020robust,hong2020outage}
				\\ \cline{2-4}
				&
				Secure MISO UAV-ground broadcast&
				{\begin{tabular}{m{9cm}}{User sum secrecy rate maximization constrained by the maximum outage probability of each user's secrecy rate}\end{tabular}} &
				Twin deep deterministic policy gradient algorithm \cite{guo2021learning}
				\\ 
				\cline{1-4}		
			\end{tabular}
		}
	\end{center}
\end{table*}

To summarize, the robust beamforming design in the deterministic model has been extensively studied in the literature for assorted scenarios via various optimization approaches, as summarized in Table \ref{ImpCSI}. It has been shown that despite the CSI error, robust passive beamforming can still dramatically improve the wireless system performance over the traditional system without IRS as well as the non-robust design which overlooks the CSI error. 

\subsubsection{Stochastic Model}
First, for the single-user MISO system, the authors in \cite{zhangjz2020robust} derived the signal MMSE in closed-form under the Gaussian CSI error model. Then, they jointly optimized the BS/user/IRS transmit/receive/passive beamforming to minimize the MSE at the user, which was efficiently solved by invoking the AO and majorization-minimization techniques. The authors in \cite{chu2020label}, on the other hand, proposed an autoencoder-based DNN and designed its activation function, loss function, and feature selection, to optimize the robust beamforming. It was shown in \cite{chu2020label} that the proposed approach can achieve comparable performance to the SDR, as well as good robustness against various types of CSI errors.  

Nonetheless, in the more general system setup, various convex approximation methods (e.g., constrained stochastic SCA, sphere bounding, and Bernstein-type inequality) have been applied in the literature to relax the intricate probabilistic objective functions or constraints in their design problems, jointly with the standard AO and SDR techniques. {Specifically, in \cite{zhao2021outage}, the authors considered a MISO broadcast system and proposed a novel constrained stochastic SCA algorithm to tackle the difficulty due to the outage probability, which can reliably guarantee the non-outage performance of all users. The MISO broadcast system was also studied in \cite{zhou2020framework,wang2020robust,le2020robust}, where the authors applied a different technique of Bernstein inequality or central limit theorem to approximate/relax the probabilistic outage constraints, which guarantees the non-outage performance of all users as well.} Instead of considering the generic MISO channel as in the above works, the authors in \cite{zhou2021stochastic} studied the robust beamforming design in the mmWave MISO broadcast system under the geometric channel model. Assuming a Bernoulli distributed blockage parameter for each path, they minimized the sum outage probability of all users by jointly optimizing the hybrid beamforming at the BS and passive beamforming at the IRS. To solve this stochastic optimization problem, a low-complexity stochastic block gradient descent method was proposed, where a set of sensible blockage patterns were learned to facilitate the optimization. 

In addition to the MISO broadcast systems considered above, the authors in \cite{zhang2020robust} investigated the robust beamforming design in the MISO cognitive radio network with multiple SUs and a single PU. Assuming imperfect CSI on the PU-associated links, they utilized the sphere bounding to derive a tractable upper bound on the outage probability and solved the approximate problem via the AO method. Furthermore, the authors in \cite{hong2020robust} and \cite{hong2020outage} studied the robust beamforming design in a secure single-user MISO system overheard by multiple eavesdroppers, with imperfect CSI on all cascaded BS-IRS-eavesdropper links. The Bernstein inequality was invoked to relax the outage probability into a tractable form. It was found in \cite{hong2020robust} and \cite{hong2020outage} that when the CSI uncertainty is high, more power should be allocated to artificial noise instead of the user's information signal. Moreover, a twin deep deterministic policy gradient algorithm was proposed in \cite{guo2021learning} for a secure UAV-ground broadcast system, accounting for the effect of outdated CSI. 

To summarize, as compared to its counterpart in the deterministic model, the robust beamforming design in the stochastic model is less studied due to the more challenging probabilistic constraints involved. As such, the main focus of existing works is on seeking feasible convex approximation techniques to recast them into a tractable form, as summarized in Table \ref{ImpCSI}. It is anticipated that more research efforts for the robust beamforming design under the stochastic model will be given to investigate new setups and optimization techniques.

\subsection{IRS Passive Beamforming Design with Statistical/Hybrid CSI}\label{statistical}
Although the CSI error is taken into account in the aforementioned studies, the robust active and passive beamforming design  requires real-time channel estimation to obtain the instantaneous CSI on all links. In practice, this approach may incur prohibitively high signal processing complexity and large training/feedback overhead, especially when the sizes of IRS and BS antenna array both become large. To balance the trade-off between the channel estimation overhead and system performance, the statistical or hybrid CSI (e.g., hybrid instantaneous and statistical CSI) in the IRS-aided communication systems have been exploited in recent works to facilitate the IRS passive beamforming design. 

\begin{figure*}[!t]
	\centering
	\includegraphics[width=5.5in]{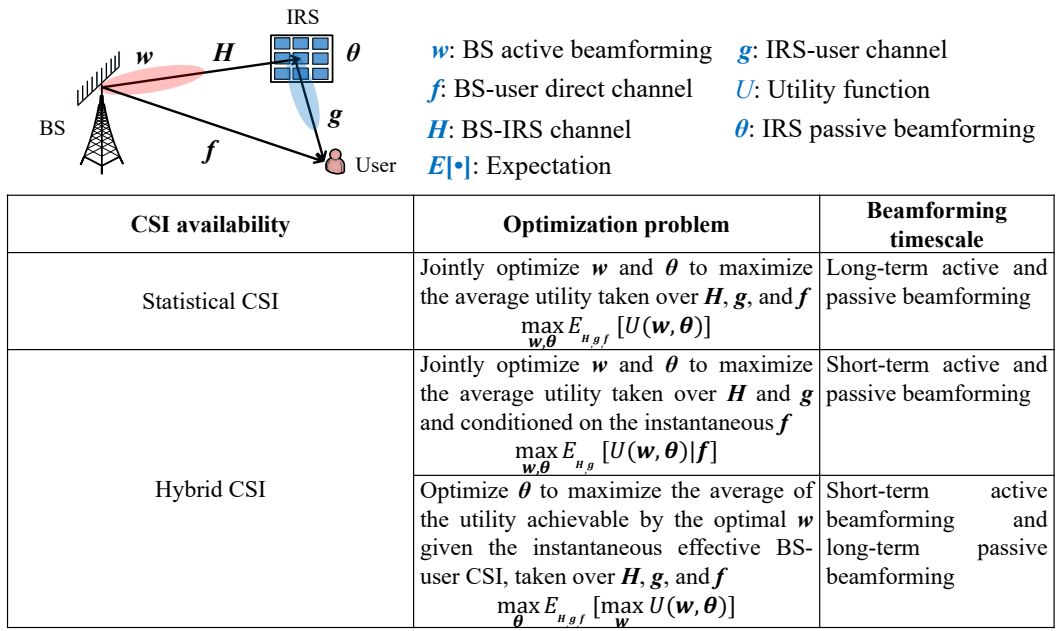}
	\DeclareGraphicsExtensions.\vspace{-6pt}
	\caption{IRS passive beamforming design with statistical and hybrid CSI.}\label{RefDesign}
	\vspace{-9pt}
\end{figure*}

In particular, the beamforming design based on statistical CSI aims for the long-term performance (e.g., ergodic rate, coverage probability, etc.) and thus only requires the statistics of the channels, such as their distributions, means, second moments, etc., which vary much slower than the instantaneous CSI and thus are easier to be obtained in practice. This thus greatly saves the channel training time, at the cost of real-time performance due to the lack of instantaneous CSI. To further improve the overhead-performance trade-off, leveraging both the statistical and instantaneous CSI (termed hybrid CSI) turns out to be an appealing solution, e.g., by only estimating a subset of all channels in real-time, while leaving the other channels  that are more difficult or time-consuming to estimate (e.g., IRS-user and IRS-eavesdropper channels) statistically known only. Then, the joint BS and IRS beamforming is optimized to maximize the average utility over the statistical CSI but conditioned on the available instantaneous CSI. 
An alternative strategy is the two-timescale beamforming, where the IRS passive beamforming is designed in long term based on the statistical CSI on all links, while the BS's active beamforming is dynamically tuned based on its effective channels with all users in real-time. {For example, for the single-user MISO system shown in Fig.\,\ref{RefDesign}, the BS's active beamforming can be set as the maximum ratio transmission (MRT) based on its effective channel with the user (direct plus IRS-reflected channels), i.e., ${\bf f}+{\bf L}{\bm{\theta}}$, so as to maximize the achievable rate of the user with the fixed IRS passive beamforming $\bm{\theta}$. The associated optimization problems for the above beamforming designs are illustrated in Fig.\,\ref{RefDesign}.} It is noted from Fig.\,\ref{RefDesign} that all cases result in challenging stochastic optimization problems, which are generally more difficult to solve as compared to the passive beamforming design under perfect CSI or imperfect CSI with deterministic errors. More detailed discussions are given as follows.

\subsubsection{Statistical CSI} 
The authors in \cite{papazafeiropoulos2021coverage} optimized the long-term IRS passive beamforming to maximize the coverage probability of a multi-IRS aided single-user SISO system under the correlated Rayleigh channel model. Their results showed that the channel correlation among different IRS elements may be beneficial to enhance the coverage probability. Furthermore, the authors in \cite{hu2020statistical} investigated the joint long-term active and passive beamforming design in the downlink single-user MISO system under the Rician-fading channel model. A tight upper bound of the ergodic capacity was derived and then maximized by designing customized IRS passive beamforming methods for the scenarios with Rician- and Rayleigh-fading BS-user channels, respectively. As compared to \cite{hu2020statistical}, the authors in \cite{jia2020analysis} further considered the presence of a co-channel BS and investigated the corresponding long-term IRS passive beamforming design. To maximize the ergodic rate of the user, they derived a deterministic upper bound and then maximized it using the parallel coordinate descent algorithm. 

Moreover, the authors in \cite{zhang2021large,wang2021joint,zhang2021largescale,dang2020joint} optimized the joint long-term active and passive beamforming in the IRS-aided single-user MIMO system, with a goal to maximize its ergodic capacity. In particular, in \cite{zhang2021large}, by assuming independent Rician fading for all channels, the authors drew upon random matrix theory and utilized the replica method to derive a large-scale approximation of the ergodic capacity. Instead of assuming independent channels as in \cite{zhang2021large}, the authors in \cite{wang2021joint} assumed spatially correlated Rayleigh and Rician fading for BS-user and IRS-associated links to capture the close inter-antenna/element spacing at the BS/IRS, respectively. They provided a tight upper bound of the ergodic rate via Jensen inequality. Instead of assuming the rich-scattering channels, i.e., Rayleigh and Rician fading, the authors in \cite{zhang2021largescale} considered a more practical double-scattering channel model to account for rank deficiency and spatial correlation in IRS-assisted wireless systems. A deterministic approximation of the ergodic capacity was derived by leveraging random matrix theory. Without assuming any prior knowledge on the channel model, the long-term active and passive beamforming in \cite{dang2020joint} was optimized based only on the second moments of all channels, which suffice to provide a tractable upper bound of the MIMO ergodic rate. It was shown that in terms of ergodic rate performance, the IRS-aided system may not outperform the conventional system without IRS, unless the IRS is deployed sufficiently near either the transmitter or receiver. 

The long-term beamforming design has also been studied under other system setups, such as interference channel and physical-layer security. Specifically, the authors in \cite{peng2021analysis} optimized the long-term passive beamforming design in the SISO interference channel by using the genetic algorithm, which was shown able to achieve near-optimal performance. Furthermore, the authors in \cite{liu2021secrecy} maximized the ergodic secrecy rate of a MIMO secure communication system with blocked direct BS-user/eavesdropper link and statistical CSI on IRS-associated links. Random matrix theory was exploited to derive a deterministic approximation to this problem, which was then solved by combining the AO and projected gradient ascent methods. 

\begin{table*}[!t]
	\begin{center}
		\renewcommand{\arraystretch}{1.0}
		\caption{IRS Reflection Design with Hybrid and/or Statistical CSI}\label{StaCSI}\vspace{-6pt}
		\resizebox{\textwidth}{!}{%
			\large\begin{tabular}{|c|l|l|l|l|}
				\hline
				\bf{CSI uncertainty} &
				\quad\quad\;\;\bf{System setup} &
				\qquad\qquad\qquad\bf{Channel model} &
				\quad\qquad\quad\qquad\bf{Optimization problem} &
				\;\;\begin{tabular}[c]{@{}c@{}} {\bf Approximation and optimization}\\ {\bf method} \end{tabular}  \\ \hline\hline
				\multirow{14}{*}{{\bf Statistical CSI}}
				&
				Single-user SISO&
				{\begin{tabular}{m{9cm}}{Spatially correlated Rayleigh-fading IRS-associated channels and independent Rayleigh-fading BS-user channel}\end{tabular}}&
				Coverage probability maximization&
				{\begin{tabular}{m{8cm}}{Deterministic equivalent and projected gradient ascent method\cite{papazafeiropoulos2021coverage}}\end{tabular}}\\
				\cline{2-5}
				&
				Single-user MISO&
				{\begin{tabular}{m{9cm}}{Rician-fading IRS-associated channels, Rician/Rayleigh fading BS-user channel}\end{tabular}}&
				\multirow{9}{*}{Ergodic capacity/rate maximization}&
				{\begin{tabular}{m{8cm}}{Jensen inequality and SDR\cite{hu2020statistical}}\end{tabular}}\\
				\cline{2-3}\cline{5-5}
				&
				\begin{tabular}[c]{@{}c@{}}Single-user MISO with\\ co-channel interference\end{tabular}&
				{\begin{tabular}{m{9cm}}{Rician-fading IRS-associated channels, Rayleigh fading BS-user channel}\end{tabular}}&
				&
				{\begin{tabular}{m{8cm}}{Jensen inequality and parallel coordinate descent\cite{jia2020analysis}}\end{tabular}}\\
				\cline{2-3}\cline{5-5}
				&
				\multirow{6}{*}{Single-user MIMO} &
				{\begin{tabular}{m{9cm}}{Rician fading for all channels}\end{tabular}}&
				&
				{\begin{tabular}{m{8cm}}{Random matrix theory, replica method, and AO\cite{zhang2021large}}\end{tabular}}	\\
				\cline{3-3}\cline{5-5}	
				&&
				{\begin{tabular}{m{9cm}}{Spatially correlated Rayleigh/Rician fading for all links}\end{tabular}}&
				&
				{\begin{tabular}{m{8cm}}{Jensen inequality and SDR\cite{wang2021joint}}\end{tabular}}\\
				\cline{3-3}\cline{5-5}
				&&
				{\begin{tabular}{m{9cm}}{Double-scattering channels for all links}\end{tabular}} &
				&
				{\begin{tabular}{m{8cm}}{Standard interference function theory and AO\cite{zhang2021largescale}}\end{tabular}}\\
				\cline{3-3}\cline{5-5}
				&&
				{\begin{tabular}{m{9cm}}{Independent but arbitrary channels for all links}\end{tabular}} &
				&
				{\begin{tabular}{m{8cm}}{Jensen inequality and AO\cite{dang2020joint}}\end{tabular}}\\
				\cline{2-5}
				&
				SISO interference channel&
				{\begin{tabular}{m{9cm}}{Rician-fading IRS-associated channels and blocked BS-user channels}\end{tabular}}&
				Average user sum-rate maximization&
				{\begin{tabular}{m{8cm}}{Jensen inequality and genetic algorithm\cite{peng2021analysis}}\end{tabular}}\\
				\cline{2-5}
				&
				Secure single-user MIMO&
				{\begin{tabular}{m{9cm}}{Rician-fading IRS-associated channels, blocked BS-user and -eavesdropper channels}\end{tabular}}&
				Ergodic secrecy rate maximization &
				{\begin{tabular}{m{8cm}}{Random matrix theory and AO with projected gradient ascent method\cite{liu2021secrecy}}\end{tabular}}\\
				\cline{1-5}
				\multirow{22}{*}{{\bf Hybrid CSI}}
				&
				\multirow{4}{*}{MIMO multiple access}&
				\multirow{3}{*}{\begin{tabular}{m{9cm}}{Rician-fading IRS-associated channels, instantaneous CSI on the BS-user link}\end{tabular}}&
				Average global energy efficiency maximization&
				\begin{tabular}{m{8cm}}{Random matrix theory, AO, fractional programming, and majorization-minimization\cite{you2021reconfigurable}}\end{tabular}\\
				\cline{4-5}
				&&&
				\begin{tabular}[l]{@{}l@{}} Weighted sum of average energy/spectral\\ efficiency maximization \end{tabular}&
				\begin{tabular}{m{8cm}}{Deterministic equivalent, BCD, homotopy optimization, penalty dual decomposition, and majorization-minimization\cite{you2020energy}}\end{tabular}\\
				\cline{2-5}
				&
				Secure single-user MISO&
				\begin{tabular}{m{9cm}}{Rician-fading eavesdropper-associated channels, instantaneous CSI on other links}\end{tabular}&
				Average secrecy rate maximization&
				\begin{tabular}{m{8cm}}{Jensen inequality and penalty dual convex approximation\cite{liu2020reconfigurable}}\end{tabular}\\
				\cline{2-5}
				&
				\multirow{4}{*}{Single-user MISO}&
				\multirow{3}{*}{\begin{tabular}{m{9cm}}{Rician-fading IRS-associated channels and Rayleigh-fading BS-user channel}\end{tabular}}&
				\multirow{6}{*}{Ergodic capacity/rate maximization}&
				\begin{tabular}{m{8cm}}{Jensen inequality\cite{han2019large}}\end{tabular} \\
				\cline{5-5}
				&&&&
				\begin{tabular}{m{8cm}}{Jensen inequality and generalized Rayleigh quotient\cite{gao2020distributed}}\end{tabular}\\
				\cline{3-3}\cline{5-5}
				&&
				\begin{tabular}{m{9cm}}{Historical channel observations of all links}\end{tabular}
				&&
				\begin{tabular}{m{8cm}}{Stochastic gradient decent\cite{guo2020intelligent}}\end{tabular}\\
				\cline{2-3}\cline{5-5}
				&
				mmWave single-user MISO&
				\begin{tabular}{m{9cm}}{Geometric BS-user and IRS-user channels and LoS BS-IRS channel}\end{tabular}&&
				\begin{tabular}{m{8cm}}{Jensen inequality and AO\cite{yang2020intelligentreflecting}}\end{tabular}\\
				\cline{2-5}
				&
				\multirow{2}{*}{MISO broadcast}&
				\begin{tabular}{m{9cm}}{Generic channels for all links}\end{tabular}&
				\begin{tabular}[l]{@{}l@{}} BS transmit power minimization problem constrained\\ by individual average QoS for all users\end{tabular}&
				\begin{tabular}{m{8cm}}{Primal-dual decomposition and deep unfolding technique\cite{zhao2021two}}\end{tabular}\\
				\cline{3-5}
				&&
				\begin{tabular}{m{9cm}}{Spatially correlated Rician fading for all links}\end{tabular}&
				\multirow{6}{*}{User average sum rate maximization}&
				\begin{tabular}{m{8cm}}{Stochastic SCA\cite{zhao2021intelligent}}\end{tabular}\\
				\cline{2-3}\cline{5-5}
				&
				\multirow{3}{*}{Massive MISO broadcast}&
				\begin{tabular}{m{9cm}}{Rician-fading BS-IRS and IRS-user channels, blocked BS-user channel}\end{tabular}&
				&
				\begin{tabular}{m{8cm}}{Jensen inequality and genetic algorithm\cite{zhi2020power}}\end{tabular}\\
				\cline{3-3}\cline{5-5}
				&&
				\begin{tabular}{m{9cm}}{Rician-fading BS-IRS and IRS-user channels, Rayleigh-fading BS-user channel}\end{tabular}&
				&
				\begin{tabular}{m{8cm}}{Jensen inequality and genetic algorithm\cite{zhi2021statistical}}\end{tabular}\\
				\cline{2-3}\cline{5-5}
				&
				MIMO broadcast&
				\begin{tabular}{m{9cm}}{Random user locations, antenna positions, cluster distribution, and multi-path channel gains}\end{tabular}&
				&
				\begin{tabular}{m{8cm}}{Generalized weighted MMSE algorithm\cite{abrardo2021intelligent}}\end{tabular}\\
				\cline{2-5}
				&
				Multi-cell MISO&
				\begin{tabular}{m{9cm}}{Rayleigh fading for all links}\end{tabular}&
				\begin{tabular}[l]{@{}l@{}} User minimum average SINR maximization\\ (for long-term IRS-user association optimization)\end{tabular} &
				\begin{tabular}{m{8cm}}{Jensen inequality, branch-and-bound algorithm, and successive refinement algorithm\cite{mei2020performance}}\end{tabular}\\
				\cline{2-5}
				&
				SISO-NOMA&
				\begin{tabular}{m{9cm}}{Rician-fading BS-IRS and IRS-user channels, blocked BS-user channel}\end{tabular}&
				\begin{tabular}[l]{@{}l@{}} User sum-rate maximization (for long-term IRS\\ deployment optimization) \end{tabular}&
				\begin{tabular}{m{8cm}}{Exhaustive search and local region optimization method\cite{mu2021joint}}\end{tabular}\\
				\cline{1-5}
			\end{tabular}
		}
	\end{center}
\end{table*}
The above works are summarized in Table \ref{StaCSI}. As noted from Table \ref{StaCSI}, they mainly focused on deriving a tractable bound or deterministic equivalent of the optimization objective, e.g., ergodic capacity and coverage probability, to facilitate the beamforming design. Some common approaches such as random matrix theory and Jensen inequality, and their effectiveness have been evaluated therein. It is interesting to note that some works, e.g., \cite{jia2020analysis} and \cite{dang2020joint}, have revealed that the IRS-aided system may not outperform the conventional system without IRS in terms of ergodic rate performance, under some specific conditions or setups. It is also noted from Table \ref{StaCSI} that existing works have mostly considered the basic single-user system setup; thus, more general system setup may be explored in the future.

\subsubsection{Hybrid CSI}
As previously discussed, with the hybrid CSI (i.e., combined instantaneous and statistical CSI), some works aimed to maximize the average system utility over the statistical CSI on a subset of links, conditioned on the instantaneous CSI on the other links. {Specifically, the authors in \cite{you2021reconfigurable} and \cite{you2020energy} considered a MIMO multiple access communication system and assumed statistical CSI on the IRS-associated links and instantaneous CSI on the IRS-BS link. They aimed to jointly optimize the transmit covariance matrices of all users and IRS passive beamforming to maximize the global energy efficiency of the system\cite{you2021reconfigurable} and the weighted sum of ergodic energy efficiency and ergodic spectral efficiency\cite{you2020energy}.} To tackle the stochastic optimization problems, the authors first developed an asymptotically deterministic equivalent of the objective functions and solved the resultant deterministic problem by combining various methods, including block coordinate descent (BCD), homotopy optimization, penalty dual decomposition, majorization-minimization, etc. Moreover, in \cite{liu2020reconfigurable}, the authors considered an ergodic secrecy rate maximization problem in a secure MISO system, with instantaneous CSI on legitimate user's links and statistical CSI on the eavesdropper-associated links. To solve this problem, they first invoked the Jensen inequality to derive a tractable lower bound of the ergodic secrecy rate and then maximized it by using a penalty dual convex approximation algorithm.

Different from \cite{you2021reconfigurable,you2020energy,liu2020reconfigurable}, the second line of research designed the two-timescale active and passive beamforming, as detailed below. First, the authors in \cite{han2019large,gao2020distributed,guo2020intelligent} investigated the two-timescale beamforming design for the IRS-aided single-user MISO system. In this case, the optimal BS short-term active beamforming was obtained as the MRT based on the effective BS-user channel, while the IRS long-term passive beamforming was designed based on the statistical CSI on all links. As such, the main focus of \cite{han2019large,gao2020distributed,guo2020intelligent} lies in solving the corresponding stochastic problem for the long-term IRS passive beamforming. Specifically, by assuming Rician fading for all IRS-associated channels and Rayleigh fading for the BS-user channel, the authors in \cite{han2019large} derived an upper bound of the system ergodic rate and then obtained the optimal long-term IRS passive beamforming in closed-form that maximizes the upper bound. In \cite{gao2020distributed}, the authors assumed a large-scale system and derived a tractable approximation to the ergodic capacity, accounting for both statistical CSI and stochastic error of effective CSI at the BS. Then, the optimal long-term passive beamforming was obtained via projection by solving a relaxed problem. {Unlike \cite{han2019large} and \cite{gao2020distributed} where the channel model is known {\it a priori}, the authors in \cite{guo2020intelligent} proposed two learning-based approaches to design the IRS passive beamforming based only on historical channel observations. On the other hand, the authors in \cite{yang2020intelligentreflecting} focused on the two-timescale hybrid/passive beamforming at the BS/IRS in a mmWave MISO system under the geometric channel model, where the BS is aware of the angle information of each path but only the statistics of its complex gain.}

{The authors in \cite{zhao2021two} and \cite{zhao2021intelligent} focused on the two-timescale beamforming designs in the more challenging MISO broadcast system to minimize the BS transmit power subject to individual average QoS constraints for all users and maximize their ergodic sum-rate, respectively, under the spatially correlated Rician fading channel model. Due to the presence of inter-user interference, they proposed some more sophisticated algorithms, such as deep unfolding technique and iterative two-timescale stochastic SCA, to solve the associated problems.} Furthermore, the authors in \cite{zhi2020power} and \cite{zhi2021statistical} studied the two-timescale beamforming design in the uplink of a massive MIMO system. By assuming Rician-fading BS-IRS and IRS-user channels as well as the Rayleigh-fading BS-user channel, they optimized the long-term IRS passive beamforming to maximize the ergodic sum rate of all users, while the BS's short-term active beamforming was set as the matched filter based on the effective BS-user channel. An approximate ergodic sum rate was obtained in closed-form, based on which a genetic algorithm was applied to obtain a high-quality passive beamforming design. The two-timescale beamforming design in the more general MIMO broadcast system was studied in \cite{abrardo2021intelligent}, where the authors jointly optimized the short-term BS transmit precoding matrix and users' receive combining matrices, as well as IRS long-term passive beamforming to maximize the average sum-rate of all users, accounting for the randomness in user locations, antenna positions, multi-path cluster distribution, and the multi-path channel gains. The active precoding and combining at the BS and users are set based on the effective BS-user MIMO channel via the weighted MMSE algorithm. A generalized weighted MMSE algorithm was then proposed to solve the resulting stochastic optimization problem.

Finally, it is worth noting that in addition to the two-timescale beamforming design, there is another line of research that designs the active/passive beamforming in the short term, while optimizing other resource allocation in the long term (see, e.g., \cite{mei2020performance,mu2021joint}). Specifically, in \cite{mei2020performance}, the authors optimized the long-term IRS-user associations in a multi-cell MISO wireless network, given the short-term BS/IRS active/passive beamforming. Different from \cite{mei2020performance}, the authors in \cite{mu2021joint} investigated the long-term IRS deployment design in a SISO-NOMA system based on the deterministic components of Rician-fading channels. Given the IRS deployment, the IRS passive beamforming and BS power allocation were designed in real time based on the instantaneous CSI. 

As summarized in Table \ref{StaCSI}, compared to the beamforming design with statistical or instantaneous CSI only, the design based on hybrid CSI is able to flexibly balance the trade-off between the performance and overhead. However, the latter approach usually requires a tractable approximation to the system utility and incurs a higher complexity in optimization. For example, in the two-timescale beamforming, the optimal (short-term) active beamforming is coupled with the (long-term) IRS passive beamforming, thus making the associated problems difficult to solve. To circumvent this difficulty, some of the above works simplified the two-timescale beamforming design by considering suboptimal active beamforming, which thus suffers performance loss in general; while the others developed more sophisticated algorithms to tackle this difficulty by using, e.g., deep unfolding \cite{zhao2021two} and stochastic SCA techniques \cite{zhao2021intelligent}. 

\subsection{IRS Passive Beamforming Design with No Explicit CSI}
\begin{table*}[!t]
	\begin{center}
		\renewcommand{\arraystretch}{1.4}
		\caption{IRS Reflection Design with No Explicit CSI}\label{Table_est}
		\resizebox{\textwidth}{!}{%
			\begin{tabular}{|c|c|c|l|}
				\hline
				\bf{Approach} &
				\bf{Channel model} &
				\bf{Used information} &
				\qquad\qquad\qquad\qquad\qquad\qquad\qquad\bf{Proposed IRS passive beamforming design method}  \\ \hline\hline
				\multirow{10}{*}{\begin{tabular}[c]{@{}c@{}}\bf{Beam training} and \\ \bf{channel tracking}\end{tabular}} &
				\multirow{5}{*}{LoS channel}&
				\multirow{2}{*}{\begin{tabular}[c]{@{}c@{}} Received beam  power\end{tabular}}&
				\begin{tabular}[c]{@{}l@{}} $\bullet$ Propose a ternary-tree hierarchical beam training method for IRS-aided multi-user systems \cite{ning2021terahertz}\end{tabular} \\ \cline{4-4}
				& 
				&
				&
				\begin{tabular}[c]{@{}l@{}} $\bullet$ Propose a multi-beam training method for IRS-aided multi-user systems with small training overhead \cite{you2020fast}\end{tabular}  \\ \cline{3-4}
				&
				&
				\multirow{3}{*}{\begin{tabular}[c]{@{}c@{}} Received signals\end{tabular}} &
				\begin{tabular}[c]{@{}l@{}} $\bullet$ Propose a random beam training method to estimate AoD/AoA angles at each individual IRS-aided user \cite{wang2021joint}\end{tabular}  \\ \cline{4-4}
				
				&
				\multirow{7}{*}{Geometric channel}&
				&
				\begin{tabular}[c]{@{}l@{}} $\bullet$ Propose an extended Kalman filter algorithm to predict channel parameters for IRS-aided mobile users by using \\ a few candidate IRS beams according to estimated AoD/AoAs \cite{zegrar2020general}\end{tabular}  \\ \cline{2-4}
				
				&
				&
				Received signals &
				\begin{tabular}[c]{@{}l@{}} $\bullet$ First estimate the (angular) speed of mobile user based on received signals  and then predict the AP/IRS-user\\ AoA/AoD  for designing the IRS beamforming \cite{you2021enabling} \end{tabular}  \\ \cline{3-4}
				
				&
				&
				\begin{tabular}[c]{@{}c@{}} Received beam power\end{tabular} &
				\begin{tabular}[c]{@{}l@{}}  $\bullet$ Update the candidate IRS beam pairs for beam training  when the received power of mobile user is less than \\ a threshold \cite{tian2021fast} \end{tabular}  \\ \cline{1-4}

				\multirow{10}{*}{\begin{tabular}[c]{@{}c@{}} \bf{End-to-end} \\ \bf{IRS passive}\\ \bf{ beamforming learning}\end{tabular}} 				
				&
				Rayleigh fading channel &
				\begin{tabular}[c]{@{}c@{}} User location,  \\ optimal beamforming\\ at reference locations\end{tabular} &
				\begin{tabular}[c]{@{}l@{}}  $\bullet$ Design a DNN to learn the mapping from the user location to the optimal IRS reflection \cite{huang2019indoor} \end{tabular}  \\ \cline{2-4}
				
				&
				\multirow{3}{*}{Geometric channel}&
				\begin{tabular}[c]{@{}c@{}} User location, \\ reflect beamforming \end{tabular} &
				\begin{tabular}[c]{@{}l@{}}  $\bullet$ Predict the achievable rate of the user based on user location and candidate reflect beamforming vectors by \\using  unsupervised training \cite{sheen2021deep} \end{tabular}  \\ \cline{3-4}

				&
				&
				\begin{tabular}[c]{@{}c@{}} Received pilots\end{tabular} &
				\begin{tabular}[c]{@{}l@{}}  $\bullet$ Use the received pilot signals reflected through the IRS to train the deep feedforward network for  IRS-aided\\ single-user systems \cite{ozdogan2020deep} \end{tabular}  \\ \cline{2-4}
				
				&
				\multirow{3}{*}{Rician fading channel}&
				\begin{tabular}[c]{@{}c@{}} Received pilots, \\ user location\end{tabular} &
				\begin{tabular}[c]{@{}l@{}}  $\bullet$ Use the received pilots and user location information to learn the IRS reflection design for IRS-aided\\ multi-user systems \cite{jiang2021learning} \end{tabular}  \\ \cline{3-4}
				
				&
				&
				\begin{tabular}[c]{@{}c@{}} Received pilots\end{tabular} &
				\begin{tabular}[c]{@{}l@{}}  $\bullet$ Parameterize the mapping from the received pilots to the optimal beamforming  by tuning  a DNN  based\\  on unsupervised training \cite{jiang2020learning} \end{tabular}  \\ \cline{1-4}
				
				\begin{tabular}[c]{@{}c@{}}  {\bf User-location} \\  {\bf based method} \end{tabular}			
				&
				\begin{tabular}[c]{@{}c@{}}Angle-domain\\ Rician fading channel \end{tabular} &
				\begin{tabular}[c]{@{}c@{}} User location\end{tabular} &
				\begin{tabular}[c]{@{}l@{}}  $\bullet$  Exploit the user location information to estimate the effective angles from the IRS to users  for designing the \\ joint active and passive beamforming \cite{hu2020location,hu2021angle} \end{tabular}  \\ \cline{1-4}
				
				\multirow{3}{*}{\begin{tabular}[c]{@{}c@{}} {\bf Random}\\  {\bf beamforming} \end{tabular}}			
				&
				\multirow{3}{*}{Rayleigh fading channel} &
				\multirow{3}{*}{\begin{tabular}[c]{@{}c@{}} No information \end{tabular}} &
				\begin{tabular}[c]{@{}l@{}} $\bullet$ Propose to randomly change IRS reflection pattern multiple times in each channel coherence interval \\without CSI  in multicast systems \cite{tao2020intelligent} \end{tabular}  \\ \cline{4-4}	
				&
				&
				&
				\begin{tabular}[c]{@{}l@{}} $\bullet$ IRS randomly sets its reflection, while the BS employs the proportional fair  scheduling to exploit the multi-user\\ diversity gain\cite{chaaban2020opportunistic} \end{tabular}  \\ \cline{1-4}	
				
				\begin{tabular}[c]{@{}c@{}}  {\bf Heuristic method} \end{tabular}			
				&
				Rician fading channel &
				\begin{tabular}[c]{@{}c@{}} Received power/SNR \end{tabular} &
				\begin{tabular}[c]{@{}l@{}} $\bullet$ Use the particle swarm optimization method to find the near-optimal IRS reflection based on the received  \\SNR at the user \cite{souto2020beamforming} \end{tabular}  \\ \cline{1-4}

			\end{tabular}
		}
	\end{center}
\end{table*}
Besides the IRS passive beamforming designs above, there also exist other approaches that do not need explicit CSI, which are discussed in this subsection.

\subsubsection{Beam Training and  Channel Tracking}

\begin{figure}[!t]
	\centering
	\includegraphics[width=3.2in]{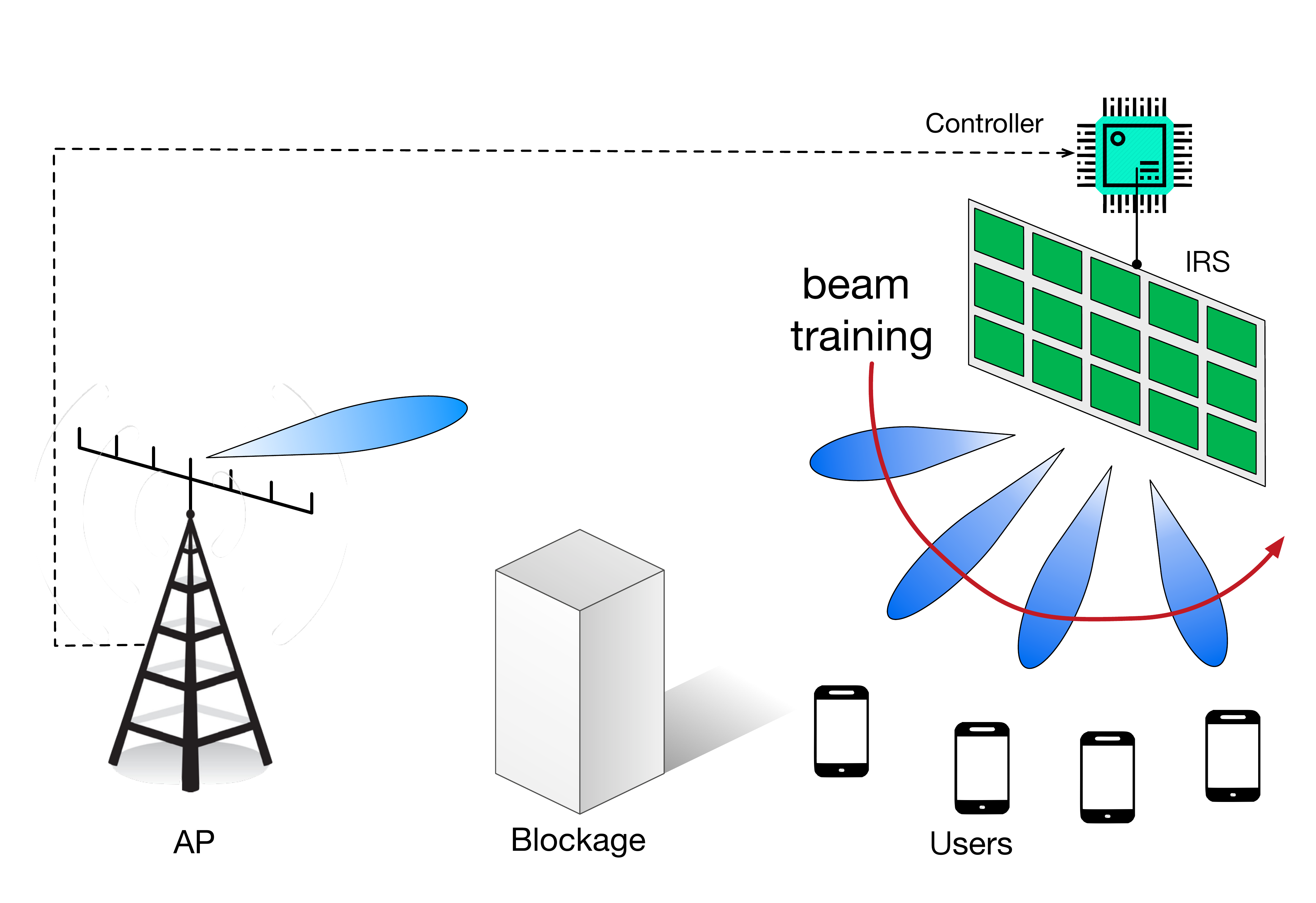}
	\caption{{IRS beam training and channel tracking.}}
	\label{beam_training}
\end{figure}

For IRS-aided communication systems with sparse channels e.g., in high-frequency bands, beam training is an efficient method to establish an initial high-SNR link from the transmitter to the receiver through the IRS, yet without requiring
explicit CSI. Specifically, the beam training method aims to 
select the best IRS beam from a predefined codebook 
that yields the strongest signal power at the receiver. 
Among others, the most straightforward beam training method is 
the \emph{sequential} single-beam training that exhaustively searches over all possible beam directions at the AP/IRS. This method, however, may incur prohibitively high training overhead, due to the large number of IRS reflecting elements that generate pencil-like beams. To reduce the training overhead, a \emph{hierarchical} IRS beam training method was proposed in \cite{ning2021terahertz}  that 
first locates the best beam sector using wide beams and then resolves the fine-grained IRS beam in the sector using narrow beams. 
Nevertheless, the hierarchical training method requires frequent user feedback to refine the beam selection, hence incurring linearly scaling training overhead with the number of users. 
To address this issue, several new IRS beam training methods have been recently proposed  \cite{you2020fast,wang2021joint}. Specifically, a novel \emph{multi-beam training} method was devised in \cite{you2020fast}, where the IRS reflecting elements are divided into multiple sub-arrays to steer different beam directions simultaneously and each user can find its optimal IRS beam direction with high likelihood via simple received signal power/SNR comparisons over time. 
This multi-beam training method needs neither user feedback as in the hierarchical beam training, nor the exhaustive search as in the sequential beam training, thus significantly reducing the training overhead for IRS-aided multi-user systems. 
Besides, the authors in \cite{wang2021joint} proposed a \emph{random} beam training method for IRS-aided mmWave systems, where the multi-antenna BS, user, and IRS perform random beamforming, and the ML estimation method is used to resolve the AP/IRS-user angles-of-departure (AoDs)/angles-of-arrival (AoAs) from the received signals at each individual user, thereby avoiding frequent user feedback and reducing training overhead. 
While the above works considered the system with a single or multiple distributed IRSs, the authors in \cite{mei2021distributed} focused on a general multi-IRS system with inter-IRS signal reflection, for which the optimal beam training may result in formidably high complexity. By leveraging the quasi-static BS-IRS and inter-IRS channels as well as cooperative training, the authors proposed a new distributed beam training scheme with combined offline and online beam training, thus greatly reducing the complexity for practical implementation.

Although the channel between the BS and IRS remains largely static due to their fixed locations, the IRS/BS-user channels are generally dynamic and correlated over time due to user mobility. To avoid frequent beam training over time, various channel/beam tracking methods have been proposed in the literature for IRS-aided wireless systems, which can be roughly divided into the following three main categories \cite{zegrar2020general,you2021enabling,tian2021fast}. The first one is the extended Kalman filter based algorithm that models the IRS channels in adjacent time slots by the Markov process and uses a series of channel measurements to update channel estimation parameters over time  \cite{zegrar2020general}. 
The main issue of this algorithm lies in its potential beam misalignment due to the large number of IRS reflecting elements and hence sharp beams. The second method is the speed-estimation based channel tracking  \cite{you2021enabling} that firstly estimates the user's (angular) speed based on received signals and then predicts the AP/IRS-user AoA/AoD for designing the IRS passive beamforming. This channel tracking method may become inaccurate if the user's (angular) speed changes dramatically over time. The third category is the shortlisted beam-training method that selects a small number of candidate beam pairs for fast beam training. For example, the authors in \cite{tian2021fast} proposed to update a few candidate IRS beams for beam training when the user's received power is less than a threshold, by exploiting both the received signal strength and received signal angle difference in different IRS reflection configurations. This beam tracking method is mainly designed for the IRS-aided point-to-point communication system in high-frequency bands, which, however, may incur high training overhead in the multi-user setup.

\subsubsection{Deep-learning Based Reflection Design}
\begin{figure*}[!t]
	\centering
	\includegraphics[width=5in]{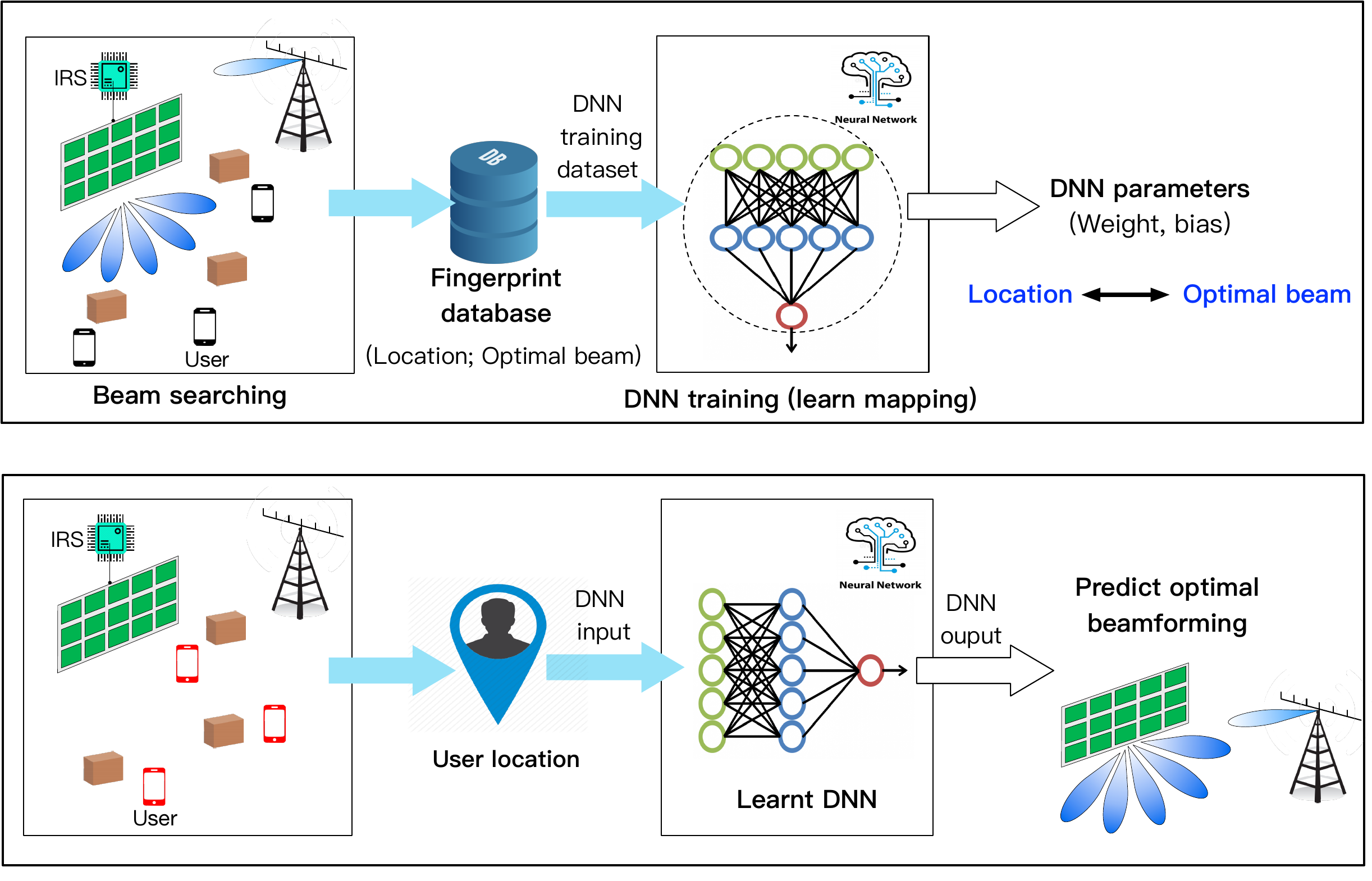}
	\caption{{End-to-end IRS reflection design.}}
	\label{ML}
\end{figure*}

Deep learning  techniques have been recently leveraged to design the IRS passive beamforming without  explicit CSI, by exploiting its advantages in learning the non-linear mapping from training data. In particular, the conventional approach is by treating the IRS channel estimation and passive beamforming design as two sequential and separate phases, and applying deep learning techniques for one or both phases (see, e.g., \cite{taha2021enabling,song2020unsupervised,feng2020deep}).
However, this two-phase design method may not be efficient for the IRS-aided wireless system due to the following reasons. First, 
deep-learning based IRS channel estimation aims to minimize the channel estimation error, which does not necessarily improve the IRS passive beamforming performance. Second, 
the channel estimation/learning phase may cause error propagation and hence performance degradation in the subsequent passive beamforming learning. Third, the two-phase deep learning method in general has a high computational complexity due to the massive number of IRS channel parameters  to be learned. To address the above issues, an alternative promising approach, called \emph{end-to-end IRS reflection learning}, has been recently proposed to directly learn the IRS passive beamforming design without channel estimation/learning (see, e.g., \cite{huang2019indoor,sheen2021deep,ozdogan2020deep,jiang2021learning,jiang2020learning,yang2020deep,yang2020intelligentlearning}).

Specifically, user location information was exploited in \cite{huang2019indoor} to directly learn the optimal IRS passive beamforming in the indoor environment. To this end, as shown in Fig.~\ref{ML}, a fingerprinting database was first constructed that collects the optimal IRS passive beamforming at prescribed user locations based on the exhaustive search. Then, the dataset was used to train a properly designed DNN for learning the mapping from user location to the optimal IRS passive beamforming. This method was further extended in \cite{sheen2021deep} to predict the achievable rate at any user location. However, the performance of IRS passive beamforming in practice is not merely determined by user location, but also other parameters such as the small-scale fading that cannot be fully characterized by the locations of the transmitter and receiver. Motivated by the above, the authors in \cite{ozdogan2020deep}  advocated to exploit the received pilots to learn the optimal IRS passive beamforming design for IRS-aided single-user systems, where the BCD optimization method was adopted to obtain the training data of IRS passive beamforming. Numerical results showed that this method can achieve a higher rate than that based on the LS channel estimation, because the former can directly learn the features of both the channel and IRS passive beamforming from the training dataset, while the latter lacks prior information of the channel and suffers rate performance loss due to channel estimation errors. This end-to-end IRS passive beamforming learning method was further extended in \cite{jiang2021learning,jiang2020learning}, where both the received pilots and user location information were utilized to learn the optimal IRS passive beamforming for IRS-aided multi-user systems.  
Specifically, the permutation invariant graph neural network (GNN) architecture was used in \cite{jiang2021learning} to capture the interactions among different users and directly learn to optimize both the BS's active beamforming and IRS's passive beamforming. As the joint beamforming for multi-user systems is difficult to handle,  unsupervised learning techniques were utilized to learn the GNN weights for network utility maximization.
The proposed method was shown to achieve comparable rate performance with the method based on perfect CSI and the BCD optimization, and also significantly reduce the training overhead as compared to the linear MMSE estimator. This indicates that the neural network can extract even more useful information than the explicit channel estimation for designing the IRS passive beamforming.

\subsubsection{Other Approaches}
Besides the above two approaches, there are also other methods that can be used to design the IRS passive beamforming without explicit CSI. For example, random IRS passive beamforming is an efficient approach that does not need any CSI
and thus is free of heavy channel estimation overhead, while it sacrifices the full passive beamforming gain 
\cite{tao2020intelligent,chaaban2020opportunistic}.
Specifically, the authors in  \cite{tao2020intelligent} considered an IRS-aided multicast system and proposed to generate random IRS reflections over time for reshaping the distributions of all users' channels. This method was shown to achieve a lower outage probability than the CSI-based passive beamforming scheme that requires large channel estimation overhead.
In \cite{chaaban2020opportunistic}, the authors proposed an IRS-aided opportunistic beamforming scheme, where the IRS reflecting elements induce time-varying random phases. Given the user feedback on their individual downlink SNRs, the BS employs a proportional fair  scheduling to maximize the average sum-rate by exploiting the multi-user diversity gain.
Besides, location information can also be utilized to design the CSI-free passive beamforming when the IRS is properly deployed to establish LoS paths with both the BS and users \cite{hu2020location,hu2021angle}. 
With imperfect user location information due to user mobility, the effective angles from the IRS to users can be estimated and used to design the BS's active beamforming  and IRS's passive beamforming. Moreover, heuristic algorithms can be adopted to design low-complexity IRS passive beamforming without CSI. For example,
the particle swarm optimization method was used in \cite{souto2020beamforming} to gradually search for the near-optimal IRS passive beamforming based on the received SNR at the user.

{In Table~\ref{Table_est}, we summarize the main approaches for the IRS passive beamforming design with no explicit CSI. In this case, various side information provides inference on the optimal IRS beam direction, such as the received SNRs over different beams, received training signals, as well as user location. Despite these initial works, several key design issues need to be tackled in future work. For example, it is interesting to study the deep-learning based reflection design in more complex scenarios, e.g., multi-cell networks, multipath environment, and multi-IRS aided networks. Second, how to efficiently utilize all side information (e.g., received beams, user location, training signals) for designing more efficient IRS beam training has not yet been investigated in the existing literature. Moreover, it is important to characterize the performance limit for different IRS reflection design approaches with no explicit CSI, and devise efficient methods to approach their respective limit.}

\section{IRS Hardware Constraints and Imperfections}\label{Hardware}
The early works on IRS have mostly assumed the ideal IRS/transceiver hardware models to simplify the designs of IRS channel estimation and passive beamforming. However, such ideal IRS hardware models may result in considerable performance loss in practice due to various hardware constraints and imperfections/impairments at both the IRS and transceiver. This has motivated substantial research efforts recently to study the practical IRS channel estimation and passive beamforming designs subject to different hardware constraints and imperfections/impairments, which are overviewed in this section.
\subsection{Discrete Reflection in Phase/Amplitude} 

While the ideal IRS reflection model with continuously adjustable phase-shift/amplitude is convenient for optimization and provides useful performance bounds, it is practically difficult to realize due to the high implementation cost for building high-resolution phase shifters/amplitude controllers. 
As such, it is more cost-effective to implement the IRS with discrete and finite phase-shift/amplitude levels 
that require only a small number of control bits for each element, e.g., two-level ($0$ or $\pi$) phase-shift control and/or two-level (reflecting or absorbing) amplitude control. {Let  $b_{\beta}$ and $b_{\theta}$ denote the number of bits for controlling the number of reflection amplitude and phase-shift levels, which are denoted by $K_{\beta}$ and $K_{\theta}$, respectively, with  $K_{\beta}=2^{b_{\beta}}$ and $K_{\theta}=2^{b_{\theta}}$. Then  the sets of discrete reflection amplitudes and phase shifts at each IRS element can be respectively expressed as
	\begin{align}
	\mathcal{F}'_{\beta} &= \{ \bar{\beta}_1, \cdots,  \bar{\beta}_{K_{\beta}}  \},  \label{eq:14}\\
	\mathcal{F}'_{\theta} &= \{\bar \theta_1, \cdots, \bar \theta_{K_{\theta}} \}, \label{eq:15}
	\end{align}
	where $0 \leq \bar{\beta}_m< \bar{\beta}_{m'}\leq 1$ for $1 \leq m<{m'}\leq K_{\beta}$ and $0\leq \bar \theta_l< \bar \theta_{l'}<2\pi$ for $1 \leq l<l'\leq K_{\theta}$.
	Compared to the ideal continuous reflection amplitude/phase-shift models, their quantized versions in \eqref{eq:14} and \eqref{eq:15}  greatly complicate the IRS channel estimation and passive beamforming designs
	due to its combinatorial nature.}
In \cite{you2020channel} and \cite{you2020intelligent}, the IRS  cascaded channel estimation problem was studied under the constraint of IRS discrete phase-shifts, where a near-orthogonal DFT-Hadamard based training reflection matrix was constructed by using proper quantization techniques to minimize the channel estimation error.
This work was later extended in \cite{yang2021channel}, where the authors used the element-wise BCD optimization method to refine the initial 
DFT-Hadamard based training reflection matrix for reducing channel estimation error. 
Besides,  the ON/OFF IRS channel estimation and training reflection design based on the two-level amplitude control were studied in \cite{yang2020intelligent,mishra2019channel}, which, however, generally suffer from substantial reflection power loss, as compared to that of the full-ON IRS with full signal reflection.

On the other hand, for the IRS passive beamforming design with discrete phase shifts, a straightforward approach is to exhaustively search over all possible phase-shift levels for optimizing the communication performance. However, this may incur prohibitively high computational complexity for IRSs with high-resolution phase shifts and a large number of reflecting elements \cite{xu2019discrete}.  More efficient approaches thus have been proposed in the literature to address this issue (see, e.g., \cite{wu2019beamforming,di2020hybrid,you2020channel,wu2019beamformingICASSP,xiu2021secrecy,zhao2021exploiting}). For example, the branch-and-bound method was applied in \cite{wu2019beamforming,di2020hybrid} to obtain a high-quality suboptimal solution with reduced computational complexity on average, but it still incurs exponential complexity in the worst case. To further reduce the complexity,  the relax-and-quantize technique was proposed in \cite{wu2019beamforming,you2020channel} to design suboptimal IRS reflections by firstly solving an approximate problem with relaxed continuous-phase constraints
and then applying the nearest phase-quantization method to the optimized phase shifts. Nevertheless, this approach may suffer performance loss arising from the round-off errors, especially when the resolution of each phase shift is not high. Besides, the element-wise BCD method was proposed in \cite{wu2019beamformingICASSP,xiu2021secrecy} to sub-optimally solve the NP-hard IRS discrete-phase optimization problem,
which was shown to achieve comparable rate performance with the above relax-and-quantize method.
Moreover, to address the difficulty in addressing the constraints of both discrete phase-shifts and amplitudes, a penalty-based method was proposed in \cite{zhao2021exploiting} that introduces auxiliary continuous variables for their discrete versions and imposes a penalty term for controlling their difference in optimization. In addition, the effects of low-resolution phase shifters on the passive beamforming performance were studied in \cite{xu2020reconfigurable,zhang2020reconfigurable}. It was shown that  
a $3$-bit  phase-shifter is able to achieve the full diversity order \cite{xu2020reconfigurable}, while 
a $2$-bit  phase-shifter is enough for achieving close rate performance to the continuous-value baseline  when the IRS size becomes large \cite{zhang2020reconfigurable}.

\begin{figure}[!t]
	\centering
	\includegraphics[width=0.4\textwidth]{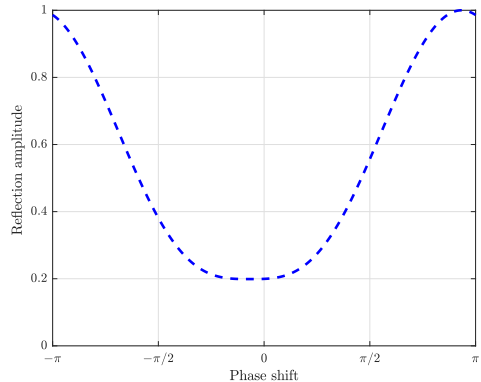}  
	\caption{{Reflection amplitude versus phase shift for the practical IRS reflecting element} \cite{abeywickrama2020intelligent}. } \vspace{-0.0cm}\label{amplitude:depend:phaseshift}
\end{figure}

\subsection{Reflection with Phase-dependent Amplitude}
Although the existing works on IRS have mostly assumed independent control between the IRS reflection amplitude and phase shift, it was shown in \cite{abeywickrama2020intelligent,costa2021electromagnetic} that the reflection amplitude of each IRS reflecting element is practically non-constant and non-linear with respect to (w.r.t.) its phase shift. {Specifically, as shown in Fig.~\ref{amplitude:depend:phaseshift} \cite{abeywickrama2020intelligent},  the IRS reflection amplitude typically attains its minimum value at the phase shift of zero, while it monotonically increases and asymptotically approaches the maximum value of one when the phase shift tends to $-\pi$ or $\pi$.} Such a phase-dependent amplitude control renders the practical IRS channel estimation and passive beamforming designs highly challenging, as the conventional approach
that independently optimizes the IRS amplitude and phase-shift is inapplicable. To address the difficulties in IRS channel estimation, 
the authors in \cite{yang2021channel} proposed a customized IRS training reflection pattern 
to minimize the channel estimation error by using the BCD optimization method, which was shown to achieve better performance in practice than the conventional training design assuming independent phase and amplitude control.

Given the CSI, different optimization methods have been proposed to design the IRS passive beamforming under the phase-dependent amplitude control model. Specifically, the element-wise BCD method was used in \cite{abeywickrama2020intelligent} to iteratively optimize each phase shift with the others being fixed, with its effect on the reflection amplitude  taken into account. This method was further applied in \cite{cai2020practical,li2021intelligent} to design the IRS phase-shift and phase-dependent amplitude for both the single- and multi-user MIMO-OFDM systems. Besides, the penalty-based method was applied in \cite{abeywickrama2020intelligent} to deal with the difficulty in the phase-dependent IRS amplitude control, by adding a penalty term associated with the IRS reflection values. Numerical results showed that for IRS-aided single-user systems, both the element-wise BCD and penalty-based methods achieved appealing communication performance, yet with low computational complexity.  

\subsection{Mutual Coupling Effect Among Reflecting Elements}

The sub-wavelength short distance between IRS reflecting elements inevitably causes mutual (circuit) coupling, where the impedance of each element is affected by those of its neighboring elements, leading to the intricately coupled reflection coefficients among reflecting elements \cite{di2020smart,bjornson2021reconfigurable}. 
This is in sharp contrast to the conventional IRS hardware model that assumes independent reflection control among different reflecting elements, thus making the IRS channel estimation and passive beamforming designs more involved. For the IRS channel estimation with mutual coupling, a key challenge  is how to acquire the element-wise cascaded channels separately and accurately. This problem has not yet been addressed in the existing literature, to the best of our knowledge. One possible and practical solution 
is by resorting to the element-grouping strategy \cite{yang2020intelligent,zheng2019intelligent}
that groups adjacent elements (typically with strong mutual coupling effect) into a subsurface, under which the subsurfaces may have mild mutual coupling with each other. As a result, 
only the subsurface-level channels (instead of element-level channels) 
need to be estimated, thus effectively circumventing the mutual coupling effect and reducing training overhead as well. However, more in-depth investigation is required to fundamentally understand how to properly group the adjacent elements for suppressing undesired mutual coupling while achieving satisfactory performance in the IRS channel estimation and passive beamforming design.

To facilitate the IRS passive beamforming design in the presence of mutual coupling, the authors in \cite{gradoni2021end} proposed an end-to-end electromagnetic-compliant communication channel model based on Maxwell's equations, which incorporates the effects of mutual coupling at the transmitter, receiver, and IRS. This impedance-based channel model resembles the communication-theoretic models \cite{wu2021intelligent} in terms of the received SNR, while it is more complicated to deal with due to the phase-dependent amplitude as well as the existence of self-impedance and mutual coupling.  
To tackle this difficulty, the authors in \cite{qian2021mutual} first considered the case with negligible mutual coupling and obtained a closed-form expression for the optimal tunable impedance. Then, the element-wise BCD method was adopted to design a high-quality suboptimal IRS passive beamforming for the general case with non-negligible mutual coupling. Numerical results revealed that the mutual coupling among IRS scattering elements significantly deteriorates the end-to-end SNR when their inter-distance is less than half wavelength. Moreover, the element-wise BCD method was later extended in \cite{abrardo2021mimo} to design the mutual-coupling-aware IRS passive beamforming in MIMO interference channels.

\subsection{Other Hardware Imperfections/Impairments}

Similar to the conventional wireless systems without IRS, there exist various transceiver/IRS hardware impairments in IRS-aided systems that may cause distortions in the system performance, such as IRS phase noise \cite{xing2021achievable,hu2018capacity,badiu2019communication}, transmitter/receiver RF impairments \cite{papazafeiropoulos2021intelligent,liu2020beamforming,shen2020beamforming,liu2020energy,boulogeorgos2020much,hemanth2020outage}, analog imperfection and quantization errors \cite{zhang2020cascaded,li2021passive,xiu2021uplink, xiu2021secrecy,zhi2020uplink},  amplifier non-linearity \cite{shaikh2021performance}, etc. In particular, the IRS phase noise caused by  IRS discrete phase and/or intrinsic hardware imperfection can be modeled as 1) uniformly distributed random noise at each element \cite{xing2021achievable} or 2) an additive Gaussian noise with the noise power increasing with its distance to the center of IRS accounting for the calibration effect \cite{hu2018capacity}. Besides, the joint effects of transmitter/receiver RF impairments, oscillator phase noise, and AGC noise can be characterized by the extended error vector magnitude model, where the transmitter/receiver hardware impairment is modeled as the zero-mean Gaussian noise with its variance proportional to the undistorted transmitted/received signal power \cite{schenk2008rf}. The practical IRS channel estimation and passive beamforming designs under the transceiver/IRS hardware impairments have been recently investigated in the literature (see, e.g., \cite{papazafeiropoulos2021intelligent,liu2020beamforming,zhang2020cascaded,li2021passive,xing2021achievable,badiu2019communication,shen2020beamforming,liu2020energy,boulogeorgos2020much,hemanth2020outage,zhou2020spectral,shaikh2021performance,xiu2021uplink, xiu2021secrecy,zhi2020uplink}).
Specifically, for channel estimation, the authors in \cite{papazafeiropoulos2021intelligent,liu2020beamforming} proposed 
a linear-MMSE based cascaded channel estimation scheme,
where the transceiver distortions were modeled by the Gaussian distribution and IRS phase-shift errors were modeled by a circular distribution.
In \cite{zhang2020cascaded,li2021passive}, the authors considered an IRS-aided receiver with low-resolution ADCs 
and proposed an efficient scheme to estimate the cascaded channel with the quantization error taken into account. However, the effects of various hardware imperfections/impairments have not been well characterized in the existing works on IRS channel estimation and thus deserve further investigation in the future.

As for the IRS passive beamforming under hardware impairments, 
the authors in \cite{xing2021achievable} analyzed the achievable rate of single-user systems in the presence of IRS phase noise, and showed that IRS hardware impairment degrades the achievable rate more severely when more reflecting elements are equipped. Moreover, it was revealed in \cite{badiu2019communication} that 
despite IRS phase-shift errors,  the IRS-aided communication system  can still achieve the square-scaling order in SNR and the linear-scaling order in diversity w.r.t. the number of reflecting elements, while the rate performance is deteriorated by the phase uncertainty.  
For the transceiver hardware impairments, a non-convex optimization problem was formulated and solved in \cite{shen2020beamforming} to maximize the achievable rate of an IRS-aided communication system by treating the transceiver hardware impairments as interference.
Besides, the ergodic and outage capacities of IRS-aided communication systems based on the extended error vector magnitude model were analyzed in \cite{liu2020energy,boulogeorgos2020much,hemanth2020outage}, which revealed that  the system capacity tends to saturate when the number of reflecting elements exceeds a threshold due to transceiver hardware impairments. Furthermore, the authors in \cite{zhou2020spectral} showed that the performance degradation at high SNR is mainly affected by the BS's hardware impairment rather than the phase noise arising from IRS discrete phase-shifts, since the IRS passive beamforming simultaneously affects the desired signal and distortion noise. In \cite{shaikh2021performance}, 
the authors considered the effects of the non-linear high power amplifier and showed that the outage capacity can be increased by mitigating the nonlinear distortion via operating the high power amplifier with back-off. 
In \cite{xiu2021uplink, xiu2021secrecy}, the effects of ADC quantization errors on the rate performance of IRS-aided systems were studied, where the authors jointly optimized the ADC quantization bits, IRS passive beamforming, and beam selection matrix for maximizing the user's achievable rate.
It was revealed that there is no need to use a large number of RF chains thanks to the enormous passive beamforming gain of IRS. 
In \cite{zhi2020uplink}, the authors aimed to derive the uplink achievable rate in the presence of  BS's quantization error and IRS phase noise. The analysis showed that if the number of IRS reflecting elements is large, the uplink rate performance is limited by the resolution of ADCs at the BSs, while the IRS phase noise only causes a constant rate loss.

\begin{table*}[!t]
	\centering
	\caption{{IRS Channel Estimation and Passive Beamforming Design under Hardware Constraints/Imperfections}}
	\vspace{-0.3cm}
	\resizebox{\textwidth}{!}{
		\begin{tabular}{|m{3.4cm}<{\centering}|m{4cm}<{\centering}|m{2cm}<{\centering}|m{19cm}|}
			\hline
			\textbf{Hardware Constraint/ Imperfection} &
			\textbf{Description} &
			\textbf{Practical Issue} &
			\textbf{\qquad\qquad\qquad\qquad\qquad\qquad\qquad\qquad Solution Approach} \\ \hline
			\multirow{6}{*}{\textbf{\begin{tabular}[c]{@{}c@{}}Discrete reflection \\ in phase/amplitude \end{tabular}}} &
			\multirow{6}{*}{\begin{tabular}[c]{@{}l@{}}Discrete and finite \\ amplitude/phase-shift\\ levels control.\end{tabular}} &
			IRS channel estimation &
			\begin{tabular}[c]{@{}l@{}}$\bullet$ Near-orthogonal IRS training reflection design subject to the discrete phase-shift constraint in narrowband \\
				system \cite{you2020channel,you2020intelligent}\\ $\bullet$ Cascaded channel estimation accounting for discrete phase-shift model in broadband system \cite{yang2021channel}\\ $\bullet$ ON/OFF training reflection pattern \cite{yang2020intelligent,mishra2019channel}\end{tabular} \\ \cline{3-4} 
			&
			&
			IRS reflection design &
			\begin{tabular}[c]{@{}l@{}}$\bullet$ Branch-and-bound method \cite{wu2019beamforming,di2020hybrid}. \\ $\bullet$ Relax-and-quantize technique \cite{wu2019beamforming,you2020channel}\\ $\bullet$ Element-wise BCD method \cite{wu2019beamformingICASSP,xiu2021secrecy}\\ $\bullet$ Penalty-based optimization method \cite{zhao2021exploiting}\\ $\bullet$ The effects of low-resolution phase shifters on the passive beamforming performance \cite{xu2020reconfigurable,zhang2020reconfigurable}\end{tabular} \\ \hline
			\multirow{4}{*}{\textbf{\begin{tabular}[c]{@{}c@{}}Phase-dependent\\  amplitude\end{tabular}}} &
			\multirow{4}{*}{\begin{tabular}[c]{@{}l@{}}Reflection amplitude\\ of each reflecting \\ element is a non-linear\\ function of its phase \\ shift.\end{tabular}} &
			IRS channel estimation &
			$\bullet$ AO-based IRS training reflection design subject to non-linear phase-dependent amplitude variation \\ \cline{3-4} 
			&
			&
			IRS reflection design &
			\begin{tabular}[c]{@{}l@{}}$\bullet$ BCD method \cite{abeywickrama2020intelligent,cai2020practical,li2021intelligent}\\ $\bullet$ Penalty-based method \cite{abeywickrama2020intelligent}\end{tabular} \\ \hline
			\multirow{4}{*}{\textbf{\begin{tabular}[c]{@{}c@{}}Mutual coupling\\ among elements\end{tabular}}} &
			\multirow{4}{*}{\begin{tabular}[c]{@{}l@{}}The impedance of each\\  element is affected \\ by those of its \\ neighboring elements\end{tabular}} &
			IRS channel estimation &
			$\bullet$ Circumvent the mutual coupling issue by resorting to the element-grouping strategy \cite{yang2020intelligent,zheng2019intelligent} \\ \cline{3-4} 
			&
			&
			IRS reflection design &
			$\bullet$ Element-wise BCD method \cite{qian2021mutual,abrardo2021mimo} \\ \hline
			\multirow{5}{*}{\textbf{\begin{tabular}[c]{@{}c@{}}Other hardware\\ imperfections/\\impairments\end{tabular}}} &
			\multirow{5}{*}{\begin{tabular}[c]{@{}l@{}}IRS phase noise, \\ transmitter/receiver \\ RF impairments, analog\\ imperfectness, and \\ quantization errors, etc.\end{tabular}} &
			IRS channel estimation &
			\begin{tabular}[c]{@{}l@{}}$\bullet$ Linear MMSE based cascaded channel estimation accounting for the IRS and transmitter hardware \\
				impairments \cite{papazafeiropoulos2021intelligent,liu2020beamforming}\\ $\bullet$ Cascaded channel estimation accounting for quantization error due to low-resolution ADCs at the receiver\\
				in IRS-aided MISO system \cite{zhang2020cascaded,li2021passive}\end{tabular} \\ \cline{3-4} 
			&
			&
			IRS reflection design &
			$\bullet$ Treat hardware impairments as distortion noise/interference in the reflection design and analyze the rate performance in the presence of IRS phase noise \cite{xing2021achievable,hu2018capacity,badiu2019communication}, transmitter/receiver RF impairments \cite{papazafeiropoulos2021intelligent,liu2020beamforming,shen2020beamforming,liu2020energy,boulogeorgos2020much,hemanth2020outage}, analog imperfectness and quantization errors \cite{zhang2020cascaded,li2021passive,xiu2021uplink, xiu2021secrecy,zhi2020uplink}, and amplifier non-linearity \cite{shaikh2021performance} \\ \hline
		\end{tabular}
	}
\end{table*}

\section{New IRS Architectures and Other Applications}\label{app}
In this section, we overview the practical designs for  new IRS architectures and other applications, 
which are important and deserve further investigation.

\subsection{New IRS Architectures}
\begin{figure}[!t]
	\centering
	\includegraphics[width=3.0in]{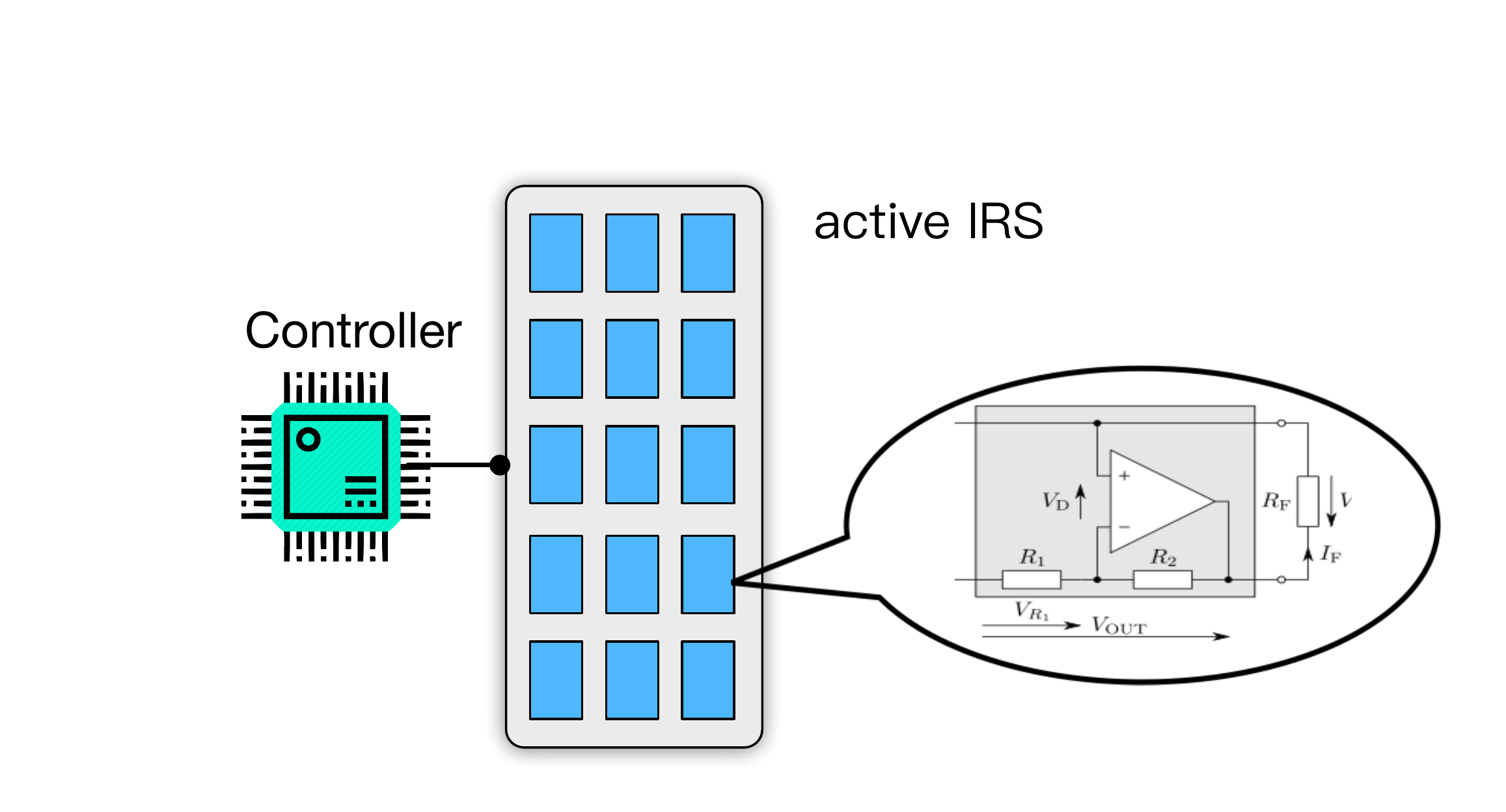}
	\caption{{Hardware architecture of active IRS.}}
	\label{active}
\end{figure}

\subsubsection{Active IRS}
The communication performance of passive-IRS aided  systems may be practically constrained by the well-known high \emph{(product-distance)} path-loss \cite{wu2021intelligent}, which can be compensated for by equipping the IRS with a large number of passive reflecting elements or reduced by deploying the passive IRSs close to the transmitter and/or receiver.
Alternatively, a new type of IRS, referred to as \emph{active} IRS as shown in Fig.~\ref{active}, has been recently proposed (see, e.g., \cite{long2021active,zhang2021active,You2021active,Khoshafa2021}) to address the issue of passive IRS, by amplifying the reflected signal with low-cost
\emph{negative resistance} components (e.g., tunnel diode and negative impedance converter), albeit at a modestly higher hardware and energy cost. 
To exploit both the reflective beamforming gain and power amplification gain offered by the active IRS, it is indispensable to acquire the CSI associated with the active IRS, which is  challenging for IRSs without sensing devices. 
This is because the active IRS requires not only the CSI of the cascaded transmitter-IRS-receiver link as the passive IRS, 
but also  the additional statistical information of the amplification-induced noise at the receiver. This makes the conventional cascaded channel estimation methods for passive IRS inapplicable, thus calling for new approaches catered to active IRS. Moreover, to achieve superior communication performance, the IRS active beamforming should be designed to strike the balance between increasing the amplified signal power and reducing the amplification noise power, which is worthy of further investigation in future work.

\subsubsection{Relaying IRS}
\begin{figure}[!t]
	\centering
	\includegraphics[width=3.0in]{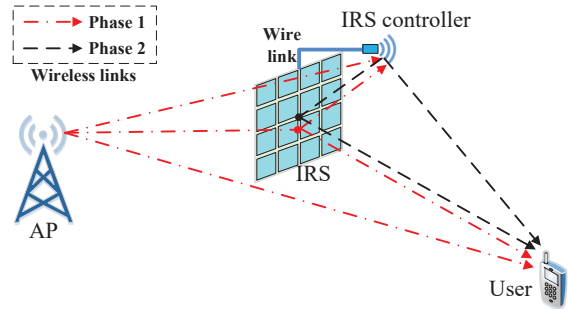}
	\caption{{A relaying IRS-assisted communication system \cite{zheng2021irs}.}}
	\label{RelayIRSsystem}
\end{figure}
The passive IRS and active relay in general have their own pros and cons in terms of energy/spectral efficiency, hardware/software complexity, coverage/serving range, etc. \cite{Emill2020Intelligent,Renzo2020Reconfigurable,Ye2021Spatially,gu2021performance,bazrafkan2021simple}. For instance, the passive IRS can cost-effectively improve the communication performance in its local coverage; in contrast, the active relay consumes more energy while achieving a broader coverage. To reap the complementary advantages of the passive IRS and active relay instead of treating them as two competing technologies, {the authors in \cite{zheng2021irs} proposed a novel relaying IRS architecture as shown in Fig.~\ref{RelayIRSsystem},} where the IRS controller is further exploited to actively relay the information for extending the coverage and enhancing communication performance. Note that this new relaying IRS architecture significantly differs from other recent works (see, e.g., \cite{obeed2021joint,Yildirim2021Hybrid,ying2020relay,kang2021irs,wang2021jointbeamforming,nguyen2021hybrid}) that consider adding  the active relay to the IRS-aided communication system, thus resulting in higher deployment and energy cost. However, the channel estimation for the relaying-IRS-aided communication system is more complicated than the conventional IRS-aided systems due to more channel coefficients to be estimated. To address this issue, a practical channel estimation scheme was proposed in \cite{zheng2021irs}, where the cascaded/direct CSI of the BS-relay (IRS controller) link and the relay-user link are estimated in parallel at the BS and user, respectively; while the CSI of the IRS-relay link is easily obtained by modelling it by the near-field LoS channel model due to the short distances between the relay (IRS controller) and reflecting elements. 
As the study on the relaying IRS is still in its infancy, more research efforts along this direction are needed for devising efficient channel estimation schemes as well as practical IRS passive beamforming and relaying designs, which are interesting problems to address in future work.

\subsubsection{Intelligent Refracting/Transmitting Surface (IRS/ITS)}
\begin{figure}[!t]
	\centering
	\includegraphics[width=3.0in]{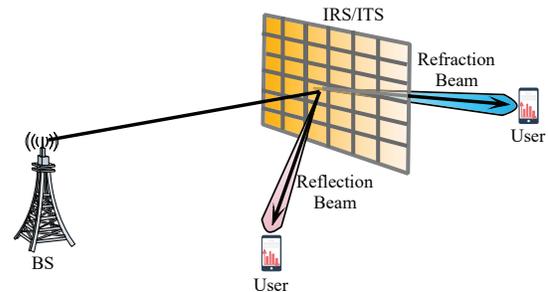}
	\caption{{An IRS/ITS-assisted communication system.}}
	\label{IRS_refraction}
\end{figure}
Most of the existing works on metasurface aided communication have considered the reflection-type metasurface due to its high reflection efficiency as well as low hardware complexity and cost. However, the reflection-type IRS can enhance the communication performance only when both the transmitter and receiver reside at the same side (i.e., the reflection half-space of metasurface). To expand the communication coverage, {the \emph{refraction}-type metasurface, also called \emph{intelligent refracting/transmitting surface}(IRS/ITS), can be employed to serve the transmitters and receivers located at its opposite sides as shown in Fig.~\ref{IRS_refraction},} albeit suffering a non-negligible signal penetration loss. Several new practical issues arise when deploying the refraction-type metasurface in the network. For example, it remains open whether the channel reciprocity still holds for signal refractions from different sides, which may have a significant impact on the ITS channel estimation design. Next, it is interesting to investigate how to deploy both the reflection- and refraction-type metasurfaces in the network to maximize the communication coverage and achieve optimal rate performance. Moreover, it is practically important to study the model of signal penetration loss w.r.t. different incident and refraction angles, as well as its impact on the communication performance.

\subsection{Other IRS Applications}
\subsubsection{IRS-Aided Wireless Power Transfer} 

RF wireless power transfer (WPT) is a promising approach to prolong the battery lives of IoT devices, whose efficiency, however,  is practically constrained by the small receiving aperture of IoT devices as well as the high power loss over distance. One efficient approach to tackle these issues is by properly deploying IRSs in the network to establish LoS links between the IRSs and the transmitter/receiver for reducing the power loss. Moreover, the  large IRS aperture and high passive beamforming gain can be exploited to enhance the received energy. Despite these appealing advantages, several practical issues need to be addressed in the IRS-aided WPT system. For example, it is crucial to acquire the accurate CSI of the IRS associated channels for designing efficient power transfer. Although some initial efforts have been made for estimating the channels in single-/multi-user WPT systems in frequency-flat channels \cite{mishra2019channel,mishra2020passive}, the IRS channel acquisition method in frequency-selective channels remains uncharted. Moreover, it is essential to balance the trade-off between the time for channel estimation and downlink power transfer to maximize the amount of harvested energy at the receiver. On the other hand, beam training is a practical IRS passive beamforming scheme  for WPT that does not require explicit CSI, while it may involve 
energy measurement feedback from the energy receiver \cite{jung2021rss}. 
Besides, under the practical non-linear energy harvesting model, IRS passive beamforming should be jointly optimized with the signal waveform to maximize the WPT efficiency \cite{zhao2021irs}, which needs further investigation.

\subsubsection{IRS-Aided Spatial Modulation}
Apart from passive beamforming,
IRS can also be used to transfer additional low-rate information by embedding/encoding implicit information onto its reflection pattern, which shares a similar concept with  ``spatial modulation" and thus referred to as ``reflection modulation" \cite{lin2020reconfigurable}. 
At the receiver side, the information from both the transmitter and IRS needs to be detected for achieving enhanced rate performance. Specifically, for coherent detection, it is indispensable to acquire the explicit CSI for differentiating the information mapped to different reflection states \cite{lin2020reconfigurable,karasik2020adaptive}. Attentive to this, 
the authors in \cite{karasik2020adaptive} proposed an MMSE-based cascaded channel estimation scheme  that  jointly designs the encoder of the transmitted signal and IRS reflection pattern.
On the other hand, for non-coherent detection, the information carried by the transmitted signal and IRS reflection pattern can be detected without CSI. 
For example,  differential reflection modulation schemes can be devised to jointly encode the permutation order of IRS reflection patterns and the phases of transmitted signals  \cite{guo2020differential}. 
Despite low complexity, the differential reflection modulation generally suffers some performance loss as compared to the reflection modulation with coherent detection. Thus, efficient schemes need to be designed to balance the trade-off between complexity and rate performance. 

\subsubsection{IRS-Aided Non-terrestrial Communications}
{As mentioned early, non-terrestrial communications, such as UAV communication and satellite communication, can be considered as
a promising solution to complement terrestrial communications.}
The performance of UAV communication systems is practically constrained by its size, weight, and power (SWAP) limitations, as well as occasional blockages in the UAV-ground channels. To overcome these drawbacks, a promising approach is by deploying terrestrial IRSs in the network to assist the UAV-ground communication. This helps bypass environmental obstacles more effectively by creating LoS links between the UAV and ground users through the IRS reflected links. Moreover, it allows the UAV to serve IRS-aided users without flying close to them and hence saving its propulsion energy consumption.  However, the IRS-aided UAV communication also faces new challenges in the designs of IRS channel estimation and passive beamforming. Specifically, the high-mobility of UAV renders its channel with the IRS much more dynamic, thus calling for efficient approaches to track UAV-IRS channels over time within a short channel coherence time \cite{cao2021reconfigurable}. Besides, the UAVs at high altitude may cause severe pilot contamination to the IRS-aided ground  users that reuse the same pilot, which thus inevitably deteriorates the channel estimation performance. On the other hand, with terrestrial IRSs, the UAV placement/trajectory needs to be jointly designed with the IRS passive beamforming \cite{you2021enabling}. For example, the UAV altitude can be greatly reduced if it can establish LoS links with the terrestrial IRS for serving its nearby users. Moreover, for high mobility UAVs with imperfect CSI, it is practically important to design the robust IRS passive beamforming and UAV trajectory \cite{guo2021learning}, which needs further investigation. {On the other hand, besides the UAV communication, the application of IRS to other types of airborne systems such as satellite communication is also a very new and interesting direction, which deserves further studies. 
For example, in \cite{zheng2022Satellite}, the authors considered a new IRS-aided satellite communication system with two-sided cooperative IRSs, where the communications between the satellite and various ground nodes in different applications/scenarios are aided by distributed IRSs deployed near them. Accounting for the high-mobility, the authors proposed an efficient cooperative beamforming design and a practical transmission protocol to conduct distributed channel estimation and beam tracking, demonstrating the substantial performance gain and great potential of the IRS-aided satellite communication.}

\subsubsection{IRS-Aided Physical-Layer Security}
For physical layer (PHY) security, IRS has emerged as a promising technology to enhance the secrecy rate by smartly reconfiguring the channels of both the eavesdropping and legitimate users with passive beamforming. This is particularly useful  in the challenging scenario of conventional systems without IRS, where the eavesdropping channel is stronger than the legitimate channel and/or they are highly correlated. In this case, IRS can be properly deployed with adaptively tuned passive reflection to increase/reduce the achievable rate of the legitimate/eavesdropping user. To achieve secure communication in IRS-aided PHY systems, CSI acquisition is indispensable but practically challenging, since the CSI of eavesdroppers may not be easy to obtain if they intentionally remain covert in channel estimation. Moreover, when conducting the IRS channel estimation for legitimate users, the eavesdroppers may inwardly learn/intercept the legitimate CSI and/or launch pilot spoofing/contamination attack to impair the channel estimation of legitimate users. These issues have been recently addressed in \cite{liu2021detect,huang2020intelligent,bereyhi2020secure,zheng2020uplink2}. For example, the authors in \cite{liu2021detect} proposed a three-step training scheme to detect the pilot spoofing attack as well as
acquire the cascaded CSI of both the legitimate and eavesdropping users. To combat the interception of eavesdroppers,  a cooperative channel estimation scheme was devised in \cite{bereyhi2020secure} to acquire the IRS-related channels, which were then used for designing the zero-forcing beamforming to reduce the information leakage to eavesdroppers. As IRS CSI acquisition (especially the CSI of eavesdroppers) may not be accurate in practice, it is necessary to design the robust IRS passive beamforming in IRS-aided PHY systems under the statistical (cascaded) CSI error model (see, e.g.,  \cite{hong2020robust,lu2020robust,yu2020robust,wang2020energy}). Besides, the authors in \cite{yan2021intelligent} employed the IRS to deliberately introduce extra randomness in the wireless propagation environment for hiding active wireless transmissions, and revealed the trade-off between the secrecy performance and CSI accuracy. Nevertheless, several practical issues in IRS-aided PHY systems have not been well tackled, e.g., the optimal design for detecting eavesdroppers, the trade-off between channel estimation and achievable secrecy performance, which deserve future studies.

\subsubsection{IRS-Aided Cognitive Radio}
IRS can be employed to improve the communication performance of cognitive radio networks by smartly steering IRS passive beamforming to suppress/cancel the interference from the secondary transmitter to the primary receiver as well as enhance the rate performance of the secondary receiver. To achieve this, it is essential for the secondary transmitter to acquire the CSI associated with the PU, which is practically challenging as there may be limited or even no dedicated feedback channel from the primary network to the secondary network. To tackle this challenge, the existing works (see, e.g., \cite{yuan2020intelligent,xu2020resource,zhang2020robust}) have assumed the TDD operation mode and thus all PU-associated CSI (including both direct and reflecting links) can be estimated by the secondary transmitter based on their observed signals from PUs thanks to the channel reciprocity. However, this approach is inapplicable to FDD systems, which needs further investigation. Moreover, another open problem is how to utilize IRS to perform spectrum sensing in  cognitive radio networks for detecting the spectrum availability and interference strength.

\subsubsection{IRS-Aided RF Sensing}
In conventional RF sensing systems, the multi-antenna BS can be employed as a MIMO radar to sense target locations in the network, which, however, may suffer a limited sensing range due to the severe round-trip path-loss over long distance and hence a small received power of the reflected echo signal. To expand the RF sensing range, an efficient approach is by densely distributing IRSs in the network to assist the RF sensing at the BS, where the IRSs can provide additional LoS reflection paths to sense the targets and enhance the power of echo signals. On the other hand, IRS can be directly utilized as a sensing infrastructure, where IRS reflecting elements reflect the signals 
transmitted by its nearby anchor nodes to the targets, and dedicated receiving sensors are installed at the IRS to receive echo signals from the targets. For IRS-aided RF sensing systems, one practical challenge is how to design the IRS passive beamforming to improve the sensing performance (e.g., increasing the target detection and AoA estimation accuracy) in the presence of direct echo links and various interferences from environment clutters \cite{buzzi2021foundations,jiang2021intelligent,aubry2021reconfigurable}. Moreover,  the performance limits of the IRS-aided sensing system remain unknown, especially under practical constraints, e.g., IRS discrete phase shifts. Furthermore, it is worth investigating how to exploit the dual role of IRSs to realize the integrated sensing and communication (ISAC) functions in future 6G networks.

\section{Conclusions}\label{con}
In this paper, we provide a comprehensive survey on the up-to-date research results for the new and emerging IRS-aided wireless communication systems, by focusing on the main practical challenges in their implementation, including IRS channel estimation/acquisition, passive beamforming/reflection design, and hardware constraints/imperfections.
Specifically, we first review the state-of-the-art results on IRS channel estimation under different IRS architectures and system setups, as well as the main signal processing methods used.
Next, for different practical scenarios of CSI availability, namely, imperfect instantaneous CSI, statistical/hybrid CSI, and no explicit CSI, we present a detailed overview of research results on their pertinent IRS reflection design and optimization.
Furthermore, practical hardware constraints and imperfections/impairments at both the IRS and wireless transceiver are reviewed   and their effects on the IRS channel estimation and reflection design are discussed in detail.
Finally, new IRS architectures and other IRS applications for wireless networks are outlined and their practical design challenges are highlighted to inspire future research.
It is hoped that this survey paper on IRS-aided wireless communications will provide a timely and useful guide for researchers and practitioners working on this innovative wireless technology for its efficient practical design and implementation in future wireless systems. 

\ifCLASSOPTIONcaptionsoff
  \newpage
\fi

\bibliographystyle{IEEEtran}
\bibliography{IRS_survey_Rev}

\end{document}